\documentclass[%preprint, 
 amsmath,amssymb,
 aps, physrev, prfluids,
]{revtex4-2}

\usepackage{graphicx}% Include figure files
\usepackage{dcolumn}% Align table columns on decimal point
\usepackage{bm}% bold math
\usepackage{newtxtext}
\usepackage{newtxmath}
\usepackage{mathrsfs}

\usepackage{graphicx}
\usepackage{subcaption}
\usepackage{booktabs}
\usepackage{multirow}
\usepackage{array}
\usepackage{tikz}
\usepackage[justification=raggedright,singlelinecheck=false]{caption}
\usepackage{natbib}

\usepackage{xspace}
\usepackage{siunitx}
\usepackage{comment}
\usepackage{xpatch}
\usepackage[normalem]{ulem}   % for \sout
\usepackage{xcolor}

\usepackage{hyperref}
\hypersetup{
    colorlinks = true,
    linkcolor  = blue,
    citecolor  = blue,
    urlcolor   = blue
}

\newcommand{\mainref}[2]{\hyperref[#1]{#2}}

\newcommand{\PDfirst}[2]{\frac{\partial #1}{\partial #2}}

\newcommand{\Dfirst}[2]{\frac{\mathrm{d} #1}{\mathrm{d} #2}}

\newcommand{\MATLAB}{\textsc{Matlab}\xspace}

\newcommand{\RNum}[1]{\uppercase\expandafter{\romannumeral #1\relax}}

\usepackage[mathlines]{lineno}% Enable numbering of text and display math

\begin{document}

\preprint{APS/123-QED}

\title{\textbf{The evolution of a gas current injected into a curved
axisymmetric porous channel} 
}% 

\author{Peter Castellucci}
\email{peter.castellucci@manchester.ac.uk}
\affiliation{%
Department of Mathematics, University of Manchester, Manchester M13 9PL, UK
}%

\author{Radha Boya}
\affiliation{%
Department of Physics and Astronomy, University of Manchester, Manchester M13 9PL, UK
}%

\author{Lin Ma}
\affiliation{%
Department of Chemical Engineering, University of Manchester, Manchester M13 9PL, UK
}%

\author{Igor L. Chernyavsky}
\affiliation{%
Department of\phantom{x}Mathematics, University of Manchester, Manchester M13 9PL, UK
}%

\author{Oliver E. Jensen}
\affiliation{%
Department of\phantom{x}Mathematics, University of Manchester, Manchester M13 9PL, UK
}%


\begin{abstract}
We investigate gas injection into axisymmetric water-saturated porous channels with Gaussian and parabolic profiles, as idealized models of underground gas storage in dome-shaped anticlines. Exploiting the slenderness of each channel, we derive an evolution equation for the gas/liquid interface using a composite asymptotic approximation that accommodates large channel slopes and has a simplified small-slope form describing spreading in weakly curved channels. In the high gas-mobility limit, in contrast with flat planar channels, buoyancy influences the dynamics through different mechanisms in each geometry. For gas injected steadily into a Gaussian channel, buoyancy can continually affect the flow due to the attenuation of the gas velocity caused by axisymmetry. In parabolic channels, the increasing channel slope ensures that buoyancy eventually influences the flow, at a timescale depending on injection rate and fluid properties. Asymptotic analysis of the parabolic-channel flow reveals five temporal regimes, each with multiple spatial regions and a distinct spreading rate, reflecting the evolving spatiotemporal competition between injection and buoyancy. Initially, a thin film of gas spreads along the upper boundary; the channel slope and elongation of the film then generate a hydrostatic pressure gradient, which strengthens until buoyancy arrests the upper contact line and thickens the film. Beneath the film, liquid then drains until the interface flattens under buoyancy. Analytical solutions of reduced-order models capture interface evolution and contact-line motion through each regime and are validated against full numerical simulations. These results have implications for subsurface hydrogen and CO$_2$ storage, where a horizontal interface that advances vertically enhances both safety and storage efficiency.
\end{abstract}

\maketitle

%\tableofcontents

\section{Introduction}
The transition to cleaner, renewable energy sources requires large-scale energy-storage solutions. Naturally occurring underground formations provide a promising option due to their extensive capacity, widespread availability, and versatility in storing different forms of energy. Energy may be stored underground in various forms: mechanical energy via compressed air \citep{kushnir_thermodynamic_2012}; thermal energy in heated fluids \citep{dudfield_periodic_2012}; and chemical energy as gaseous hydrogen \citep{zivar_underground_2021}. In the case of compressed air and hydrogen, surplus renewable electricity is used to compress air or produce hydrogen through electrolysis, with the stored energy later recovered when demand exceeds supply \citep{kushnir_thermodynamic_rev_2012, muhammed_review_2022}. Proposed subsurface storage sites include depleted oil and gas reservoirs, aquifers, and salt caverns \citep{muhammed_review_2022}. These formations are porous and saturated with brine. While horizontal or gently inclined geometries are the most common, curved anticlines offer particularly favourable conditions for the storage of buoyant gases such as hydrogen \citep{heinemann_hydrogen_2018}. Anticlines are formed by the upward flexure of sedimentary layers and are bounded above by impermeable strata known as caprocks, that act as natural traps for buoyant fluids. The location at which an anticline connects to a neighbouring aquifer is referred to as the spill point; gas that migrates beyond this location may become irrecoverable. A detailed understanding of the fluid dynamics governing current evolution through this geometry is therefore essential to prevent migration beyond the spill point. The fluid dynamics of injected hydrogen within such structures forms the main motivation for this study.

Spreading currents \citep{zheng2022} have been extensively studied in horizontally confined porous channels, particularly when the fluids are incompressible and motion is driven predominantly by injection or buoyancy, and resisted by viscous stresses. Within this setting, \citet{huppert_gravity-driven_1995}, \citet{pegler_fluid_2014}, and \citet{zheng_flow_2015} analysed injection into two-dimensional planar domains, showing that the flow structure and spreading rate depend sensitively on the viscosity ratio between the fluids. \citet{nordbotten_similarity_2006} and \citet{guo_axisymmetric_2016} demonstrated that evolution in axisymmetric configurations is further governed by a parameter quantifying the relative importance of buoyancy and viscous forces. Others have incorporated additional physical effects. \citet{zheng_self-similar_2019} investigated the influence of capillarity, showing that a sharp-interface limit is recovered for large Bond numbers or in monodisperse porous media. Inertial corrections, modelled via the Darcy–Forchheimer equation, were explored by \citet{majdabadi_farahani_darcyforchheimer_2024}, revealing modifications to the spreading dynamics at higher injection rates. \citet{Castellucci_2026} showed that the effects of gas compressibility can persist throughout the spreading process in long aquifers, thereby slowing the migration and reducing the gas pressure. Miscible gravity currents involving vertical flow have also been investigated by \citet{szulczewski_evolution_2013}.

Together, these studies provide a detailed understanding of the key physical mechanisms governing such flows across a wide range of injection scenarios. However, natural aquifers are seldom perfectly flat, and the interaction of these mechanisms with variations in geometry gives rise to richer and more complex flow behaviours. Such effects can have important implications for current evolution, and for the efficiency of trapping and storage of gases such as CO$_2$ and hydrogen. \citet{macminn_co2_2010} investigated the injection and subsequent migration of CO$_2$ in an inclined aquifer subject to background groundwater flow, incorporating the effects of capillary trapping. They showed that, following injection, the maximum storage efficiency, measured as the volume of CO$_2$ retained relative to the volume of aquifer occupied, occurs for a gentle downslope groundwater flow, owing to enhanced capillary trapping. This model was later extended in \citet{macminn_co2_2011} to account for solubility trapping, which was found to have a significant effect by slowing the current evolution and by greatly increasing the storage efficiency. \citet{gunn_flow_2011} studied continuous injection into an inclined aquifer with leakage faults located either upslope or downslope of the injection well. Their results highlighted how the geometry of the aquifer determines whether or not the gas current fills the channel height, which would influence storage efficiency. Motivated by the evolution of CO$_2$ currents beneath impermeable caprock, several studies have examined the influence of topography on the spreading of unconfined or deep flows. In this case, the motion of the resident fluid can be neglected to leading order, and the interface evolution is governed by a nonlinear diffusion equation. \citet{pegler_topographic_2013} analysed the effects of smooth topographic variations represented by a power-law profile, while \citet{di_low-dimensional_2025} considered more realistic, undulating curvilinear topographies.

Previous work on gas injection into anticlines includes a computational study by \citet{hagemann_mathematical_2015} and a theoretical analysis by \citet{mortimer_dynamic_2024}. \citet{hagemann_mathematical_2015} demonstrated that the flow behaviour depends strongly on the injection rate. At low injection rates, gravitational forces stabilise the displacement by flattening the interface. At high injection rates, viscous forces dominate, leading to fingering instabilities and rapid lateral spreading, allowing the current to extend beyond the spill point. \citet{mortimer_dynamic_2024} examined flow in the buoyancy-dominated regime, focusing on how the geometry and injection conditions influence the overall storage capacity. They considered injection into the lower layer of a two-layered anticline, where the layers are separated by a low-permeability mudstone barrier with a specified capillary entry pressure. They identified a critical injection rate which determines the storage efficiency. Below this rate, the current reaches sufficient depth that the hydrostatic pressure difference across the mudstone exceeds the capillary entry pressure. The current then begins to fill the upper layer, thus maximising storage. For injection rates above this threshold, the current spreads beyond the spill point and flows into a neighbouring aquifer. 

In the present work, we build on these studies by considering injection into a porous channel bounded by impermeable curved walls, as a simplified model of a dome-shaped anticline. We focus on a single-layered system and explore a broad range of flow regimes to examine how buoyancy, injection rate, viscous forces, and geometry govern the spreading rate and the evolution of the interface. Heterogeneity and anisotropy of the porous matrix, while undoubtedly significant in applications, are not considered here. Consideration of steady injection into Gaussian and parabolic channels reveals a variety of spreading dynamics. We exploit the slenderness of the aquifer to derive a single evolution equation for the interface location, requiring that gas--liquid interface slopes are small relative to the local aquifer orientation, or that buoyancy is strong enough to enforce a horizontal interface. To accommodate weakly-curved aquifers that may develop large slopes over long distances, we formulate this evolution equation using intrinsic coordinates, developing a composite asymptotic approximation that incorporates buoyancy effects exactly. We complement simulations of gas injection with asymptotic analysis of a simplified version of the composite model that provides explicit predictions of dominant features of the dynamics.

\section{Model formulation}
\subsection{Governing equations}\label{sec:GovEq}
\begin{figure}
    \centering
    \includegraphics[width = 0.7\linewidth]{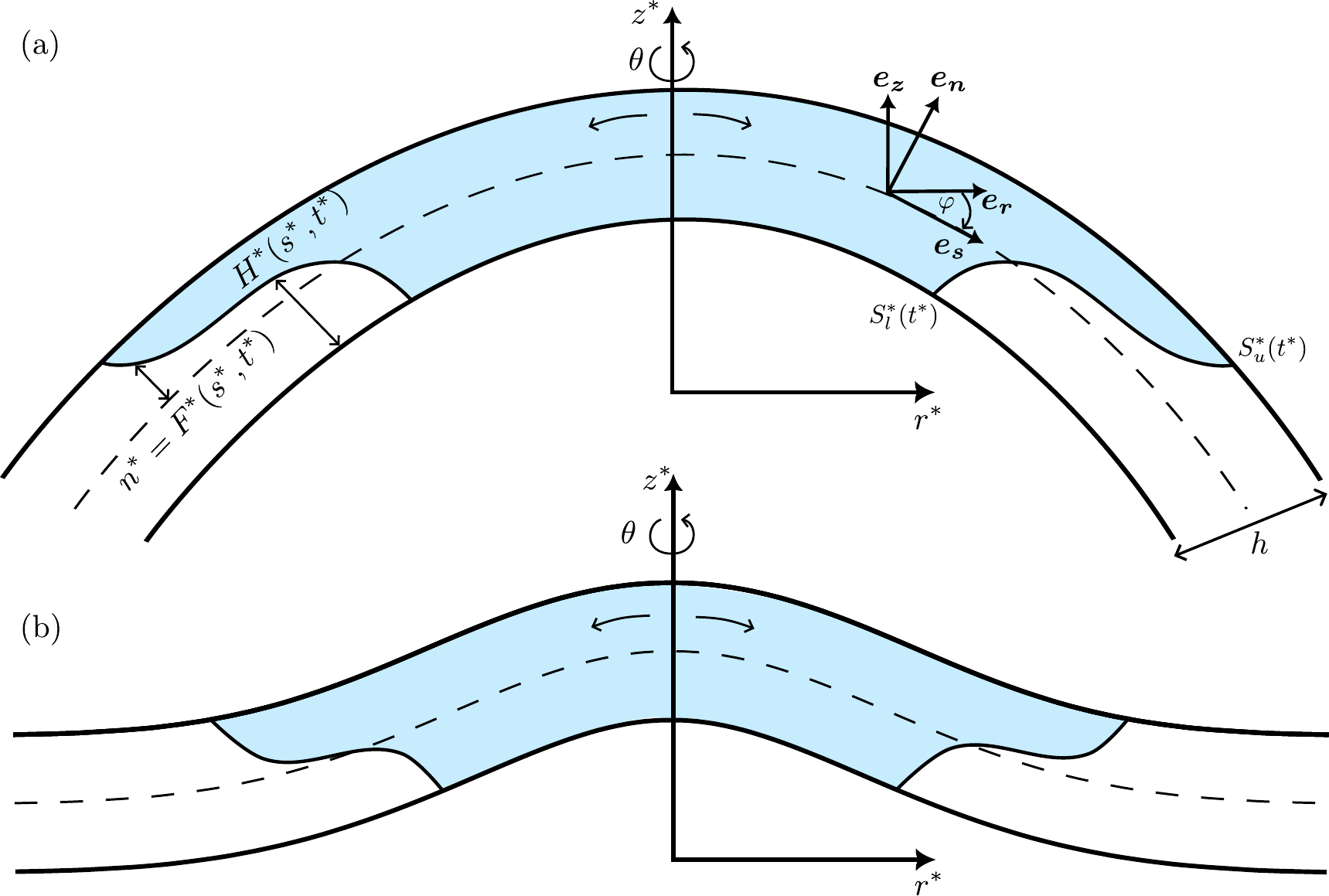}
    \caption{Schematic of the two thin channel geometries considered in this paper, defined by the curvature of their centrelines (dashed curves): (a) parabolic; (b) Gaussian. In (a), the basis vectors of the cylindrical reference frame, together with the curvilinear coordinate system $(s^*, n^*, \theta)$ used to formulate the model, are shown (the azimuthal basis vector $\boldsymbol{e_\theta}$, which points into the page, is omitted). The local inclination of the centreline to the horizontal, $\varphi$, is also indicated. Injection occurs from a point source at $r^* = 0$, $z^*=0$, and the light-blue region denotes the gas-filled domain. The intersections of the interface with the upper and lower channel boundaries occur at $s^* = S_u^*(t^*)$ and $s^* = S_l^*(t^*)$, respectively.}
    \label{fig:curvedSchematic}
\end{figure}
We consider the injection of a gas into an axisymmetric porous channel with curved boundaries, initially fully saturated with water [Fig.~\ref{fig:curvedSchematic}]. The channel has a fixed width $h$, which is assumed to be much smaller than the characteristic radial length scale $R$ of the channel. Its geometry is specified by the centreline curvature, with the boundaries located a distance $h/2$ from the centreline along the normal direction. The centreline curve is expressed as 
\begin{equation}\label{eq:shape}
    z^* = A \mathcal{F}(s^*/R),
\end{equation}
where $\mathcal{F}$ is a dimensionless shape function, $A$ is the characteristic vertical variation of the channel, and $s^*$ denotes arc length along the centreline. Stars denote dimensional variables. The curvature of the centreline in a plane of constant azimuthal angle $\theta$ is then given by
\begin{equation}\label{eq:curv}
    \kappa^*(s^*/R) = \frac{A}{R^2}
    \frac{\mathcal{F}^{\prime \prime}(s^*/R)}
    {\left(1-\left(A/R\right)^2 \mathcal{F}^{\prime}(s^*/R)^2\right)^{1/2}},
\end{equation}
for $\vert \mathcal{F}' \vert<R/A$. Below, we consider two distinguished limits: a weakly curved regime in which $A \sim h$ (where ``$\sim$'' denotes ``scales as''), so that the curvature scales as $h/R^{2}$ (the primary focus of the present study); and a finite-curvature regime in which $A \sim R$, for which the curvature scales as $1/R$.

We formulate the model in a local axisymmetric curvilinear frame $(s^*, n^*, \theta)$ aligned with the centreline of the channel [Fig.~\mainref{fig:curvedSchematic}{1(a)}], where $n^*$ is the normal distance from the centreline. The angle of the centreline from the horizontal is denoted $\varphi(s^*)$, which, for the geometries we consider, satisfies $-\pi/2 < \varphi \leq 0$ and $\varphi(0) = 0$. This curvilinear frame is obtained via a transformation from the cylindrical-coordinate description of the domain with coordinates $(r^*, \theta, z^*)$. Since the channel has uniform width $h$, the normal coordinate satisfies $-h/2 \leq n^* \leq h/2$. The curvature of the centreline is related to the angle to the horizontal via
\begin{equation}\label{eq:TurnAngle}
    \Dfirst{\varphi}{s^*} = \kappa^*(s^*).
\end{equation}
The position vector for a point in the channel, expressed in a Cartesian basis but parametrized by cylindrical coordinates $(r^*,\theta,z^*)$, is
\begin{equation}
    \boldsymbol{x}^*(r^*,\theta,z^*) = r^* \cos\theta \boldsymbol{e}_x + r^* \sin\theta \boldsymbol{e}_y + z^* \boldsymbol{e}_z.
\end{equation}
The coordinate basis vectors in cylindrical coordinates [Fig.~\mainref{fig:curvedSchematic}{1(a)}] are then defined by
\begin{equation}
    \boldsymbol{e}_r(\theta) = \PDfirst{\boldsymbol{x}^*}{r^*}, \quad  \boldsymbol{e}_\theta(\theta) = \frac{1}{r^*} \PDfirst{\boldsymbol{x}^*}{\theta}, \quad  \boldsymbol{e}_z = \PDfirst{\boldsymbol{x}^*}{z^*}.
\end{equation}
The position vector, expressed in the cylindrical basis but parametrized by the curvilinear coordinates $(s^*,n^*,\theta)$, is given by
\begin{equation}
\label{eq:Bound}
    \boldsymbol{x}^*(s^*,n^*,\theta) = r^*(s^*,n^*) \, \boldsymbol{e}_r(\theta) + z^*(s^*,n^*) \,\boldsymbol{e}_z.
\end{equation}
The elements of the Jacobian for the coordinate transformation can be written as
\begin{equation}\label{eq:Jac}
    \PDfirst{r^*}{s^*} = \cos\varphi, \quad \PDfirst{r^*}{n^*} = -\sin\varphi, \quad
    \PDfirst{z^*}{s^*} = \sin\varphi,  \quad \PDfirst{z^*}{n^*} = \cos\varphi.
\end{equation}
Integrating the normal derivatives yields
\addtocounter{equation}{1}
\begin{equation}\label{eq:JacInt}
    r^* = r^*(s^*,0) - n^* \sin\varphi, \qquad
    z^* = z^*(s^*,0) + n^* \cos\varphi,\tag{8$a,b$}
\end{equation}
(for $|n^*| \leq h/2$), which describe how the radial and vertical coordinates vary in the normal direction from the centreline. The tangential and normal unit vectors in the curvilinear frame are defined as
\begin{equation}
    \boldsymbol{e_s}(s,\theta) = \cos\varphi \boldsymbol{e_r}(\theta) + \sin\varphi \boldsymbol{e_z}, \quad \boldsymbol{e_n}(s,\theta) = -\sin\varphi \boldsymbol{e_r}(\theta) + \cos\varphi \boldsymbol{e_z},
\end{equation}
respectively, as illustrated in Fig.~\mainref{fig:curvedSchematic}{1(a)}. The volume element is $\mathrm{d} V^* = r^* \mathrm{d} s^*\mathrm{d}\theta \mathrm{d}n^*$.

The fluids are assumed to be incompressible and immiscible, separated by a sharp interface located at $n^* = F^*(s^*, t^*)$ at time $t^*$. The points of contact with the lower and upper boundaries are denoted by $S_l^*(t^*)$ and $S_u^*(t^*)$ respectively, such that $F^*(S_l^*,t^*) = -h/2$ and $F^*(S_u^*,t^*) = h/2$. The length of the interface, projected onto the centreline, is $\Delta S^* = S^*_u(t^*) - S_l^*(t^*)$. We extend the definition of $F^*$ to the whole domain by writing
\begin{equation}
    F^*(s^*,t^*)=\begin{cases} -h/2, & 0\leq s^*\leq S^*_l(t^*), \\
    h/2, & S^*_u(t^*) \leq  s^*. 
    \end{cases}
    \label{eq:f}
\end{equation}
We use the subscripts $g$ and $w$ to distinguish variables for the gas and liquid (water) phases respectively. The fluid velocities in each phase, $\boldsymbol{u}_i^* = u_i^*  \mathbf{e}_s + v_i^*  \mathbf{e}_n$ for $i = g, w$, satisfy
\begin{equation}
\label{eq:Darcy1}
\boldsymbol{u}_i^* = -\frac{k_0}{\mu_i} \left( \boldsymbol{\nabla}^* p_i - \rho_i g \sin\varphi \boldsymbol{e_s} - \rho_i g \cos\varphi \boldsymbol{e_n}\right), \qquad (i = g, w),
\end{equation}
which is Darcy’s law expressed in the curvilinear frame, where $g$ is the gravitational acceleration and $k_0$ is the intrinsic permeability of the medium. The densities $\rho_g$ and $\rho_w \equiv \rho_g + \Delta \rho$ and the viscosities $\mu_g$ and $\mu_w$ are taken to be uniform in each phase. 

Injection of gas occurs through a point source at the origin, taken to be the apex of the channel centreline. The continuity equations for the gas and liquid phases are
\addtocounter{equation}{1}
\begin{equation}\label{eq:Continuity1}
    \boldsymbol{\nabla}^* \cdot \boldsymbol{u}_g^* = q \delta^*(\boldsymbol{x}^*), \qquad \boldsymbol{\nabla}^* \cdot \boldsymbol{u}_w^* = 0, \tag{12$a,b$}
\end{equation}
where $\delta^*(\boldsymbol{x})$ is the Dirac delta function representing the point source and $q$ is the strength of the source. At the interface between the fluids, we apply the kinematic condition
\begin{equation}\label{eq:KinematicEq1}
        F^*_{t^*} + u^*_i F^*_{s^*} =  v^*_i,\quad \text{at} \quad  n^* = F^*, \quad (i = g,w),
\end{equation}
which relates the velocity of the interface to the fluid velocities for $S^*_l<s^*<S^*_u$. Subscripts $s^*$ and $t^*$ in Eq.~(\ref{eq:KinematicEq1}) denote partial derivatives, whereas subscripts $g$ and $w$ label phases. We neglect any capillary effects and assume that the pressure is continuous across the interface
\begin{equation}\label{eq:ContinuityOfPressure1}
        p^*_g = p^*_w \quad \text{at} \quad n^* = F^*, \quad\text{for}\quad S^*_l<s^*<S^*_u.
\end{equation}
The definition of $F^*$ in Eq.~(\ref{eq:f}) means that Eq.~(\ref{eq:KinematicEq1}) ensures no penetration through the lower boundary for $0 \leq s^* \leq S_l^*$ and through the upper boundary for $s^* \geq S_u^*$. No penetration through the lower boundary of the liquid-filled region and through the upper boundary of the gas-filled region requires that the velocities satisfy
\begin{equation}
\label{eq:NoPen}
    v_w^* = 0, \quad (n^* = -h/2, \, s^* > S_l^*), \qquad v_g^* = 0, \quad (n^* = h/2, \, s^* < S_u^*).
\end{equation}
Prior to injection, we assume that the initial gas bubble lies entirely above the lower boundary, so that the lower contact line has not yet formed. The initial configuration is taken to be in hydrostatic equilibrium, and hence the gas–liquid interface is initially horizontal. We defer specification of the precise functional form of the initial condition, since this depends on the particular choice of channel shape function $\mathcal{F}$ in Eq.~(\ref{eq:shape}). The gas volume is given by
\begin{equation}\label{eq:DimVol}
    V^*(t^*) = 2\pi \int_0^{S_u^*(t^*)} \int_{F^*(s^*,t^*)}^{h/2} r^*(s^*,n^*) \,\mathrm{d}n^* \mathrm{d}s^* .
\end{equation}

\subsection{Non-dimensionalization}\label{sec:LWL}
We rescale starred variables with $r^* = R r$, $z^* = Rz$, $s^* = Rs$, $n^* = h n$, $p^*_w = \Delta \rho g h p_w$, $p^*_g = \Delta \rho g h p_g$, $\boldsymbol{u}^*_g = \left(q/ (hR)\right) \, (u_g, \epsilon v_g)$, $\boldsymbol{u}^*_w = \left(q/ (hR)\right) \,(u_w,\epsilon v_w)$, $F^* = h F$, $t^* =  \left( h R^2 / q\right) \, t$, $V^* =  h R^2 V$, where $\epsilon = h/R$. The coordinate derivatives in Eq.~(\ref{eq:Jac})  then become 
\addtocounter{equation}{1}
\begin{equation}\label{eq:Jac2}
    \PDfirst{r}{s} = \cos\varphi, \quad \PDfirst{r}{n} = -\epsilon \sin\varphi, \quad
    \PDfirst{z}{s} = \sin\varphi,  \quad \PDfirst{z}{n} = \epsilon \cos\varphi.\tag{17$a$--$d$}
\end{equation}
Substituting the rescaled variables into Darcy's equations Eq.~(\ref{eq:Darcy1}), and projecting onto $\boldsymbol{e_s}$ and $\boldsymbol{e_n}$ directions, yields
\begin{subequations}\label{eq:Darcy2}
    \begin{align}
        u_g &= -\lambda \left( p_{g,s} + \frac{\rho_g}{\Delta \rho} \frac{\sin\varphi}{\epsilon}\right),  &&\epsilon^2 v_g = - \lambda \left(p_{g,n} + \frac{\rho_g}{\Delta \rho} \cos\varphi\right), \\
        u_w &= -\lambda \mathcal{M} \left(p_{w,s} +\frac{\rho_w}{\Delta \rho} \frac{\sin\varphi}{\epsilon} \right), &&\epsilon^2 v_w = -\lambda \mathcal{M} \left( p_{w,n} + \frac{\rho_w}{\Delta \rho} \cos\varphi \right),
    \end{align}
\end{subequations}
where $\rho_g/\Delta \rho$ and $\rho_w/\Delta \rho$ measure the hydrostatic pressure contributions in the gas and liquid phases, respectively, relative to the buoyancy force, and the remaining dimensionless parameters are
\begin{equation}
    \mathcal{M} = \frac{\mu_g}{\mu_w}, \qquad \lambda = \frac{ k_0 \Delta \rho g h^2}{q \mu_g}.
\end{equation}
Here, $\mathcal{M}$ is the viscosity ratio, and $\lambda$ is analogous to a Darcy–Rayleigh number \citep{homsy_viscous_1987}, measuring the relative strength of buoyancy to injection-induced viscous forces; similar nondimensional groups appear in other studies of axisymmetric confined flows \citep{guo_axisymmetric_2016, nordbotten_similarity_2006}. Substituting these rescaled variables into the conservation of mass equation % Eq.~
(\ref{eq:Continuity1}), yields
\begin{equation}\label{eq:Continuity2}
    (r u_i)_s + (r v_i)_n = \epsilon\, r v_i \kappa(s) + \epsilon \delta(\boldsymbol{x}) \mathcal{I}, \qquad \mathcal{I} = \begin{cases}
  1& i = g,\\
  0  & i = w.
\end{cases}
\end{equation}
The third term in Eq.~(\ref{eq:Continuity2}) accounts for the geometric variation of the coordinate volume element $ \mathrm{d}V = r(s,n)\,\mathrm{d}s\, \mathrm{d}n\, \mathrm{d}\theta$, which depends on the normal coordinate $n$ at fixed arc length $s$. In rescaled coordinates the boundary conditions at the interface Eqs~(\ref{eq:KinematicEq1}) and (\ref{eq:ContinuityOfPressure1}) become
\addtocounter{equation}{1}
\begin{equation}\label{eq:BCsF}
    F_t + u_i F_s = v_i, \quad (i = g,w), \quad  \text{and} \quad p_g = p_w \quad \text{at} \quad n = F, \tag{21$a,b$}
\end{equation}
and no penetration at the boundaries becomes
\begin{equation}
\label{eq:NoPenDimLess}
    v_w = 0, \quad \left(n = -1/2, \, s > S_l\right), \qquad v_g = 0, \quad \left(n = 1/2, \, s < S_u\right).
\end{equation}
The Darcy equations (\ref{eq:Darcy2}), conservation of mass equation (\ref{eq:Continuity2}) and boundary conditions~(\mainref{eq:BCsF}{21$a,b$}) and (\ref{eq:NoPenDimLess}) provide the complete dimensionless formulation of the model. In Eqs (\ref{eq:Darcy2}--\ref{eq:NoPenDimLess}), subscripts $s$, $n$ and $t$ denote partial derivatives, while $g$ and $w$ label phases.

\subsection{Thin-channel approximation}
In this section we consider a reduction of the dimensionless specification of the model when $\epsilon \ll 1$. For now, we proceed assuming that the channel has strong curvature, corresponding to $A\sim R$ in Eq.~(\ref{eq:shape}). The presence of the large gravitational terms in the Darcy equations renders the limit $\epsilon \ll 1$ singular. Consequently, the dynamics can be expected to exhibit multiple regimes: an early-time regime in which the channel slope and the angle to the horizontal $\varphi$ remain small and all terms balance in the tangential Darcy equations; and, in geometries where $\varphi$ does not remain small throughout the channel, a late-time regime in which buoyancy dominates the flow and flattens the interface. Motivated by this structure, we retain the gravitational terms in the reduced formulation and derive an evolution equation that admits the correct limiting behaviour in both regimes. Taking $\epsilon \ll 1$, we discard $v_g$ and $v_w$ in the normal Darcy equations (\ref{eq:Darcy2}), so that the pressure is hydrostatic across the channel; after integrating in the $n$-direction and applying continuity of pressure at the interface Eq.~(\mainref{eq:BCsF}{21$b$}), we recover the pressure fields 
\begin{equation}\label{eq:pressures}
    p_g = P(s,t) - \frac{\rho_g}{\Delta \rho} n \cos\varphi, \qquad p_w = P(s,t) +
    \left(F(s,t) - \frac{\rho_w}{\Delta \rho} n \right) \cos\varphi,
\end{equation}
for some scalar field $P(s,t)$ representing the centreline gas pressure. The tangential Darcy equations from Eqs~(\mainref{eq:Darcy2}{18$a,b$}) then become
\begin{subequations}\label{eq:Darcy3}
    \begin{align}
        u_g &= - \lambda\left[P_s - \frac{\rho_g}{\Delta \rho} \left( \kappa(s)n - \frac{1}{\epsilon}\right) \sin\varphi \right], \\
        u_w &= -\lambda\mathcal{M} \left[P_s + F_s \cos\varphi - \left(\kappa(s)\left(F-  \frac{\rho_w}{\Delta \rho}n\right) - \frac{\rho_w}{\epsilon \Delta \rho} \right) \sin\varphi   \right].
    \end{align}
\end{subequations}

Equation~(\mainref{eq:Jac2}{17$b$}) shows that the slenderness of the channel means that, to leading order, the radial coordinate is independent of $n$, so that $r(s,n) \approx r_c$, where $r_c = r(s,0)$ is the radial coordinate of the channel centreline. Neglecting terms at $O(\epsilon)$ in the conservation of mass equations (\ref{eq:Continuity2}) then yields, for $s > 0$,
\begin{equation}\label{eq:Continuity4}
    (r_c u_{i})_s + r_c v_{i,n} = 0,
\end{equation}
at leading order. In this limit the point source becomes a boundary condition applied at $s=0$. Integrating Eq.~(\ref{eq:Continuity4}) over $F \leq n \leq 1/2$ for $0 < s \leq S_u(t)$ for the gas phase and over $-1/2 \leq n \leq F$ for $S_l(t) \leq s$ for the liquid phase, applying the kinematic condition Eq.~(\ref{eq:KinematicEq1}) and no penetration condition Eq.~(\ref{eq:NoPen}), yields depth-integrated transport equations for the gas and liquid phases
\addtocounter{equation}{1}
\begin{equation}\label{eq:Transport}
    -r_c F_t + \left(\int_{F}^{1/2} r_c u_g \,\mathrm{d}n \right)_s = 0, \quad r_c F_t + \left(\int_{-1/2}^{F} r_c u_w \,\mathrm{d}n \right)_s = 0,\tag{26$a,b$}
\end{equation}
respectively. Integrating the sum of Eqs~(\mainref{eq:Transport}{26$a,b$}) then recovers a flux constraint
\begin{equation}\label{eq:Flux}
    \int_{F}^{1/2}  u_g \,\mathrm{d}n + \int_{-1/2}^{F}  u_w \,\mathrm{d}n = \frac{1}{2 \pi r_c},
\end{equation}
which balances the volume flux through the channel with the volume flux delivered by the source. Substituting Eqs~(\ref{eq:pressures}) and (\mainref{eq:Darcy3}{24$a,b$}) into the integrals in Eq.~(\ref{eq:Flux}), gives
\begin{subequations}\label{eq:FluxInt}
    \begin{align}
    \int_{F}^{1/2}  u_g \, \mathrm{d}n &= -\lambda \left[P_s + \frac{\rho_g }{\epsilon\Delta \rho} \sin\varphi - \frac{\kappa(s)\rho_g}{2\Delta \rho} \sin\varphi \left(\frac{1}{2}+F\right) \right] \left[\frac{1}{2} - F\right], \\[0.25\baselineskip]
    \int_{-1/2}^{F}  u_w \, \mathrm{d}n &= -\lambda \mathcal{M} \left[ P_s + \frac{\rho_w}{\epsilon \Delta \rho}\sin\varphi + F_s \cos\varphi - \frac{\kappa(s)\rho_w}{2\Delta \rho} \sin\varphi\left(F+\frac{1}{2}\right)\right]\left[F+\frac{1}{2}\right].
    \end{align}
\end{subequations}
Substituting Eqs~(\mainref{eq:FluxInt}{28$a,b$}) into Eq.~(\ref{eq:Flux}) and rearranging then yields an expression relating the pressure gradient $P_s$ to the location of the interface:
\begin{align}\label{eq:pressureGrad}
    P_s = & - \frac{1}{\frac{1}{2}-F + \mathcal{M}\left(F+\frac{1}{2}\right)}\Bigg\{ \frac{\sin\varphi}{\Delta\rho}\left[\rho_g \left(\frac{1}{2}-F\right)+ \rho_w \mathcal{M} \left(\frac{1}{2}+F\right) \right] \left[\frac{1}{\epsilon} - \frac{\kappa H}{2} \right]+\mathcal{M}\cos\varphi\left(\frac{1}{2}+F\right) F_s + \frac{1}{2\pi \lambda r_c} 
    \Bigg\}.
\end{align}
Combining the expression for the pressure gradient Eq.~(\ref{eq:pressureGrad}) with the depth-integrated velocity Eq.~(\mainref{eq:FluxInt}{28$b$}) and the transport equation for the liquid Eq.~(\mainref{eq:Transport}{26$b$}) yields an evolution equation for the interface position $F(s,t)$.  
A simpler formulation is obtained by introducing $H(s,t) = F(s,t) + 1/2$, which measures the normal distance of the interface from the lower channel boundary [Fig.~\mainref{fig:curvedSchematic}{1(a)}].  
In terms of $H$, the evolution equation, with the corresponding boundary and initial conditions, can be written as
\begin{subequations}\label{eq:Evo}
\begin{align}
    H_t &= \frac{\mathcal{M}}{2 \pi r_c}\left[\frac{ 2 \pi \lambda r_c H (1-H)\left(\left(\frac{1}{\epsilon} - \frac{\kappa H}{2} \right)\sin\varphi + H_s\cos\varphi \right) - H}{1-H+\mathcal{M}H} \right]_s, &&(s>S_l(t)) \\
    S_{u,t} &= \frac{1}{ 2\pi \mathcal{M} r_c} + \lambda \left[\left(\frac{1}{\epsilon} - \frac{\kappa}{2}\right) \sin\varphi +H_{s}\cos\varphi  \right],   &&(s = S_u), \\
    S_{l,t} &= \frac{\mathcal{M}}{ 2\pi r_c} - \mathcal{M} \lambda \left[\frac{\sin\varphi}{\epsilon}  + H_{s}\cos\varphi \right],  \qquad &&(s = S_l; S_l>0), \\
    2 \pi \lambda &r_c (1-H)H_s - 1 \to 0,  &&(s \to 0; H(0,t)>0), \\
    H &= \frac{1}{2} + F_0 - z(s,0),  &&(0 < s \leq S_u(0);\, t = 0).
\end{align}
\end{subequations}
 In Eq.~(\mainref{eq:Evo}{30$a$}), the term proportional to $\sin\varphi$ represents the hydrostatic pressure gradient along the channel centreline arising from variations in the vertical coordinate with arc length; the large $1/\epsilon$ factor reflects the strong variation of $z$ with $s$ when the channel slope $\varphi(s)$ is large. The term proportional to $\cos\varphi$ captures the contribution of buoyancy associated with hydrostatic pressure differences in the normal direction. The advective term accounts for the imposed injection. A local analysis of Eq.~(\mainref{eq:Evo}{30$a$}) near the contact lines yields the kinematic conditions Eq.~(\mainref{eq:Evo}{30$b,c$}), ensuring that there is no flux across the moving contact lines. 
Before the interface reaches the lower boundary at the apex of the channel ($H(0,t)>0$), the boundary condition Eq.~(\mainref{eq:Evo}{30$d$}) is applied. Once the interface moves away from the origin ($H(0,t)=0$ and $S_l {\geq } 0$), the no-flux boundary condition Eq.~(\mainref{eq:Evo}{30$c$}) is applied instead. Equation (\mainref{eq:Evo}{30$e$}) is the initial condition representing a flat interface; here $F_0$ is a constant defining the height of the interface above the channel centreline at $s=0$. Since the gas is incompressible, the total mass of gas is equivalent to its total volume; hence, volume conservation of gas ensures that
\begin{equation}
\label{eq:Mass}
    \frac{\mathrm{d} V}{\mathrm{d} t} =1, \qquad V(t) = 2 \pi \int_0^{S_u} r_c (1-H) \mathrm{d}s.
\end{equation}
We denote the initial volume of gas in the channel by $V_0$.

To close the governing equations (\ref{eq:Evo}), the geometry of the channel centreline must be specified. For a given centreline curvature $\kappa(s)$, the radial coordinate $r_c(s)$ and the local inclination angle to the horizontal $\varphi(s)$ can be computed as functions of centreline arc length, using
\begin{equation}\label{eq:GeomDerivs}
    \Dfirst{r_c}{s} = \cos\varphi, \qquad \Dfirst{\varphi}{s} = \kappa(s).
\end{equation}
Once these geometric quantities are determined, Eqs~(\mainref{eq:Evo}{30$a$--$e$}) can be solved to obtain the evolution of the interface from its prescribed initial condition. In this work, we consider two channel geometries: a weakly-curved parabolic channel which develops large slopes over long distances; and a weakly-curved Gaussian channel which becomes asymptotically flat over long distances. The corresponding geometric quantities for each case are given in Sec.~\ref{sec:Geom} below. Before specifying these, we derive a simpler version of Eqs~(\mainref{eq:Evo}{30$a$--$e$}) that is valid for small centreline curvature and shallow channel slopes.

\subsection{Small-slope limit}
When the channel slope is small --- either throughout the channel or near the apex of a channel --- a simplified form of the governing equations (\mainref{eq:Evo}{30$a$--$e$}) can be obtained. We introduce the scalings $\varphi = \epsilon \varphi^\dagger$, $z = \epsilon z^\dagger$, $\kappa = \epsilon \kappa^\dagger(s)$, under which Eq.~(\ref{eq:GeomDerivs}) shows that the radial coordinate of the centreline is approximately equal to its arc length, $r_c \approx s$. In this small-slope limit, the governing equations (\mainref{eq:Evo}{30$a$--$e$}) then reduce (at leading order) to
\begin{subequations}\label{eq:Evo2}
\begin{align}
    H_t &= \frac{\mathcal{M}}{2 \pi s} \left[\frac{ 2 \pi \lambda s H(1-H) (H_s + \varphi^\dagger) -   H}{1-H + \mathcal{M} H} \right]_s,  &&(S_l < s < S_u), \\
    S_{u,t} &= \frac{1}{ 2 \pi \mathcal{M} S_u} + \lambda (H_s + \varphi^\dagger),  &&(s = S_u), \\
    S_{l,t} &= \frac{\mathcal{M}}{ 2 \pi S_l} - \mathcal{M} \lambda (H_s + \varphi^\dagger),  &&(s = S_l;\, S_l > 0), \\
    2 \pi \lambda &s (1-H)H_s - 1 \to 0,  &&(s = 0;\, H(0,t)>0), \\
    H &= \frac{1}{2} + F_0 - z(s,0), \qquad &&(0 < s \leq S_u(0); t = 0).
\end{align}
\end{subequations}
The channel slope is prescribed via $\varphi^\dagger(s)$ in Eq.~(\mainref{eq:Evo2}{33$a$}); the term proportional to $(H_s + \varphi^\dagger)$ represents buoyancy effects. As in Eq.~(\mainref{eq:Evo}{30$a$}), the component proportional to $H_s$ arises from the normal component of buoyancy and the $\varphi^\dagger$-dependent term accounts for the tangential component of buoyancy. The advective term represents the effect of injection. In the limit $\varphi^\dagger \to 0$, Eqs~(\ref{eq:Evo2}) reduce to the governing equations for a flat axisymmetric channel studied by \citet{guo_axisymmetric_2016} and others. Details of the numerical methods used to solve Eqs~(\mainref{eq:Evo}{30$a$--$e$}) and (\mainref{eq:Evo2}{33$a$--$e$}) are provided in Appendix~\ref{sec:Numerics}.

\subsection{Geometry of the centreline}
\label{sec:Geom}
\subsubsection{Parabolic channel}
\label{sec:ParaGeom}
For the parabolic channel sketched in Fig.~\mainref{fig:curvedSchematic}{1(a)}, the shape function in Eq.~(\ref{eq:shape}) is defined by
\begin{equation}\label{eq:ParaShape}
z = -\epsilon \frac{r_c(s)^2}{2},
\end{equation}
which corresponds to the scaling $A \sim h$ in Eq.~(\ref{eq:curv}). Substituting Eq.~(\ref{eq:ParaShape}) into Eq.~(\ref{eq:curv}) yields the curvature of the channel centreline,
\begin{equation}\label{eq:ParaCurv}
\kappa(s) = - \frac{\epsilon}{\bigl(1 + \epsilon^2 r_c(s)^2\bigr)^{3/2}}.
\end{equation}
The curvature is therefore $O(\epsilon)$, even when $r_c = O(\epsilon^{-1})$. The arc length of a planar curve $z = g(r)$ is given by
\begin{equation}\label{eq:arc length}
s(r) = \int_0^r \sqrt{1 + \bigl(g^\prime(u)\bigr)^2}\,\mathrm{d}u.
\end{equation}
Substituting Eq.~(\ref{eq:ParaShape}) into Eq.~(\ref{eq:arc length}) yields the arc length for the parabolic channel,
\begin{subequations}
\begin{equation}\label{eq:ParaArc}
s = \frac{1}{2\epsilon}\left(\epsilon r_c \sqrt{1 + \epsilon^2 r_c^2} + \sinh^{-1}(\epsilon r_c)\right),
\end{equation}
whose inverse defines the radial coordinate $r_c(s)$ of the centreline. The centreline angle $\varphi(s)$ can be obtained by numerically integrating
\begin{equation}\label{eq:ParaAngle}
\Dfirst{\varphi}{s} = - \frac{\epsilon}{\bigl(1 + \epsilon^2 r_c(s)^2\bigr)^{3/2}}, \quad \text{with} \quad \varphi(0)=0.
\end{equation}
\end{subequations}
We combine Eq.~(\ref{eq:GeomDerivs}) with Eq.~(\mainref{eq:ParaArc}{37$a,b$}) to solve Eqs~(\ref{eq:Evo}) for flow through a parabolic channel. Although the curvature remains small across the entire domain, the local slope grows monotonically with arc length approaching $\varphi \to -\pi/2$ as $s \to \infty$. Consequently, the small-slope approximation Eqs~(\mainref{eq:Evo2}{33$a$--$e$}) is valid only while $\epsilon r_c \ll 1$. In this asymptotic regime, we introduce the small-slope rescalings $\varphi = \epsilon \varphi^\dagger$ and $\kappa = \epsilon \kappa^\dagger$, under which the leading-order variables specifying the geometry reduce to
\begin{equation}\label{eq:ParaSmallSlope}
z^\dagger = -\frac{r_c^2}{2}, \qquad
\kappa^\dagger \approx -1, \qquad
\varphi^\dagger \approx -s, \qquad
r_c \approx s.
\end{equation}
Substitution of $\varphi^\dagger$ into Eqs~(\mainref{eq:Evo2}{33$a$--$e$}) then completes the specification of the small-slope evolution model for a parabolic channel.

\subsubsection{Gaussian channel}
The second geometry we consider is a Gaussian channel whose centreline is everywhere weakly curved, as sketched in Fig.~\mainref{fig:curvedSchematic}{1(b)}. The channel geometry is prescribed through the shape function in Eq.~(\ref{eq:shape}), which we take to be $
z = -\epsilon \exp\left(-r_c(s)^2/2\right)$. Under the small-slope rescaling, this reduces at leading order to $z^\dagger = \exp\left(-s^2/2\right)$. With this choice, the leading-order curvature of the channel centreline may be written explicitly as
\begin{equation}\label{eq:CurvFun}
\kappa(s) \approx \mathcal{F}^{\prime\prime}(s) = \exp\left(-s^2/2\right)\left(s^2 - 1\right).
\end{equation}
The curvature is negative near the origin and changes sign at $|s|=1$, corresponding to a transition from a locally concave geometry towards an asymptotically flat channel. In contrast to the parabolic channel, the slope of the Gaussian channel grows until $s=1$, after which it decays exponentially and asymptotically approaches zero in the far field. The angle of the centreline to the horizontal is approximated by
\begin{equation}\label{eq:Bellturn}
\varphi^\dagger(s) = -s \exp\left(-s^2/2\right).
\end{equation}
Substitution of Eq.~(\ref{eq:Bellturn}) into Eqs~(\mainref{eq:Evo2}{33$a$--$e$}) completes the specification of the governing equations for spreading in a Gaussian-shaped channel.

\subsection{Summary of model formulation}
To summarise, we used a composite approximation to derive an evolution equation [Eq.~(\mainref{eq:Evo}{30$a$})] for the gas–liquid interface in a slender curved channel, expressed in coordinates local to the channel centreline. Under this approximation, the pressure is hydrostatic normal to the centreline, while gravitational forcing enters the tangential Darcy equation as a singular $O(1/\epsilon)$ term. This structure reflects the chosen scalings, in which variations in the vertical coordinate occur on the same lengthscale as the arc length.

Taking the small-slope limit of Eqs~(\mainref{eq:Evo}{30$a$--$e$}) yields the reduced system given by Eqs~(\mainref{eq:Evo2}{33$a$--$e$}), which forms the principal model analysed in this study. The accuracy of this approximation depends on the channel geometry. For a parabolic channel, as shown in Sec.~\ref{sec:ParaGeom}, the approximation requires $\epsilon r_c  \ll 1$, since large slopes develop far downstream of the injection point. Nevertheless, in a weakly-curved channel, Eqs~(\mainref{eq:Evo2}{33$a$--$e$}) continue to capture the essential physical mechanisms of the more general model in Eqs~(\mainref{eq:Evo}{30$a$--$e$}), even beyond its strict asymptotic regime. This is illustrated by Fig.~\ref{fig:FullEvo} and is discussed in more detail shortly. We therefore adopt the small-slope system Eqs~(\mainref{eq:Evo2}{33$a$--$e$}) as a reduced framework for elucidating the dominant spreading dynamics in curved channels and for interpreting the broader behaviour described by the composite model given by Eqs~(\mainref{eq:Evo}{30$a$--$e$}). In contrast, for a shallow-slope Gaussian channel, the approximation remains valid throughout the domain, since the slopes remain small.

In the remainder of the paper, we specialise to the limit $\mathcal{M} \ll 1$, characteristic of a gas, such as hydrogen, displacing a much more viscous liquid, and construct reduced-order asymptotic models for both parabolic and Gaussian channel geometries. Full details of the asymptotic analyses are provided in Appendix~\ref{App:ParaAsym} for a parabolic channel and Appendix~\ref{App:GaussianAsymp} for a Gaussian channel. In Sec.~\ref{sec:paraApproach} and Sec.~\ref{sec:GaussianApproach} we summarise the key steps of these approaches. In Sec.~\ref{sec:Results}, we compare numerical simulations of the full model Eqs~(\mainref{eq:Evo2}{33$a$--$e$}) with the predictions of the reduced-order models, highlighting the dominant physical balances.

\subsection{Parabolic channel: small-$\mathcal{M}$ limit}\label{sec:paraApproach}
\begin{figure}
    \centering
    \includegraphics[width=\linewidth]{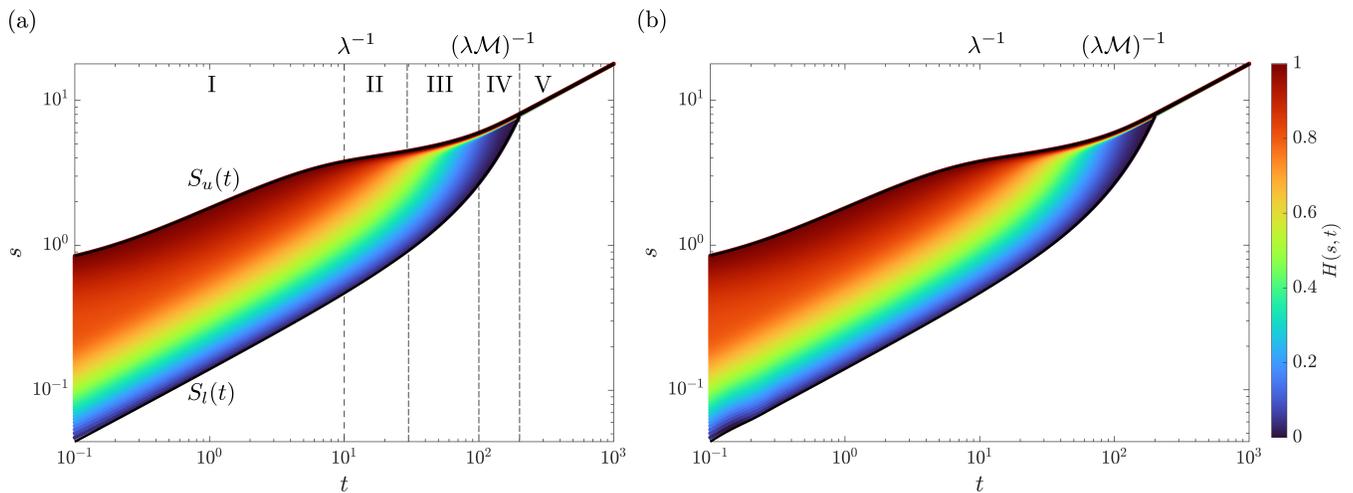}
    \caption{A numerical solution of (a) the small-slope model Eq.~(\mainref{eq:Evo2}{33$a$--$e$}) and (b) the composite model Eq.~(\mainref{eq:Evo}{30$a$--$e$}) with $\epsilon = 10^{-2}$, for $\mathcal{M}=0.1$, $\lambda=0.1$ and $F_0=0.8$ in a parabolic channel [Eq.~(\ref{eq:ParaSmallSlope})]. The evolution of the upper and lower contact lines is shown by the solid black curves. The colour map represents the interface height $H(s,t)$; for each time $t$, the vertical variation of colour indicates the spatial variation of $H$ across the interface, from $S_l$ to $S_u$. Five asymptotic regimes (\RNum{1}--\RNum{5}) are identified in (a). Times $\lambda^{-1}$ and $(\lambda \mathcal{M})^{-1}$ are also illustrated. To aid interpretation of the colour maps, film thickness and interface profiles representative of each of the five regimes are shown in Fig.~\mainref{fig:HRegions}{3}.}
    \label{fig:FullEvo}
\end{figure}

\begin{figure}
    \centering
    \includegraphics[width=0.75\linewidth]{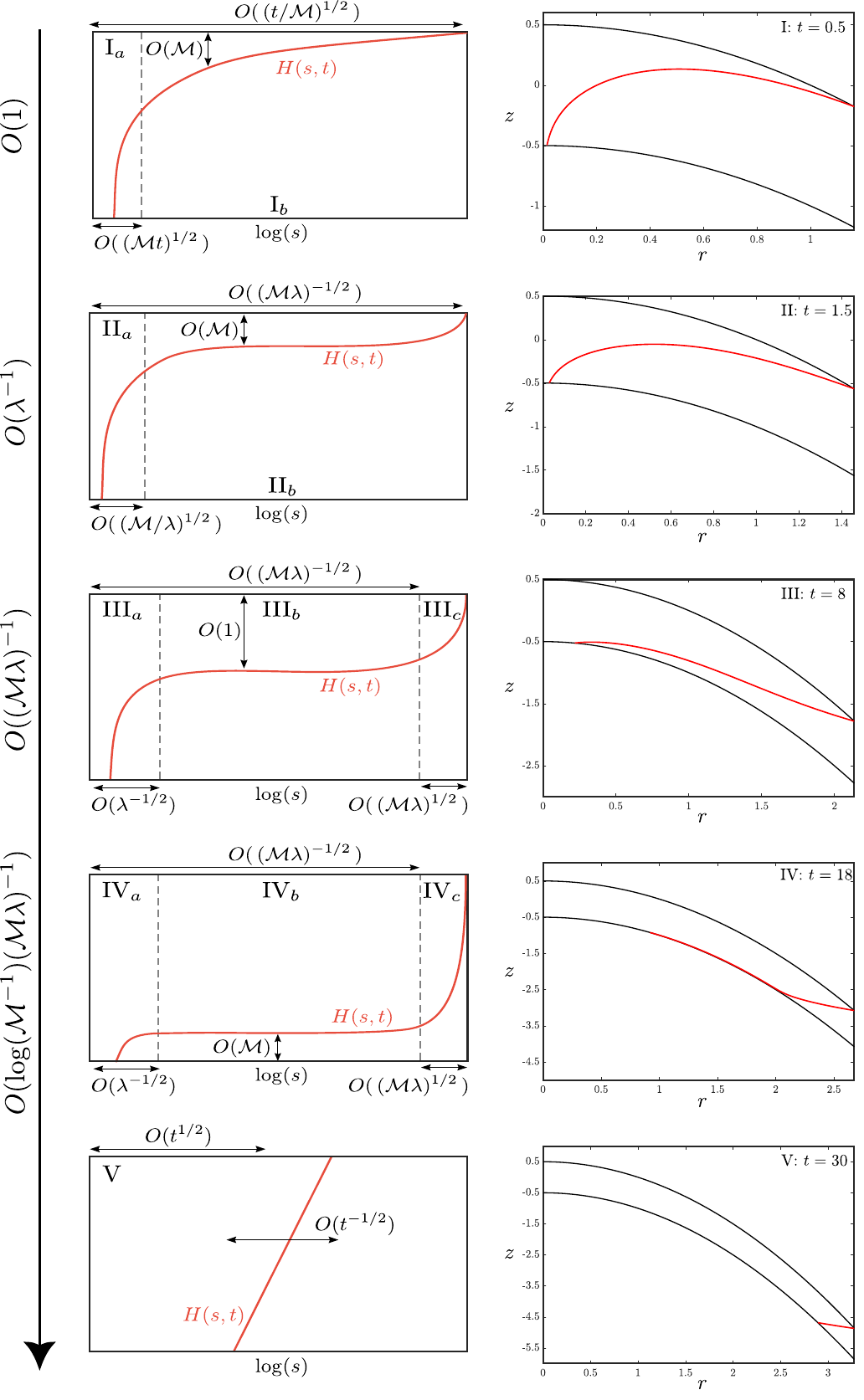}
    \caption{Illustration of five asymptotic regimes of spreading in a parabolic channel. Each row corresponds to one regime, with the timescales for each indicated by the vertical arrow. Column 1 shows schematics of the spatial regions of $H$ in the small-$\mathcal{M}$ limit with $\lambda \ll 1$, indicating the lengthscales of the asymptotic regions and the characteristic magnitudes of $H$. Column 2 shows solutions of Eqs~(\mainref{eq:Evo2}{33$a$--$e$}) for $\mathcal{M}=0.1$ and $\lambda=1$ in cylindrical coordinates, with black curves marking the parabolic channel boundaries.}
    \label{fig:HRegions}
\end{figure}
For a parabolic channel, when $\mathcal{M} \ll 1$ and $\lambda \ll 1$ in Eqs~(\mainref{eq:Evo2}{33$a$--$e$}) and (\ref{eq:ParaSmallSlope}), a number of distinguished limits arise in different spatiotemporal regions. To characterise these limits, we partition the solution into five distinct temporal regimes, labelled \RNum{1}–\RNum{5}. These regimes are illustrated by vertical dashed lines in Fig.~\mainref{fig:FullEvo}{2(a)} for a representative solution of the small-slope equations (\mainref{eq:Evo2}{33$a$--$e$}) with $\lambda = 0.1$ and $\mathcal{M} = 0.1$; Fig.~\mainref{fig:FullEvo}{2(b)} shows the corresponding solution of the full composite model [Eqs~(\mainref{eq:Evo}{30$a$--$e$})], computed using the complete geometric variables Eqs~(\ref{eq:ParaCurv}, \ref{eq:ParaArc}, \ref{eq:ParaAngle}) for $\epsilon = 10^{-2}$. The figures show the evolution of the upper and lower contact lines up to $t = 10^{3}$, with the colour map between the contact lines indicating the spatial variation of the interface height $H$ along the channel at each time. For these chosen parameter values, the solutions of the small-slope model and the full composite model are essentially indistinguishable. In regime~\RNum{1}, a long, thin gas film develops along the upper boundary, visible as a red region where $H$ is close to unity. In regime~\RNum{2}, the advance of the upper contact line $S_u$ slows and $H$ decreases within the film, indicating thickening of the gas layer. By regime~\RNum{3}, this region has thickened substantially, so that the gas- and liquid-filled layers are of comparable depth. In regime~\RNum{4}, $H$ is small across most of the domain, reflecting the formation of a thin liquid film that is draining along the channel's lower boundary. Finally, in regime~\RNum{5}, the lower contact line catches up with the upper contact line, after which both advance together; the interface then propagates as a steep front in $H$, corresponding to a flat interface in physical space.

The reduced-order models in Appendix~\ref{App:ParaAsym} reveal the physical mechanisms driving the dynamics within each regime, thereby explaining the dominant features observed in Fig.~\ref{fig:FullEvo}, as will be demonstrated in Sec.~\ref{sec:Results}. Figure~\ref{fig:HRegions} illustrates the timescales associated with regimes \RNum{1}–\RNum{5} and their hierarchical ordering. Within regimes \RNum{1}–\RNum{4}, the solution for $H$ exhibits multiple asymptotic spatial regions. We partition the domain into subregions labelled with alphabetic subscripts, ordered lexicographically from the origin; for example, \RNum{1}$_a$ denotes the region closest to the origin in regime~\RNum{1}. The first column of Fig.~\ref{fig:HRegions} shows schematics of the spatial regions, their characteristic lengthscales and magnitudes of $H(s,t)$. The second column presents a representative solution of Eqs~(\mainref{eq:Evo2}{33$a$--$e$}) mapped to cylindrical coordinates for $\mathcal{M} = 0.1$, $\lambda = 1$, and $F_0 = 0.8$, with each panel corresponding to a different time across the five temporal regimes. These parameter values were chosen to illustrate the typical interface shapes in the limits $\mathcal{M} \ll 1$ and $\lambda \ll 1$, while avoiding the long-range spreading that emerges when $\lambda \ll1$, so that the key features remain visible. Each row therefore demonstrates how the spatial structure of $H(s,t)$ (in intrinsic coordinates) maps to the interface shape in cylindrical coordinates. Within each spatiotemporal region, we derive reduced-order equations and asymptotic approximations for $H$, $S_u$, and $S_l$ (Appendix~\ref{App:ParaAsym}). Table~\ref{Tab:CLs1} summarises the resulting approximations for the contact lines across all regimes \RNum{1}–\RNum{5}, which are compared to numerical solutions of Eqs~(\mainref{eq:Evo2}{33$a$--$e$}) in Sec.~\ref{sec:Results}.

\subsection{Gaussian channel: small-$\mathcal{M}$ limit}
\label{sec:GaussianApproach}
For a shallow-slope Gaussian channel, we again focus on the small-viscosity-ratio limit, $\mathcal{M} \ll 1$, but consider a broader range of $\lambda$. Our asymptotic framework, outlined in Appendix~\ref{App:GaussianAsymp}, identifies a distinguished limit when $\beta \equiv \mathcal{M}^2 \lambda = O(1)$, in which buoyancy and injection balance; this limit serves as an organising region in parameter space. The dynamics separate into early- and late-time regimes. At early times, when $H(0,t) > 0$, the interface remains near the origin and is influenced by boundary curvature. Once the lower contact line forms ($S_l>0$) and $t \gg 1$, the solution develops an inner–outer structure: an inner region near $S_l$ and a more extended outer region containing $S_u$. In most cases, the inner region is arrested or evolves slowly, so it continues to feel the curvature of the lower boundary at late times. By contrast, the outer region consists of a rapidly spreading thin film that advances into the far field, where the channel becomes asymptotically flat.

In the outer region, the small-slope model given by Eqs~(\mainref{eq:Evo2}{33$a$--$e$}) admits a self-similar solution, reducing the governing partial differential equation to an ordinary differential equation (ODE). In Appendix~\ref{App:GaussOut} we show that the limiting cases $\beta \ll 1$ and $\beta \gg 1$ recover, respectively, the injection-dominated regime analysed by \citet{guo_axisymmetric_2016} and the buoyancy-dominated spreading regimes studied by \citet{lyle_axisymmetric_2005}, \citet{guo_axisymmetric_2016}, and others. Solutions of the resulting ODE for varying $\beta$ are presented in Sec.~\ref{sec:Results}, with numerical integration carried out using the shooting method described in Appendix~\ref{App:GaussOut}.

\section{Results} 
\label{sec:Results}
\begin{table*}
  \caption{Leading-order asymptotic approximations for the contact-line positions, and the associated timescales, in regimes \RNum{1}–\RNum{5} for spreading in a parabolic channel. The derivations of the approximations can be found in Appendix~\ref{App:ParaAsym}.
}
\begin{ruledtabular}
  \begin{tabular}{cccc} 
    \multicolumn{1}{c}{} &
    \multicolumn{1}{c}{Time} &
    \multicolumn{1}{c}{$S_l(t)$} &
    \multicolumn{1}{c}{$S_u(t)$} \\
    \cmidrule(lr){2-4}
    \RNum{1} & $t = O(1)$ &
    $\left( \frac{\mathcal{M} t}{\pi}\right)^{1/2}$ &
    $\left( \frac{t}{\mathcal{M}  \pi}\right)^{1/2}$ \\[10pt]
    \RNum{2} & $t = O(\lambda^{-1})$ &
    $\left( \frac{\mathcal{M} t}{\pi}\right)^{1/2}$ &
    $\left( \frac{1- \mathrm{e}^{-2 \lambda t}}{2\pi \mathcal{M} \lambda} \right)^{1/2}$ \\[10pt]
    \RNum{3} & $t = O((\mathcal{M}\lambda)^{-1})$ &
    $\left( \frac{\mathrm{e}^{2 \mathcal{M} \lambda t} - 1}{2\pi\lambda} \right)^{1/2}$ &
    $\left( \frac{t}{\pi (1-\mathrm{e}^{-2\mathcal{M} \lambda t})} \right)^{1/2}$ \\[10pt]
    \RNum{4} & $t = O(\log(\mathcal{M}^{-1})(\mathcal{M}\lambda)^{-1})$ &
    $\left( \frac{\mathrm{e}^{2 \mathcal{M} \lambda t} - 1}{2\pi\lambda} \right)^{1/2}$ &
    $\left(\frac{t}{\pi}\right)^{1/2} + \left(\frac{\pi}{t}\right)^{1/2}$ \\[10pt]
    \RNum{5} & $t \gg \log(\mathcal{M}^{-1})(\mathcal{M}\lambda)^{-1}$ &
    $\left(\frac{t}{\pi}\right)^{1/2} - \left(\frac{\pi}{t}\right)^{1/2} \left(\frac{\pi -V_0}{2\pi} \right)$ &
    $\left(\frac{t}{\pi}\right)^{1/2} + \left(\frac{\pi}{t}\right)^{1/2} \left(\frac{\pi +V_0}{2\pi} \right)$
  \end{tabular}
\label{Tab:CLs1}
\end{ruledtabular}
\end{table*}
% We performed a range of numerical simulations of the small-slope evolution equations~(\mainref{eq:Evo2}{33$a$--$e$}). 
In Sec.~\ref{sec:ParaResults}, we present results for a parabolic channel, focusing on the region of parameter space $\mathcal{M}<1$ and $\lambda\leq1$. In Sec.~\ref{sec:GaussianResults}, we present results for a small-slope Gaussian channel, covering the region of parameter space $\mathcal{M}<1$. For all simulations, the initial interface height was set to $F_0 = 0.8$.

\subsection{Parabolic channel}
\label{sec:ParaResults}
%The asymptotic approximations for the contact-line positions, summarised in Table~\ref{Tab:CLs1} for regimes \RNum{1}--\RNum{5}, are compared with numerical solutions of the small-slope equations~(\mainref{eq:Evo2}{33$a$--$e$}) below. Similarly, the asymptotic solutions for the interface $H(s,t)$ presented in Appendix~\ref{App:ParaAsym} are also compared with the numerical results.

\begin{figure}
    \centering
    \begin{subfigure}{0.49\textwidth}
        \caption{}
        \includegraphics[width = \textwidth]{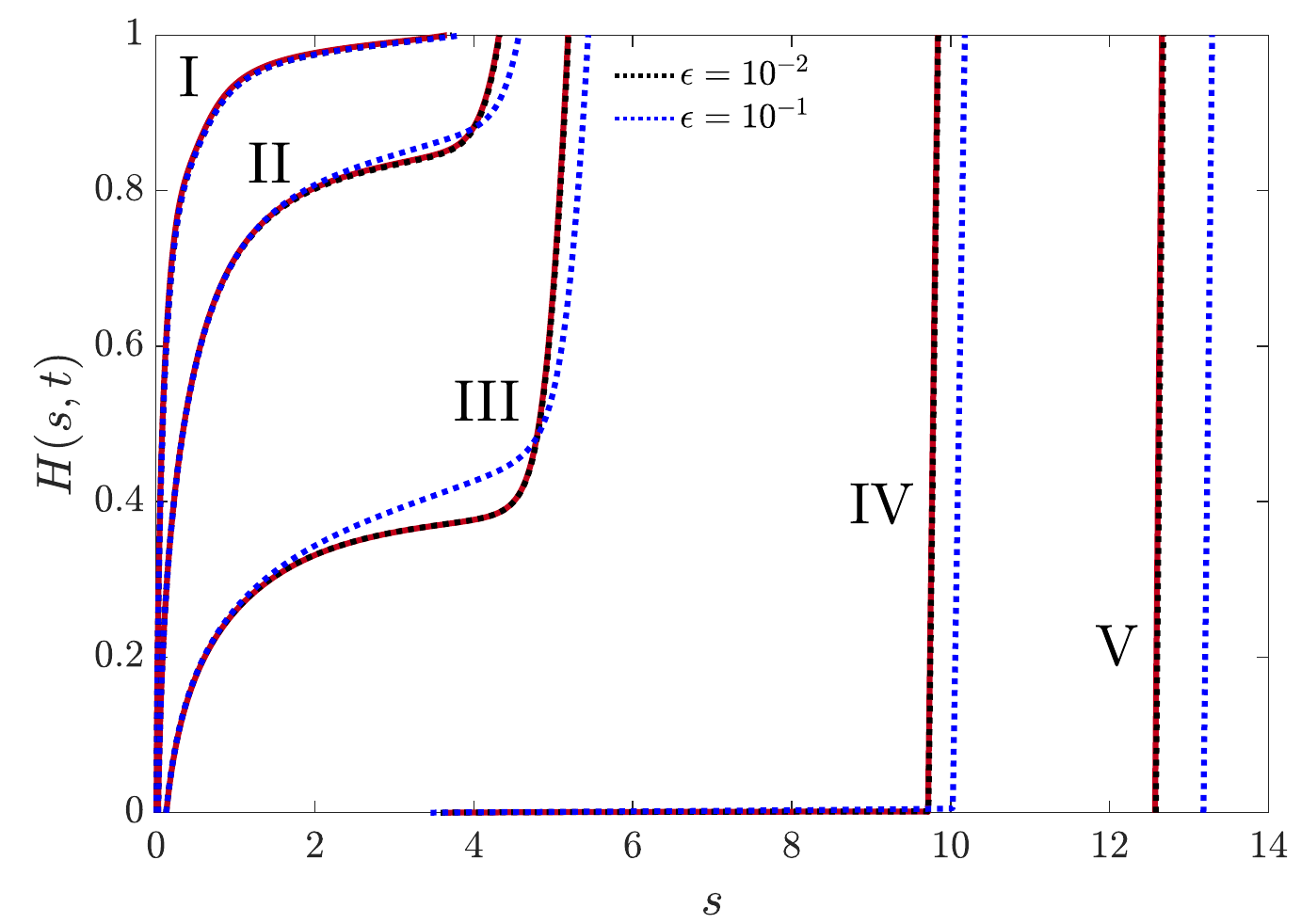}
    \end{subfigure} 
    \hfill
    \begin{subfigure}{0.49\textwidth}
        \caption{}
        \includegraphics[width = \textwidth]{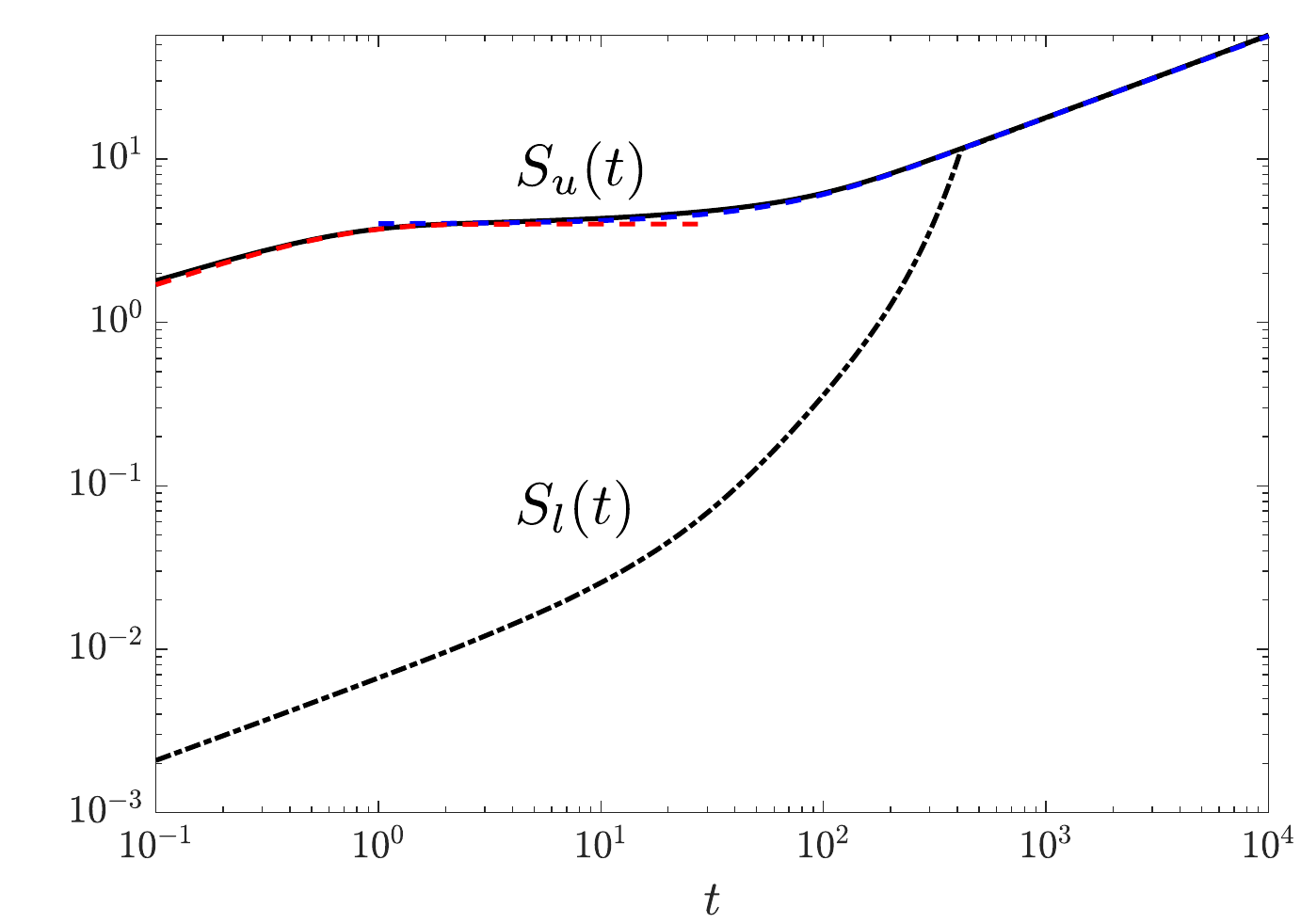}
    \end{subfigure}
    \caption{Numerical solutions of spreading in a parabolic channel for $\mathcal{M}=10^{-2}$, $\lambda=1$ and $F_0=0.8$.
(a) Evolution of the interface at times $t=\{1,10,50,300,500\}$. These profiles correspond to regimes \RNum{1}–\RNum{5}, respectively. Solid red curves show the solution of the small-slope equations~(\mainref{eq:Evo2}{33$a$--$e$}); dotted curves show the corresponding solution of the full composite equations~(\mainref{eq:Evo}{30$a$--$e$}) for $\epsilon=10^{-2}$ (black) and $\epsilon=10^{-1}$ (blue).
(b) Evolution of the upper and lower contact lines from the composite equations (\mainref{eq:Evo}{30$a$--$e$}) for $\epsilon = 10^{-2}$. Dashed red and dashed blue curves show the asymptotic predictions for regimes \RNum{2} and \RNum{3}, respectively (Table~\ref{Tab:CLs1}).}
    \label{fig:Lambda1Sol}
\end{figure}

We first present representative solutions of the small-slope model given by Eqs~(\mainref{eq:Evo2}{33$a$--$e$}) and the composite model given by Eqs~(\mainref{eq:Evo}{30$a$--$e$}), for $\lambda = 1$ and $\mathcal{M} = 10^{-2}$. Figure~\mainref{fig:Lambda1Sol}{4(a)} shows a comparison between the interface evolution predicted by the two models 
% small-slope model [Eqs~(\mainref{eq:Evo2}{33$a$--$e$})] and that of the full composite model [Eqs~(\mainref{eq:Evo}{30$a$--$e$})] 
with $\epsilon = 10^{-2}$ and $\epsilon = 10^{-1}$. For $\epsilon = 10^{-2}$, the two solutions are graphically indistinguishable over the time interval plotted. By contrast, for $\epsilon = 10^{-1}$, small discrepancies arise by $t=10$, with the error increasing at later times as the arc length $s$ grows. This behaviour is consistent with the requirement $\epsilon r_c \ll 1$ for the validity of the small-slope approximation. These comparisons confirm that Eqs~(\mainref{eq:Evo2}{33$a$--$e$}) capture the dynamics of the composite model when $\epsilon = h/R$ is sufficiently small. In particular, for small $\epsilon$, the small-slope model reliably reproduces the transition of dynamics across regimes \RNum{1}--\RNum{5}; however, errors inevitably accumulate in regime \RNum{5} once the interface extends sufficiently far that the channel slope is no longer small, as seen for $\epsilon = 0.1$.

% Having established this agreement, We now describe 
The evolution of the interface in Fig.~\mainref{fig:Lambda1Sol}{4(a)} follows the pattern illustrated in Figs~\ref{fig:FullEvo} and \ref{fig:HRegions}. Initially, a thin gas film develops along the upper boundary (regime \RNum{1}), analogous to the small-$\mathcal{M}$ spreading in flat geometries \citep{pegler_fluid_2014, guo_axisymmetric_2016}; however, unlike flat geometries, the motion is not dominated by injection alone. As the film elongates, a hydrostatic pressure gradient is established along its length, allowing buoyancy to increase the thickness of the film and slow the advance of $S_u$, which causes the nose to bulge (regime \RNum{2}; {see $t=10$ in Fig.~\mainref{fig:Lambda1Sol}{4(a)}}). In regime \RNum{3}, buoyancy begins to drive a slow, approximately spatially uniform redistribution of the interface over most of its length, progressively thinning the liquid layer beneath the gas. The injected volume flux of gas is then accommodated primarily by drainage in regions \RNum{3}$_a$ and \RNum{3}$_b$ [Fig.~\ref{fig:HRegions}], so that the flux reaching region \RNum{3}$_c$ near the upper contact line is negligible. This process continues until the injected volume is almost entirely accommodated by buoyancy-driven drainage, at which point the upper contact line becomes effectively stationary and $S_u(t)$ plateaus at 
$S_u = \left(1/(2\pi \mathcal{M}\lambda)\right)^{1/2}$, as observed in Fig.~\mainref{fig:Lambda1Sol}{4(b)}. Meanwhile, the lower contact line continues to advance under the combined effects of injection and buoyancy. 
The plateau of $S_u(t)$ is predicted by the regime-\RNum{2} approximation for $S_u(t)$ for times $t \gg \lambda^{-1}$ and by the regime-\RNum{3} solution for $\lambda^{-1} \ll t \ll (\mathcal{M}\lambda)^{-1}$ (see Table~\ref{Tab:CLs1}). 
Both approximations agree well with the full numerical solution in Fig.~\mainref{fig:Lambda1Sol}{4(b)}, within their respective ranges of validity. Once the liquid layer beneath the interface has drained, leaving only a thin film of thickness $H = O(\mathcal{M})$ along the lower boundary, the spreading transitions to regime~\RNum{4}. 

\begin{figure}
    \centering
        \includegraphics[width = 0.75\linewidth]{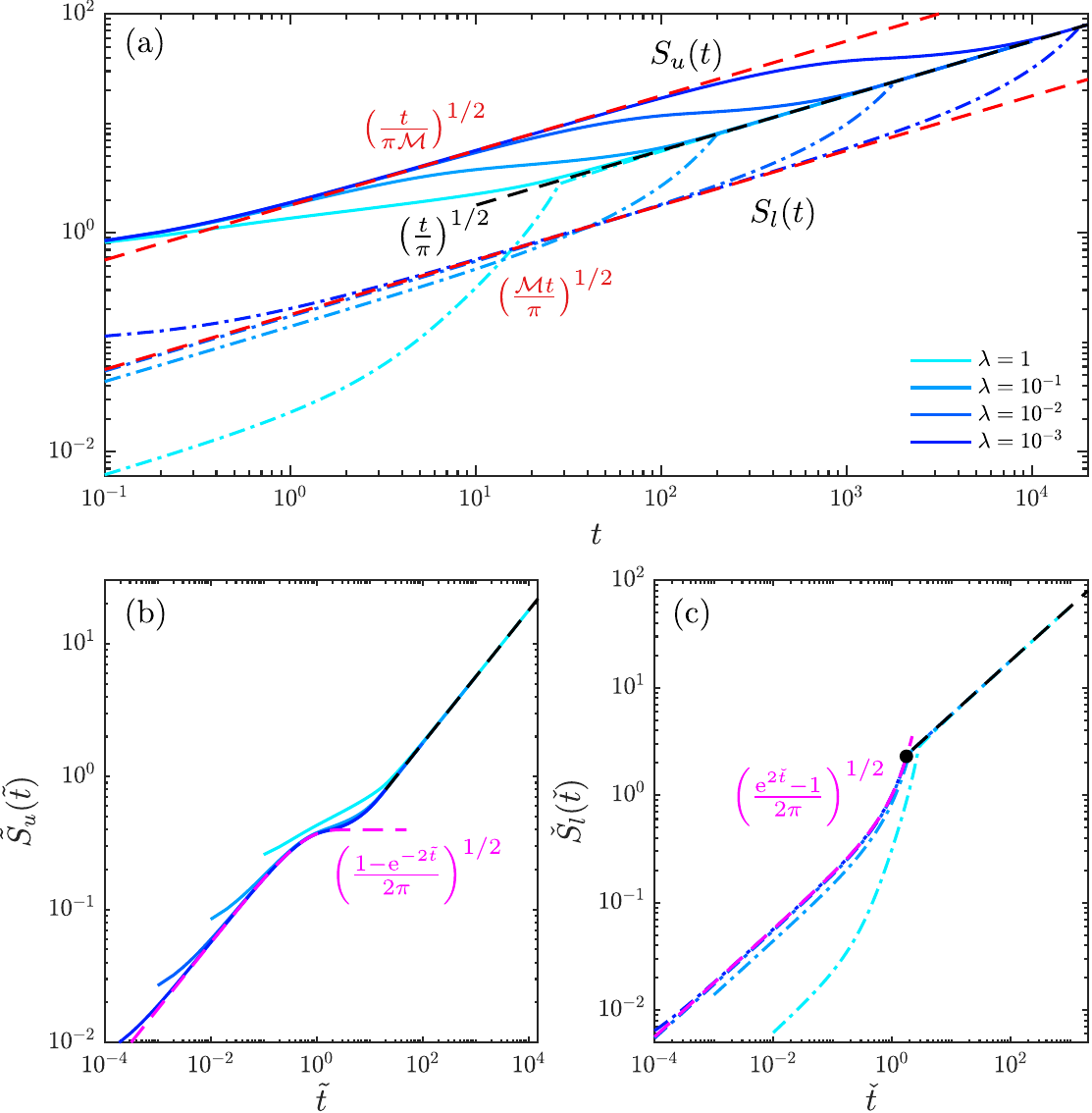}
    \caption{Evolution of the contact lines in a parabolic channel, obtained by numerically solving Eqs~(\mainref{eq:Evo2}{33$a$--$e$}) for $\mathcal{M} = 0.1$ and $F_0 = 0.8$ at decreasing values of $\lambda$. In all plots, solid lines and dashed-dotted lines correspond to the numerical solution for $S_u$ and $S_l$ respectively. Dashed lines indicate approximations from the different asymptotic regimes with each approximation labelled and matched by colour. In (a), the approximations from regime \RNum{1} for the contact lines (Table~\ref{Tab:CLs1}) are shown in dashed red, and the asymptote for both contact lines predicted by regimes \RNum{4} and \RNum{5} [Eq.~(\ref{eq:R4Su})] is shown in dashed black. Panel (b) replots the same data for $S_u$ from (a) in the coordinates of region \RNum{2}$_b$, $\tilde{t} = \lambda t$ and $\tilde{S}_u = (\lambda \mathcal{M})^{1/2} S_u$, with the dashed pink curve representing the exponential approximation for $S_u$ in this regime [Eq.~\ref{eq:R2Su}]. Panel (c) replots the same data for $S_l$ from (a) in the coordinates of regions \RNum{3}$_a$ and \RNum{4}$_a$, with $\check{t} = \mathcal{M} \lambda t$ and $\check{S}_l = \lambda^{1/2} S_l$. The dashed pink curve represents the approximation in Eq.~(\ref{eq:R3Sl}), while the black dot indicates $t_c$, the solution of Eq.~(\ref{eq:R5t}), marking when $S_l$ catches up with the moving front. Beyond this point, the lower contact line propagates with the front according to Eq.~(\mainref{eq:R5CLs}{B43$a$}), shown in dashed black.}
        \label{fig:CLSfixedM}
\end{figure}

In regime \RNum{4}, sufficient gas flux is delivered to region \RNum{4}$_c$ [Fig.~\ref{fig:HRegions}] for the upper contact line to resume its motion ({see $t=300$ in Fig.~\mainref{fig:Lambda1Sol}{4(a)}}). In this region, a steep propagating front in $H$ (corresponding to a horizontal interface in physical space) connects the thin film to the upper boundary and advances at a rate determined by the injected volume (see Table~\ref{Tab:CLs1}). Figure~\mainref{fig:Lambda1Sol}{4(b)} shows that the regime-\RNum{3} approximation for $S_u$ (see Table~\ref{Tab:CLs1}) captures the transition from effective stationarity to renewed propagation. Once the liquid from the thin film has completely drained, the lower contact line catches up with the propagating front, after which buoyancy acts to flatten the interface over its entire length (regime \RNum{5}). In this regime, the interface height $H$ is approximately linear. The spreading rate is set by the injected gas volume (recalling the ``horizontally topped'' regime described in \citet{pegler_topographic_2013}), while buoyancy remains active only over a short interfacial length scale proportional to $t^{-1/2}$, inducing a gradual tilt in $H$ and thus an effective flattening in physical space. Consequently, both contact lines propagate at the same leading-order position $t^{1/2}/(2\pi)$, with an $O(t^{-1/2})$ correction (Table~\ref{Tab:CLs1}) that accounts for the buoyancy-induced tilt (Fig.~\mainref{fig:Lambda1Sol}{4(b)}).

\begin{figure}
    \centering
    \begin{subfigure}{0.4\textwidth}
        \caption{\RNum{1}}
        \includegraphics[width = \textwidth]{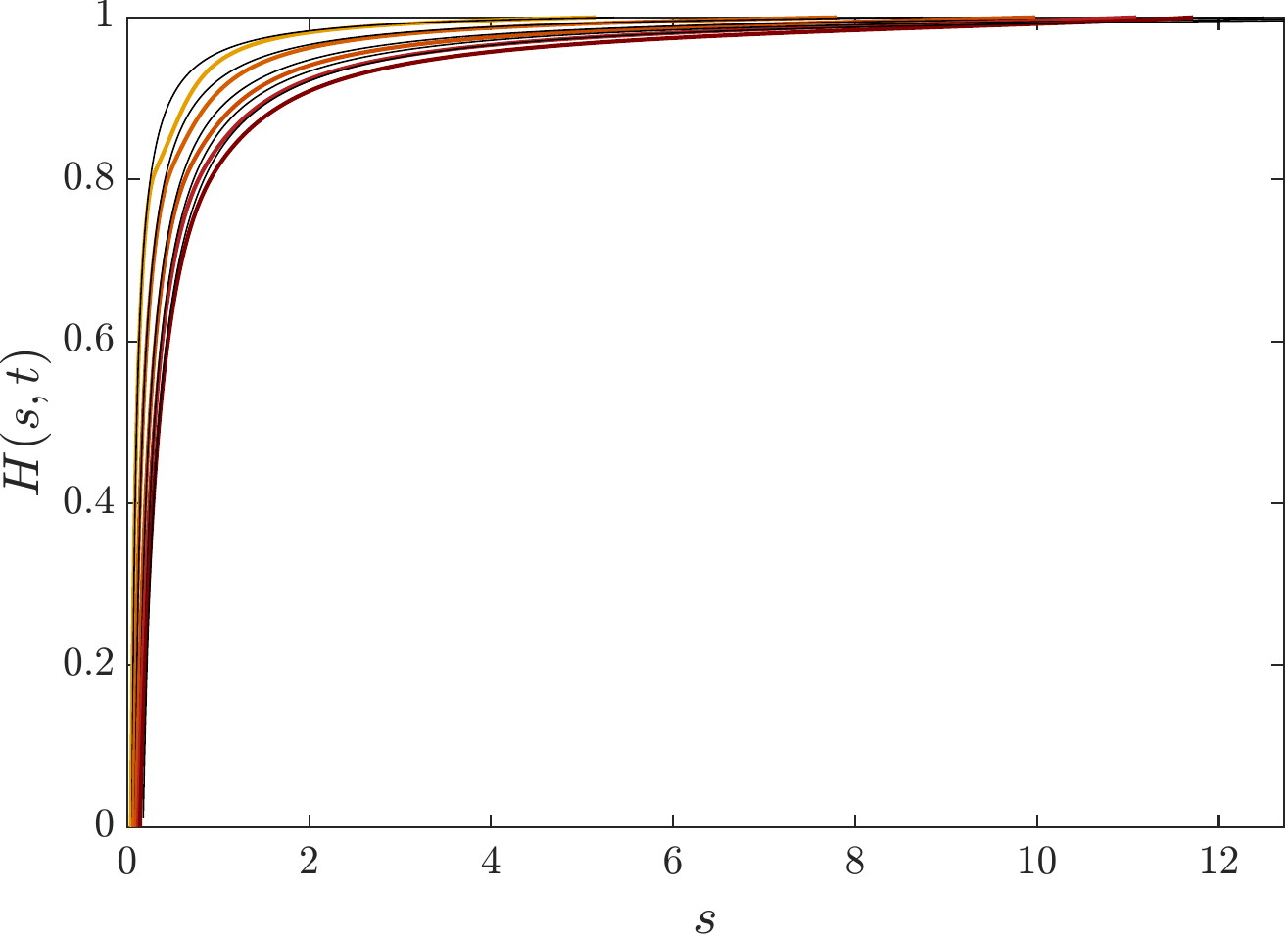}
    \end{subfigure} 
    \hspace{20pt}
    \begin{subfigure}{0.4\textwidth}
        \caption{\RNum{2}
        }
        \includegraphics[width = \textwidth]{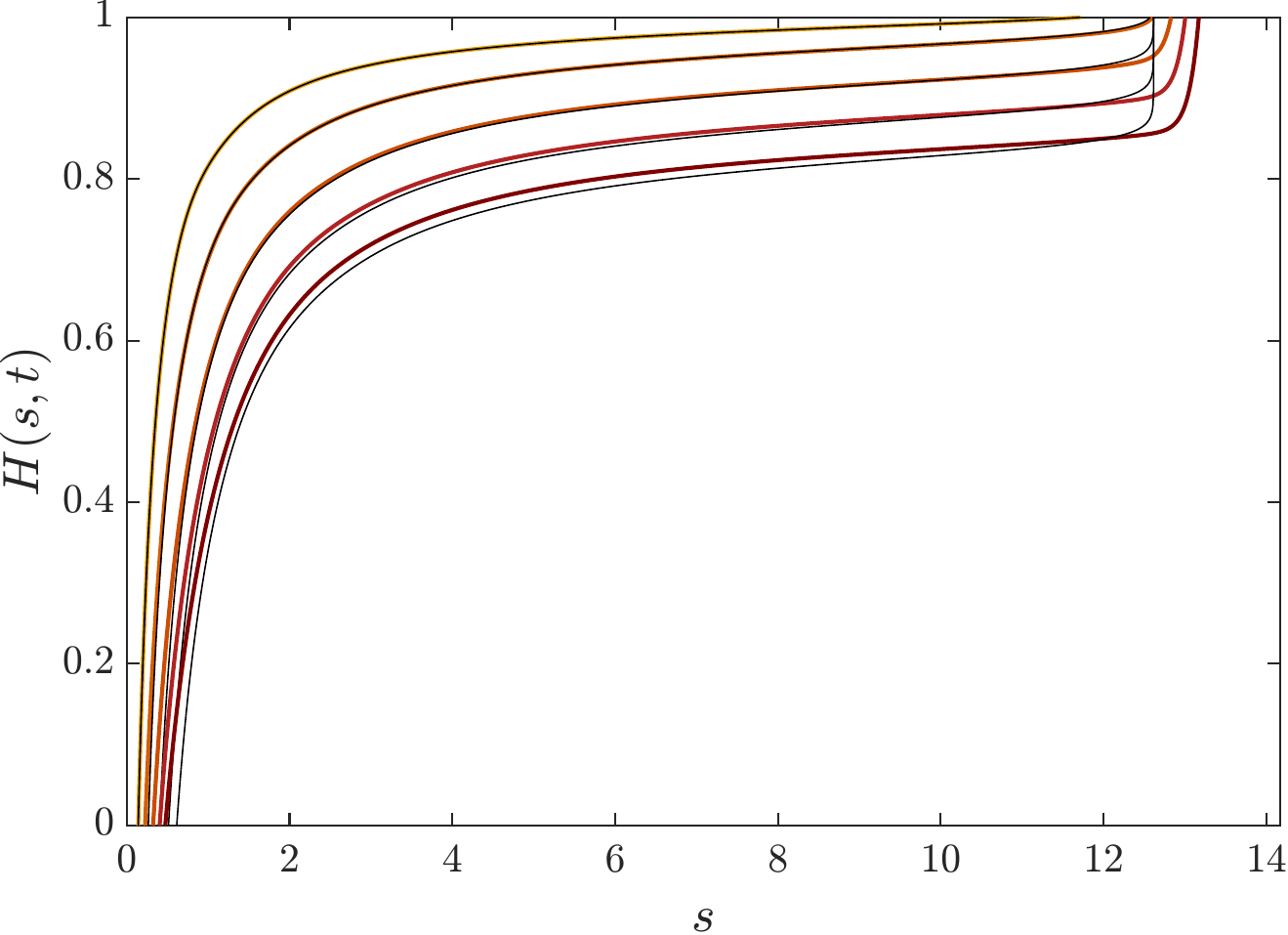}
    \end{subfigure} 
    \\
    \begin{subfigure}{0.4\textwidth}
        \caption{\RNum{3}}
        \includegraphics[width = \textwidth]{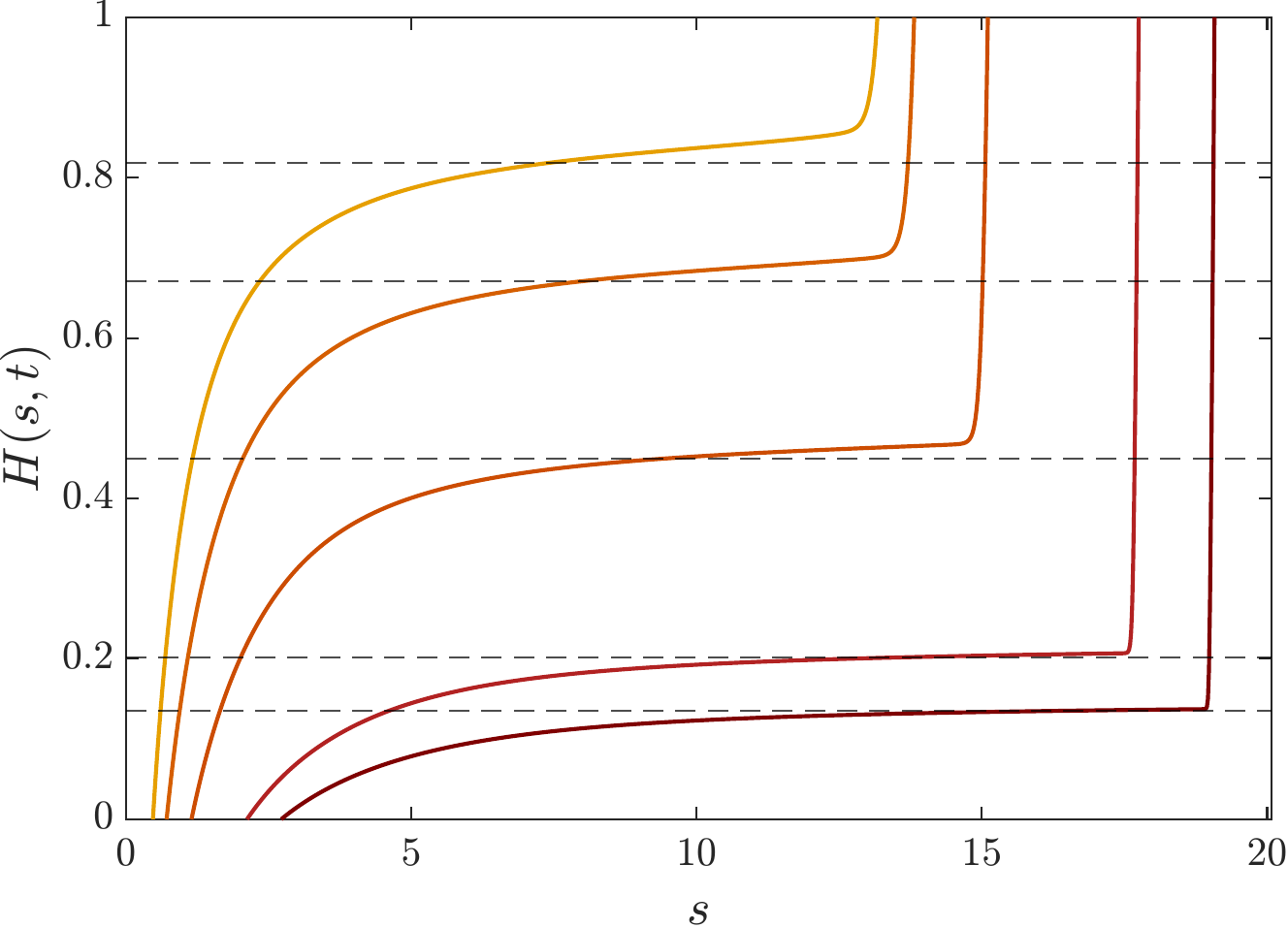}
    \end{subfigure} 
    \hspace{20pt}
    \begin{subfigure}{0.4\textwidth}
        \caption{\RNum{4}}
        \includegraphics[width = \textwidth]{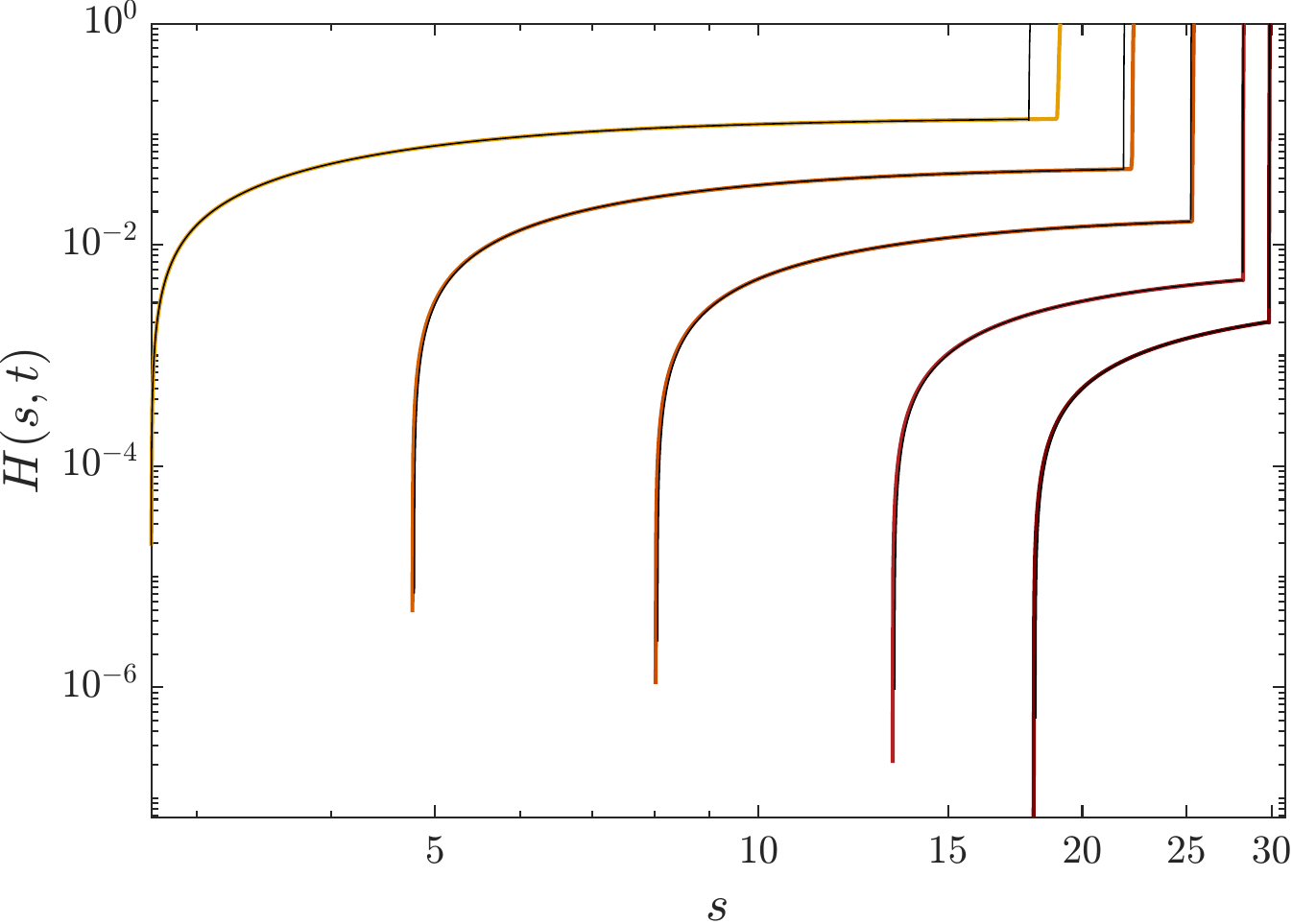}
    \end{subfigure} 
    \\
    \begin{subfigure}{0.4\textwidth}
        \caption{\RNum{5}}
        \includegraphics[width = \textwidth]{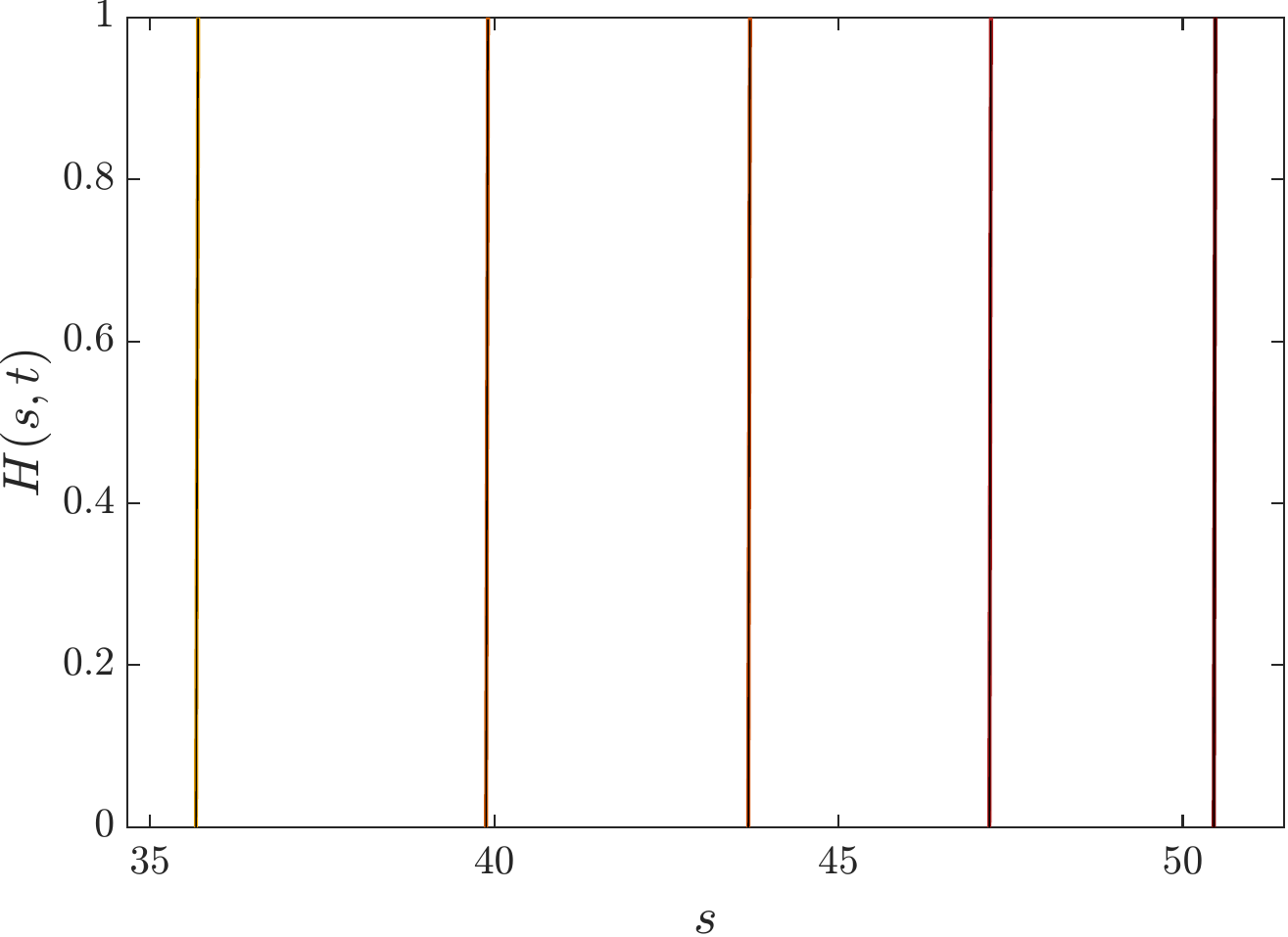}
    \end{subfigure} 
    \hspace{20pt}
    \begin{subfigure}{0.4\textwidth}
        \caption{}
        \includegraphics[width = \textwidth]{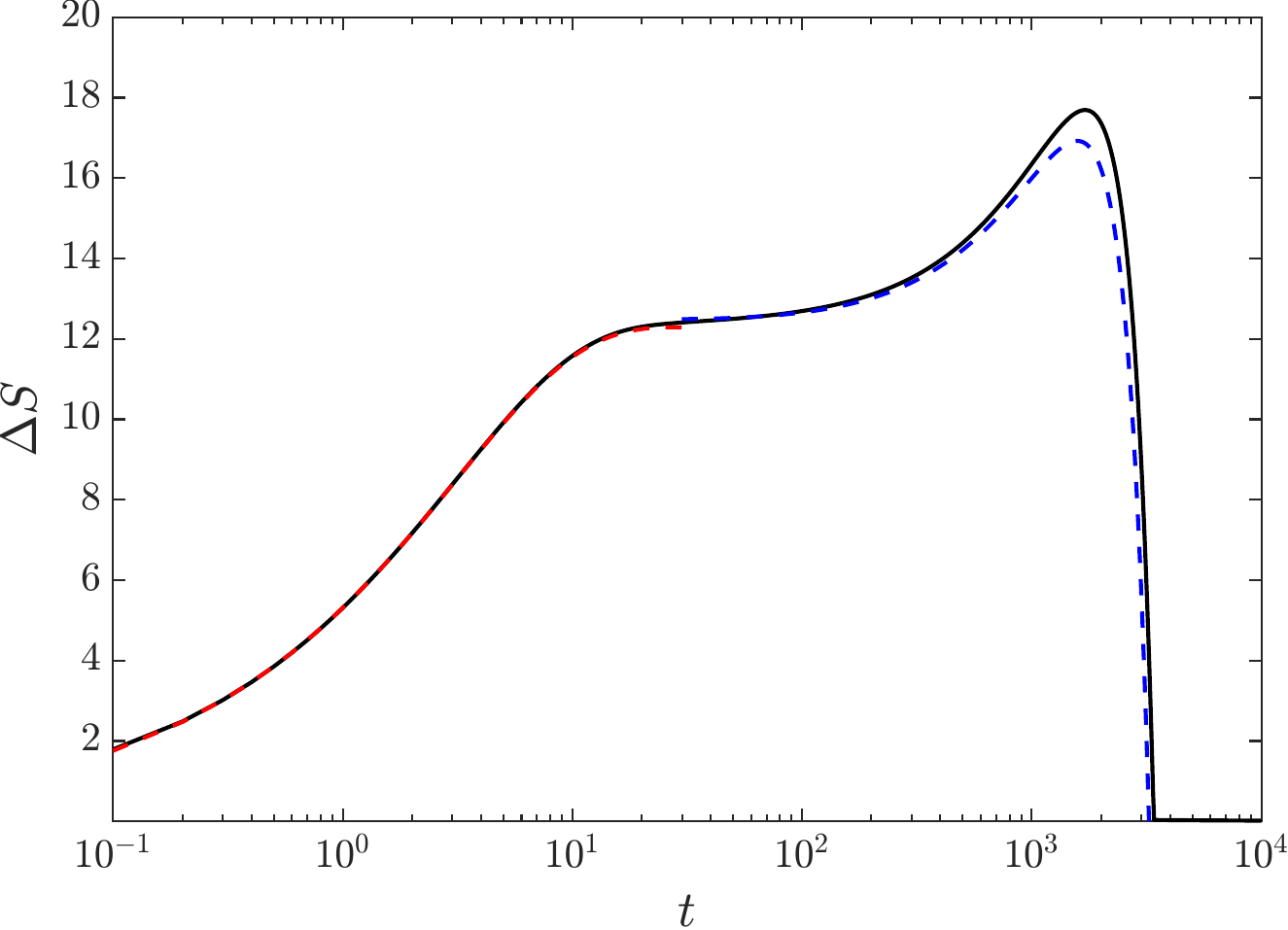}
    \end{subfigure} 
\caption{
Numerical solutions of Eqs~(\mainref{eq:Evo2}{33$a$--$e$}) for a parabolic channel with $\mathcal{M}=10^{-2}$ and $\lambda=10^{-1}$. Panels (a--e) show representative times spanning regimes \RNum{1}--\RNum{5}; coloured curves denote numerical solutions, while black curves indicate the corresponding asymptotic predictions from Appendix~\ref{App:ParaAsym}. (a) Regime \RNum{1}, $t = \{1,2.5,5,7.5,10\}$; analytical approximation Eq.~(\ref{eq:R1bSol}). 
(b) Regime \RNum{2}, $t = \{10,25,50,75,100\}$; hyperbolic approximations Eqs~(\ref{eq:R2bCharHSol}) and (\ref{eq:R2bCharsSol}). 
(c) Regime \RNum{3}, $t = \{1,2,4,8,10\}\times 10^2$; dashed curves show the predicted exponential decay Eq.~(\ref{eq:R3bEQSol}). 
(d) Regime \RNum{4}, $t = \{1,1.5,2,2.5,2.8\}\times 10^3$; composite approximation combining Eqs~(\ref{eq:R4aEQ}, \ref{eq:R4bEQSol}, \ref{eq:R4cHSol}) {(coloured curves) is almost indistinguishable from simulations (black curves)}. 
(e) Regime \RNum{5}, $t = \{4,5,6,7,8\}\times 10^3$; approximation Eq.~(\ref{eq:R5H}). 
(f) Evolution of the interface length projected onto the centreline, $\Delta S(t) = S_u(t) - S_l(t)$. Dashed red and blue curves show the regime \RNum{2} and \RNum{3} approximations, respectively, derived from the corresponding expressions for $S_u(t)$ and $S_l(t)$ (see Table~\ref{Tab:CLs1}).
}
    \label{fig:HSol}
\end{figure}

To assess the accuracy of the contact-line approximations across the different regimes and to validate the proposed scalings, in Fig.~\ref{fig:CLSfixedM} we compare the numerically computed evolution of the contact lines obtained from Eqs~(\mainref{eq:Evo2}{33$a$--$e$}) and Eq.~\eqref{eq:ParaSmallSlope} with the corresponding asymptotic predictions summarised in Table~\ref{Tab:CLs1} for a fixed viscosity ratio, $\mathcal{M}=0.1$, and varying values of $\lambda$. In Fig.~\mainref{fig:CLSfixedM}{5(a)}, {with the exception of the case $\lambda=1$,} at early times the propagation of both contact lines follows the regime-\RNum{1} solutions, during which buoyancy is subdominant to injection. The contact lines then transition through regimes \RNum{2}–\RNum{3}; a closer examination of these transitions is presented in Figs~\mainref{fig:CLSfixedM}{5(b,c)}. For all values of $\lambda$ considered, at sufficiently late times $t = O\big((\mathcal{M}\lambda)^{-1}\big)$, as shown in Fig.~\mainref{fig:CLSfixedM}{5(a)}, both contact lines approach the $(t/\pi)^{1/2}$ solution characteristic of regime \RNum{5}. {In the case $\lambda=1$, the distinction between regimes I and II is less apparent.}

The numerical solutions for $S_u$ from Fig.~\mainref{fig:CLSfixedM}{5(a)} are replotted in the region-\RNum{2}$_b$ coordinates in Fig.~\mainref{fig:CLSfixedM}{5(b)}, where, prior to the emergence of regime-\RNum{4}, the curves collapse onto the asymptotic solution Eq.~(\ref{eq:R2Su}) when $\lambda < 1$. The numerical data from Fig.~\mainref{fig:CLSfixedM}{5(a)} for $S_l$ are replotted in the regime-\RNum{3}$_a$ and regime-\RNum{4}$_b$ coordinates in Fig.~\mainref{fig:CLSfixedM}{5(c)}. During the initial propagation, the curves with $\lambda < 1$ again collapse onto the prediction by Eq.~(\mainref{eq:R3Sl}{B19}) until the lower contact line intercepts the advancing front, signalling the onset of regime-\RNum{5}. The transition time $t_c$ between regimes \RNum{4} and \RNum{5} predicted by Eq.~(\ref{eq:R5t}) is indicated by a black dot. In both cases, the close agreement between the numerical solutions for $S_u$ and $S_l$ and the asymptotic predictions Eqs~(\ref{eq:R2Su}) and (\ref{eq:R3Sl}) demonstrates that these approximations accurately capture the transition, shown in Fig.~\mainref{fig:CLSfixedM}{5(a)}, from the regime-\RNum{1} dynamics to the spreading characteristic of regime-\RNum{5}. Together, the asymptotic solutions for the contact lines from regimes \RNum{1}--\RNum{5} therefore capture the dominant behaviour over the entire evolution when $\lambda \ll 1$. For $\lambda = 1$, the numerical solution deviates noticeably from the regime~\RNum{1}--\RNum{2} solution in Fig.~\mainref{fig:CLSfixedM}{5(a)} and from the regime~\RNum{3}--\RNum{4} solution in Fig.~\mainref{fig:CLSfixedM}{5(c)}. This indicates that the asymptotic approximations do not fully capture the dynamics in regimes~\RNum{1}--\RNum{4} when $\lambda = O(1)$. The discrepancy arises because the local buoyancy term proportional to $H_s$ begins to influence the spatial regions \RNum{1}$_a$--\RNum{4}$_a$.

Figure~\ref{fig:HSol} compares the numerical solution of Eqs~(\mainref{eq:Evo2}{33$a$--$e$}) with the asymptotic approximations for $H$ derived in Appendix~\ref{App:ParaAsym} across regimes \RNum{1}--\RNum{5}, for $\mathcal{M}=10^{-2}$ and $\lambda=10^{-1}$. In Fig.~\mainref{fig:HSol}{6(a)}, corresponding to regime~\RNum{1}, the early-time dynamics are well described by the approximation Eq.~(\ref{eq:R1bSol}), which accurately captures the interface profile. As $t$ increases towards $O(\lambda^{-1})$, buoyancy effects become increasingly significant, leading to a progressive thickening of the film and a reduction in the propagation speed of the upper contact line. Consequently, as the solution approaches the transition to regime~\RNum{2}, the approximation Eq.~(\ref{eq:R1bSol}) overestimates the lateral extent of the interface and departs from the numerical solution. The evolution through regime~\RNum{2} is shown in Fig.~\mainref{fig:HSol}{6(b)}. Initialising the hyperbolic solution Eq.~(\ref{eq:R2bCharHSol}) with the numerical profile at $t=10$, the approximation captures the dominant behaviour of the interface. However, as time progresses, discrepancies emerge near the leading edge, signalling the breakdown of the regime-\RNum{2} description and the onset of regime~\RNum{3}.

Figure~\mainref{fig:HSol}{6(c)} illustrates regime~\RNum{3}, in which the interface transitions from a thin gas film adjacent to the upper boundary to a thin liquid film along the lower boundary; in region~\RNum{3}$_b$ [Fig.~\ref{fig:HRegions}], sufficiently far from the contact lines, the bulk of the interface descends at the exponential rate predicted by Eq.~(\ref{eq:R3bEQSol}). The composite approximation for regime~\RNum{4}, constructed from the solutions in regions \RNum{4}$_a$--\RNum{4}$_c$ [Eqs~(\ref{eq:R4aEQ}, \ref{eq:R4bEQSol}, \ref{eq:R4cHSol})], is shown in Fig.~\mainref{fig:HSol}{6(d)} and provides good agreement across the entire interface once the solution enters its range of validity, $H = O(\mathcal{M})$. Figure~\mainref{fig:HSol}{6(e)} shows the evolution in regime~\RNum{5}, where the interface becomes effectively flat, and the approximation Eq.~(\ref{eq:R5H}) accurately captures the subsequent propagation of the interface. Finally, Fig.~\mainref{fig:HSol}{6(f)} shows the evolution of $\Delta S = S_u(t)-S_l(t)$ alongside approximations obtained from the solutions for $S_u$ and $S_l$ from regimes \RNum{2} and \RNum{3} (see Table~\ref{Tab:CLs1}), both of which show good agreement with the numerical solution.

\subsection{Gaussian channel}
\label{sec:GaussianResults}
\begin{figure}
    \centering
    \begin{subfigure}{0.45\textwidth}
        \caption{}
        \includegraphics[width = \textwidth]{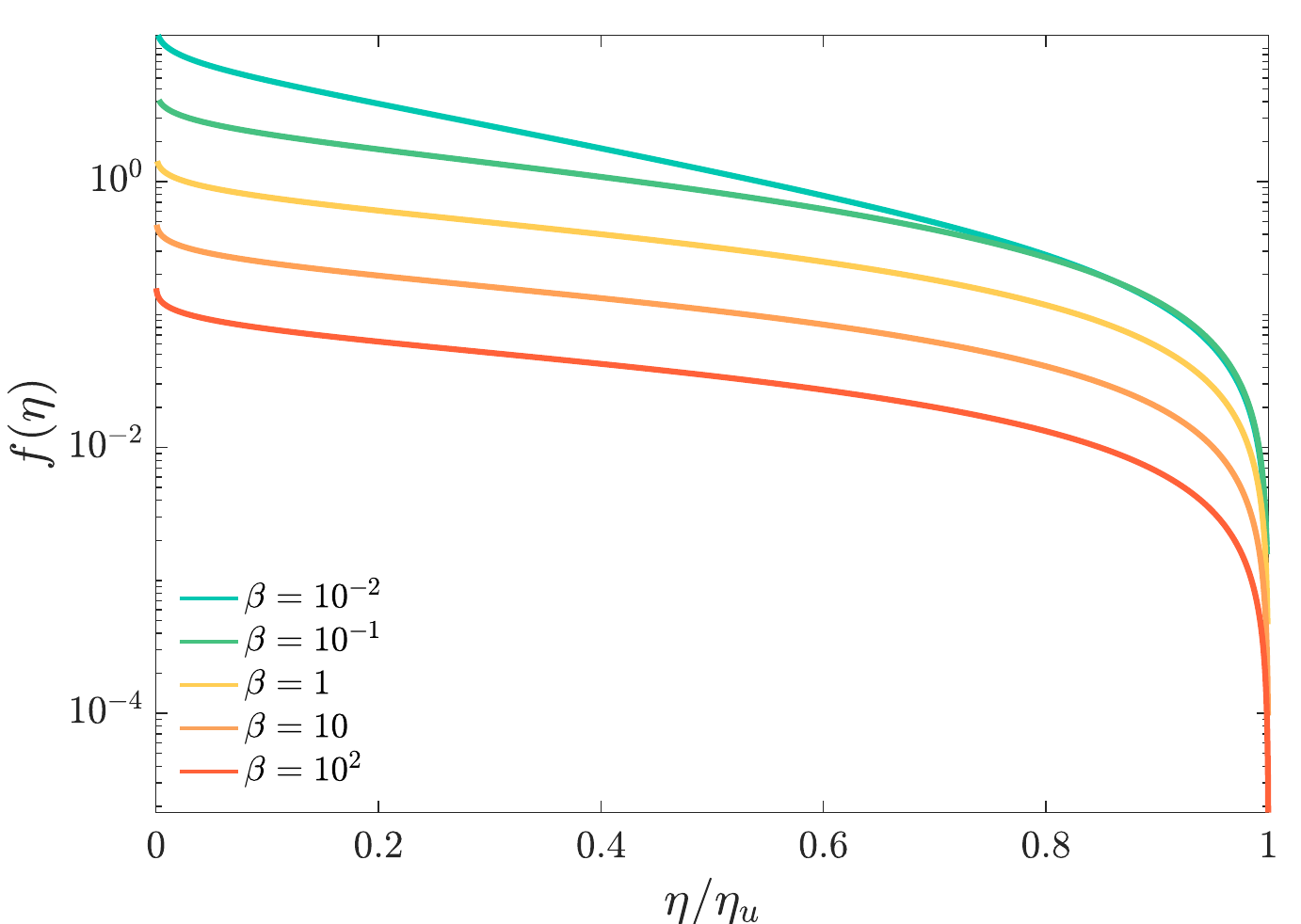}
    \end{subfigure} 
    \hspace{20pt}
    \begin{subfigure}{0.45\textwidth}
        \caption{}
        \includegraphics[width = \textwidth]{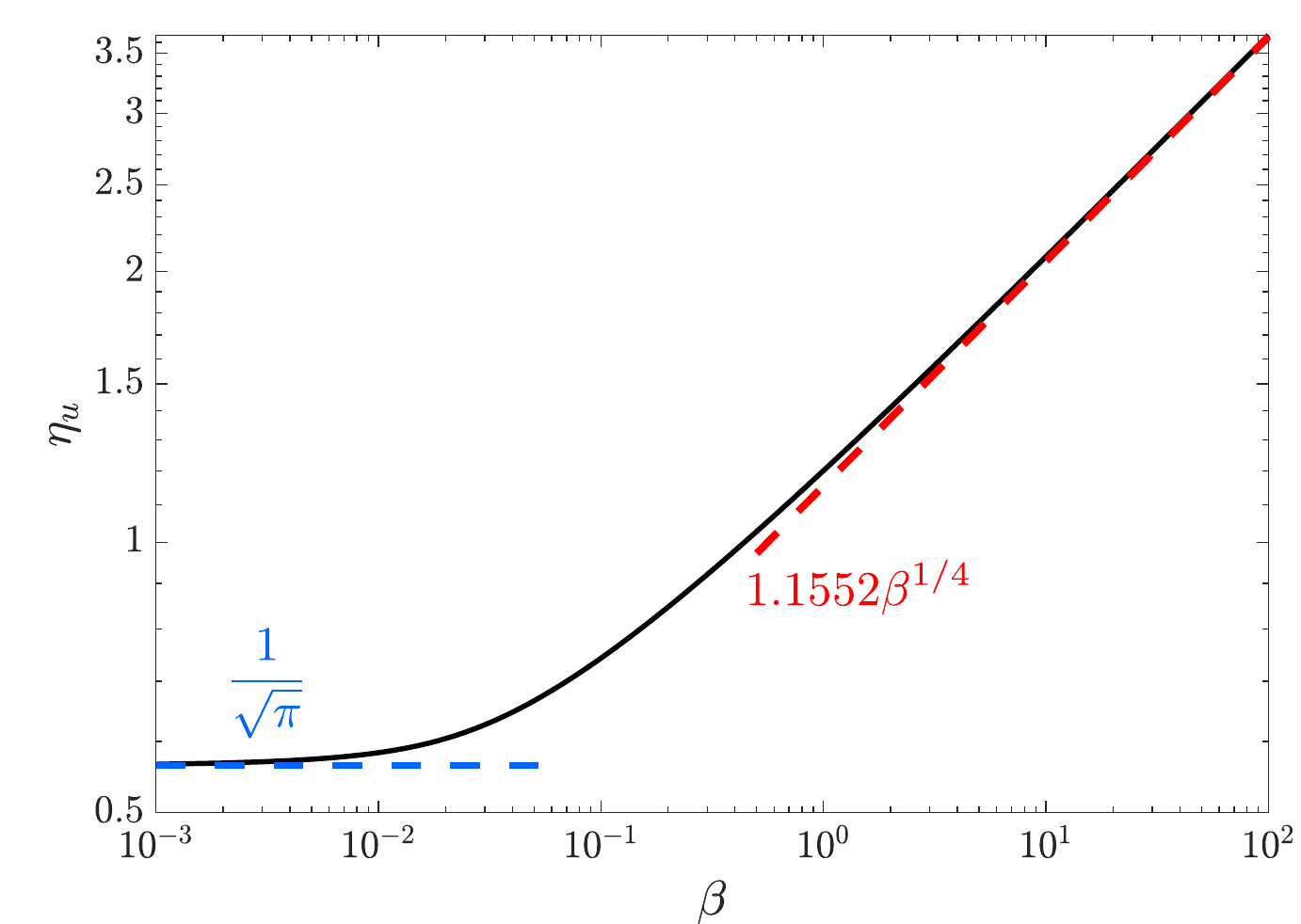}
    \end{subfigure}
    \caption{(a) Solutions of Eqs~(\mainref{eq:GaussOut}{41$a$--$d$}) describing the outer-region dynamics in a Gaussian channel, with $H = 1 - \mathcal{M} f(\eta)$ and similarity variable $\eta = s (\mathcal{M}/t)^{1/2}$, shown for increasing values of $\beta \equiv \mathcal{M}^2 \lambda$. The solutions were obtained using the shooting scheme outlined in Appendix~\ref{App:GaussianAsymp}. (b) Corresponding eigenvalue $\eta_u$ of the similarity solution as a function of $\beta$. The dashed blue and red curves denote the asymptotic approximations in the limits $\beta \ll 1$ and $\beta \gg 1$, respectively.}
    \label{fig:SimSols}
\end{figure} 
In the limit $\mathcal{M} \ll 1$, the governing equations for spreading in a Gaussian-shaped channel, Eqs~(\mainref{eq:Evo2}{33$a$--$e$}) and (\ref{eq:Bellturn}), exhibit an inner–outer structure (see Appendix~\ref{App:GaussianAsymp}). In this framework, the outer region evolves self-similarly and is governed by a single parameter $\beta \equiv \mathcal{M}^2 \lambda$, which measures the relative strength of buoyancy to injection. As shown in Appendix~\ref{App:GaussianAsymp}, for $\beta = O(1)$, introducing the similarity variables for the outer region, $\eta = s (\mathcal{M}/t)^{1/2}$, $S_u = \eta_u (t/\mathcal{M})^{1/2}$, and $H = 1 - \mathcal{M} f(\eta)$, into the governing equations (\mainref{eq:Evo2}{33$a$--$e$}) and (\ref{eq:Bellturn}), alongside the global volume constraint Eq.~(\ref{eq:Mass}), yields the boundary-value problem
\begin{subequations}\label{eq:GaussOut}
\begin{align}
    \left(\frac{1}{2\pi \eta(1+f)^2} - \frac{\eta}{2} \right) f_\eta 
    &= \frac{\beta}{\eta} \left(\frac{\eta f f_{\eta}} {1+f}\right)_\eta, 
    &&(0 \leq \eta \leq \eta_u), \\
    f &= 0, 
    &&(\eta = \eta_u),\\
    f &= \left(-\frac{\log(\eta)}{{\pi\beta }}\right)^{1/2}, \quad &&(\eta \to 0),
    \\
    \int_0^{\eta_u} \eta f \, \mathrm{d}\eta &= \frac{1}{2\pi}.
\end{align}
\end{subequations}
These equations describe the self-similar evolution of the outer region under a balance between injection and buoyancy. Figure~\mainref{fig:SimSols}{7(a)} shows solutions of Eqs~(\mainref{eq:GaussOut}{41$a$--$d$}) for the gas film thickness $f$ for increasing $\beta$, illustrating how the film thins as the influence of buoyancy is strengthened. Figure~\mainref{fig:SimSols}{7(b)} shows the corresponding eigenvalue $\eta_u$, which determines the upper contact line position via $S_u = \eta_u (t/\mathcal{M})^{1/2}$, as a function of $\beta$. The monotonic increase of $\eta_u$ with $\beta$ indicates that stronger buoyancy accelerates the spreading of the upper contact line. For small $\beta$, spreading is injection-dominated, with $\eta_u \to 1/\sqrt{\pi}$. In the limit $\beta \gg 1$, the eigenvalue asymptotes to $\eta_u \approx 1.1552 \beta^{1/4}$.

\begin{figure}
    \centering
    \begin{subfigure}{0.4\textwidth}
        \caption{}
        \includegraphics[width = \textwidth]{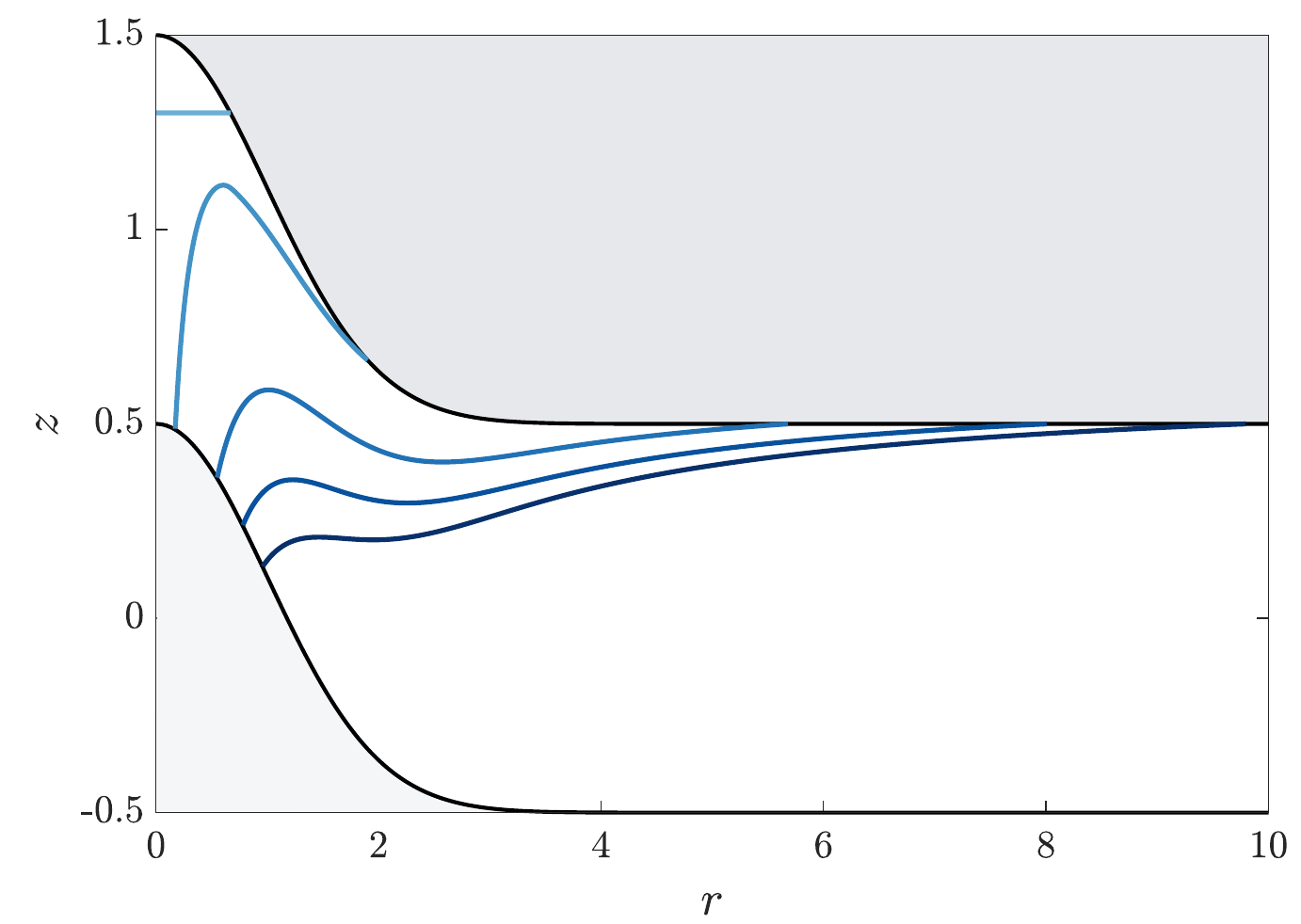}
    \end{subfigure} 
    \hspace{20pt}
    \begin{subfigure}{0.4\textwidth}
        \caption{}
        \includegraphics[width = \textwidth]{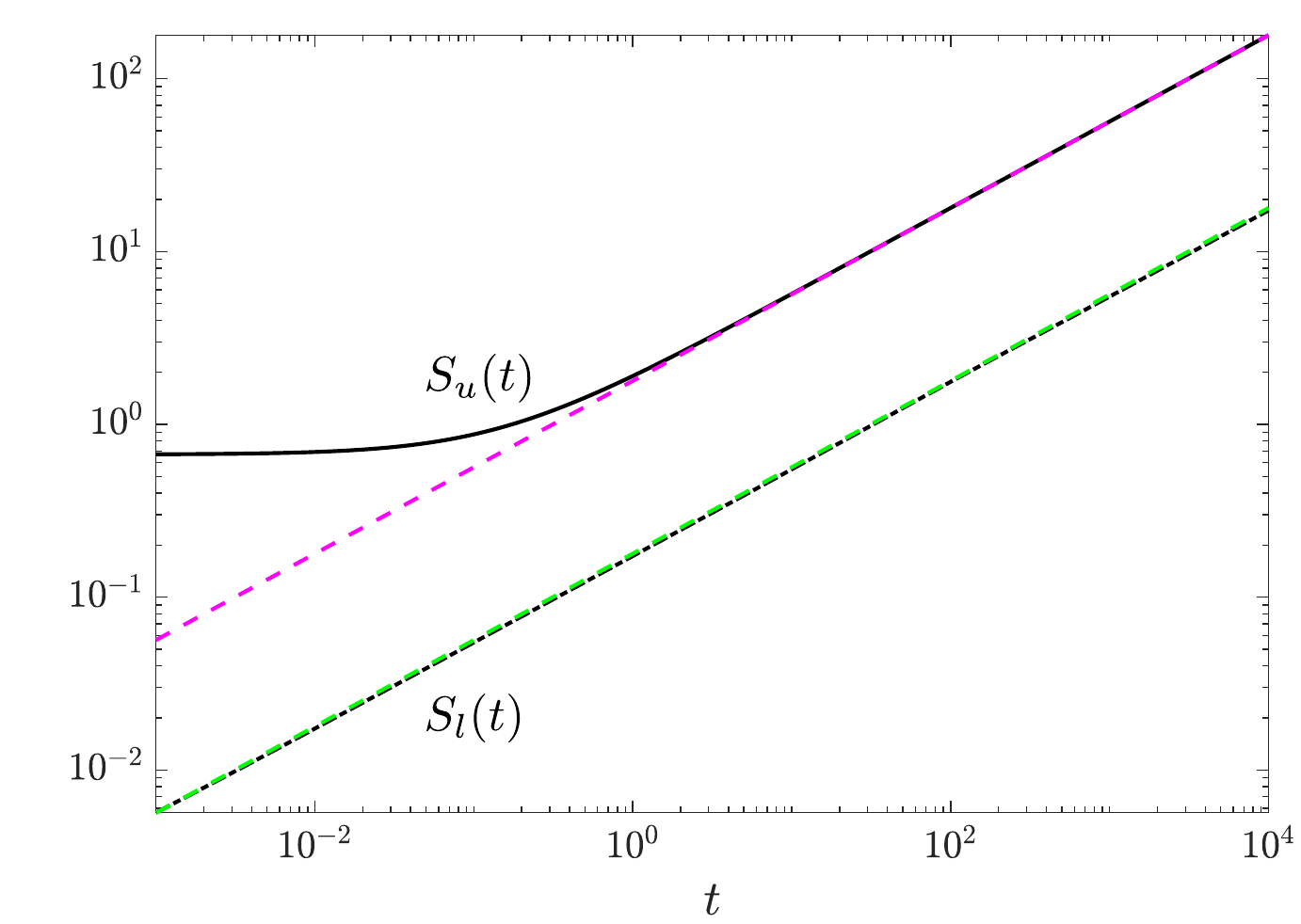}
    \end{subfigure}
    \\
    \begin{subfigure}{0.4\textwidth}
        \caption{}
        \includegraphics[width = \textwidth]{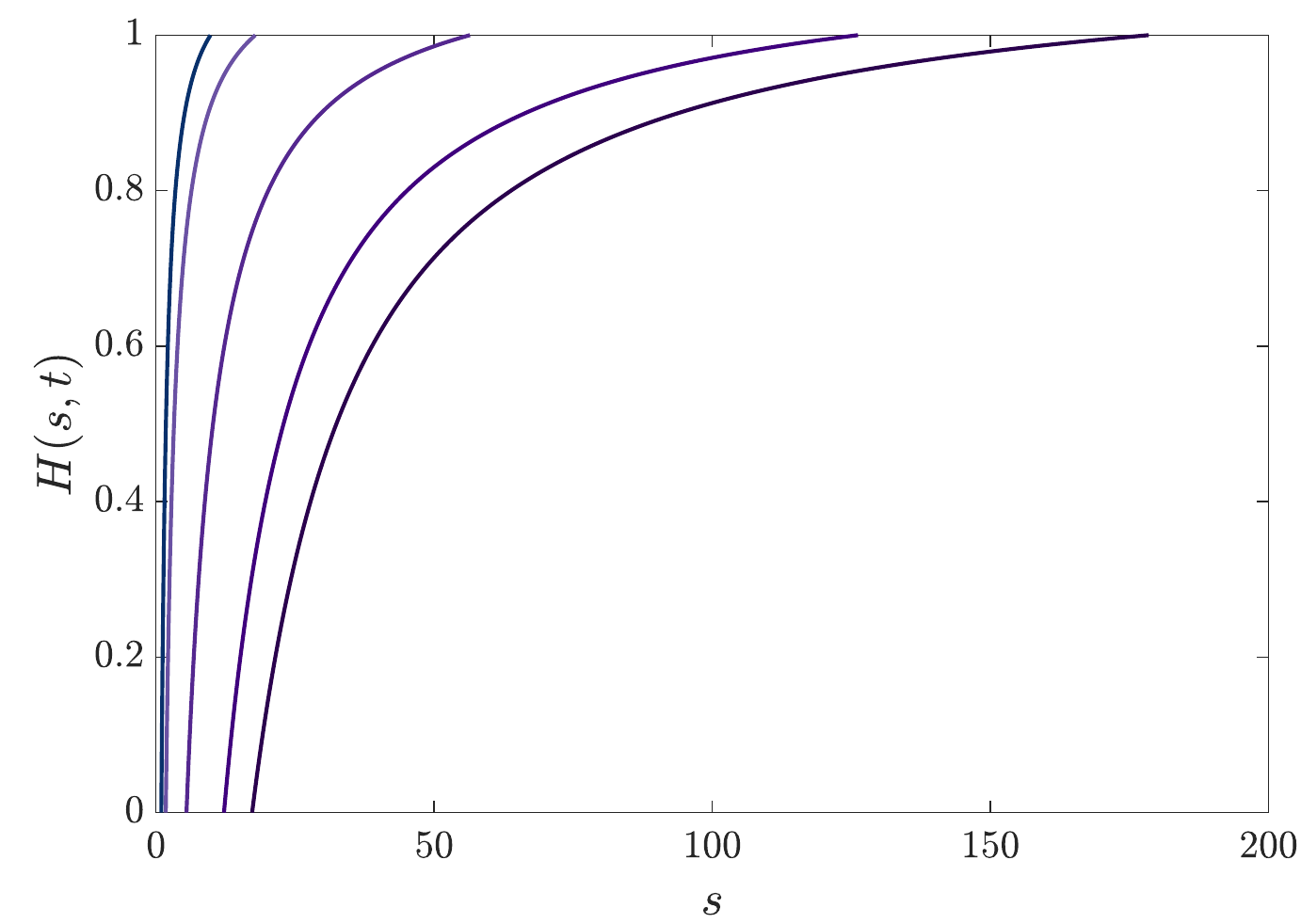}
    \end{subfigure} 
    \hspace{20pt}
    \begin{subfigure}{0.4\textwidth}
        \caption{}
        \includegraphics[width = \textwidth]{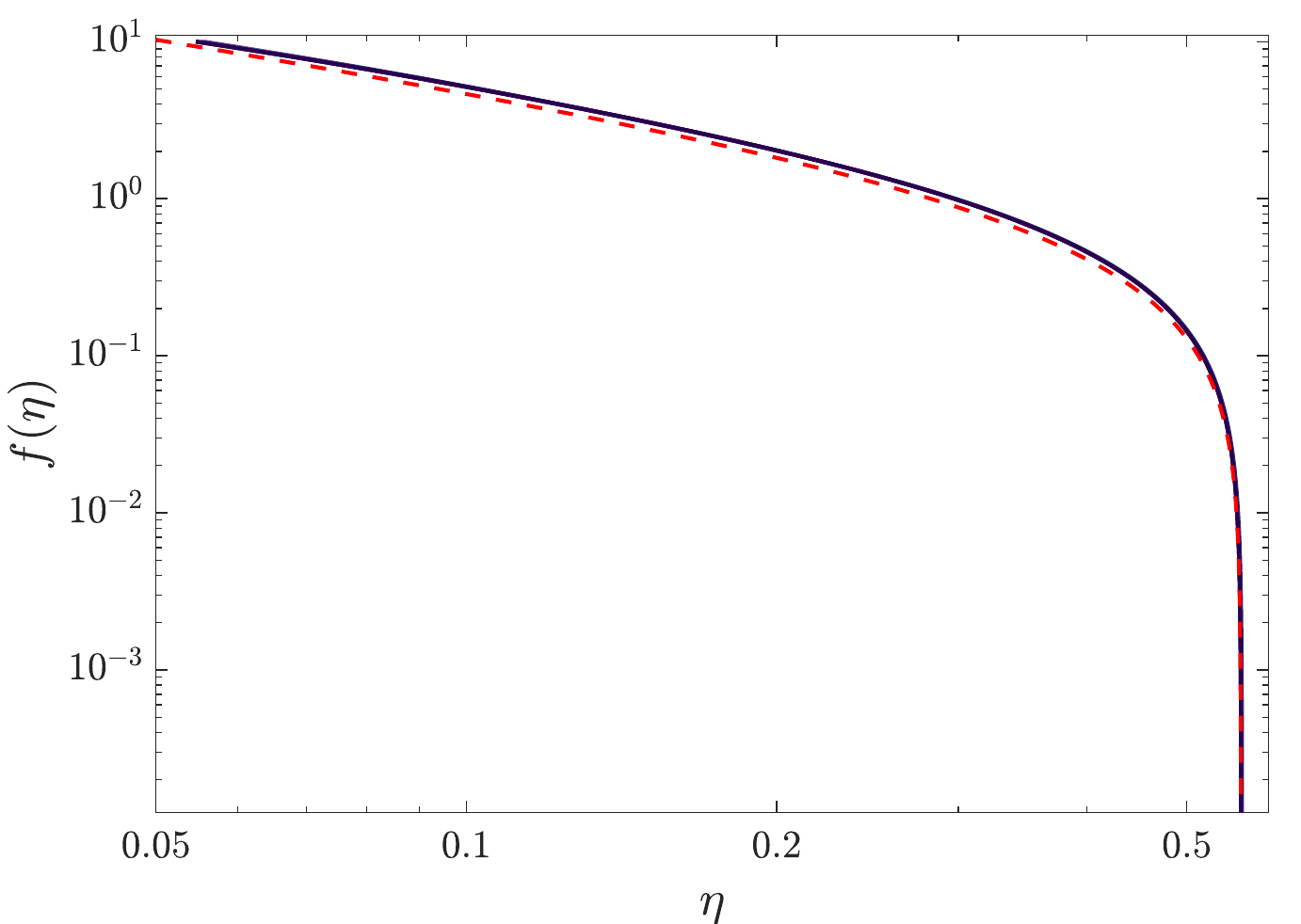}
    \end{subfigure}
    \caption{Numerical solutions of Eqs (\mainref{eq:Evo2}{33$a$--$e$}) with (\ref{eq:Bellturn}) in a Gaussian channel for $\mathcal{M}=0.1$ and $\lambda=10^{-2}$ [$\beta=\mathcal{M}^2\lambda=10^{-4}$]. (a)  The interface is shown in cylindrical coordinates at times $t=\{0,1,10,20,30\}$, with increasing time following the light-to-dark colour gradient. The channel boundaries are shown in black. 
(b) Evolution of the upper (black solid) and lower (black dash-dotted) contact lines. The late-time asymptotic prediction for $S_u$ from Eq.~(\mainref{eq:SmallBetaCLS}{42$a$}) (magenta dashed) and the late-time approximation for $S_l$ Eq.~(\mainref{eq:SmallBetaCLS}{42$b$}) (green dashed) are shown for comparison. 
(c) Interface profiles $H(s,t)$ at times $t=\{30,10^{2},10^{3},5\times10^{3},10^{4}\}$. 
(d) The profiles from (c) replotted in the self-similar variables of Eq.~(\mainref{eq:GaussLTOut2}{C6$a$}), $\eta=s(\mathcal{M}/t)^{1/2}$ and $H=1-\mathcal{M}f(\eta)$. The dashed red curve shows the analytical similarity solution Eq.~(\ref{eq:GaussInj}).
}
    \label{fig:Beta=1e-2}
\end{figure}

Below, we compare numerical solutions of the full system Eqs~(\mainref{eq:Evo2}{33$a$--$e$}) and (\ref{eq:Bellturn}) with solutions of Eqs~(\mainref{eq:GaussOut}{41$a$--$d$}) obtained using the shooting method outlined in Appendix~\ref{App:GaussianAsymp}. We also consider the analytical solution in the sublimit $\beta \ll 1$ and the reduced-order, parameter-free model in the limit $\beta \gg 1$ (see Appendix~\ref{App:GaussianAsymp}).

In the limit $\beta \ll 1$, injection dominates over buoyancy and Eqs.~(\ref{eq:GaussOut}) admits the analytical solution Eq.~(\ref{eq:GaussInj}), previously reported by \citet{guo_axisymmetric_2016}. The resulting inner–outer structure of the solution resembles that of regime~\RNum{1} in a parabolic channel. In this regime, the contact lines evolve according to
\addtocounter{equation}{1}
\begin{equation}\label{eq:SmallBetaCLS}
S_u(t) \approx \left(\frac{t}{\mathcal{M}\pi} \right)^{1/2}, \qquad
S_l(t) \approx \left(\frac{\mathcal{M}t}{\pi} \right)^{1/2}, \qquad (\beta \ll 1).\tag{42$a,b$}
\end{equation}
Figure~\ref{fig:Beta=1e-2} presents a representative solution of Eqs~(\mainref{eq:Evo2}{33$a$--$e$}) and (\ref{eq:Bellturn}) for $\beta = 10^{-4}$, implying that buoyancy effects are weak. The initial evolution of the interface is shown in Fig.~\mainref{fig:Beta=1e-2}{8(a)}, plotted in cylindrical coordinates. As expected from the results for regime~\RNum{1} in a parabolic channel, a thin gas film forms rapidly along the upper boundary. The contact lines quickly approach the asymptotic predictions Eq.~(\ref{eq:SmallBetaCLS}) after a short transient [Fig.~\mainref{fig:Beta=1e-2}{8(b)}]. Later stages of the evolution [Fig.~\mainref{fig:Beta=1e-2}{8(c)}] show the thin film elongating. Replotting the profiles in the similarity variables of Eq.~(\ref{eq:GaussLTOut2}) demonstrates collapse onto the analytical solution Eq.~(\mainref{eq:GaussInj}{C12}) [Fig.~\mainref{fig:Beta=1e-2}{8(d)}].

\begin{figure}
    \centering
    \begin{subfigure}{0.4\textwidth}
        \caption{}
        \includegraphics[width = \textwidth]{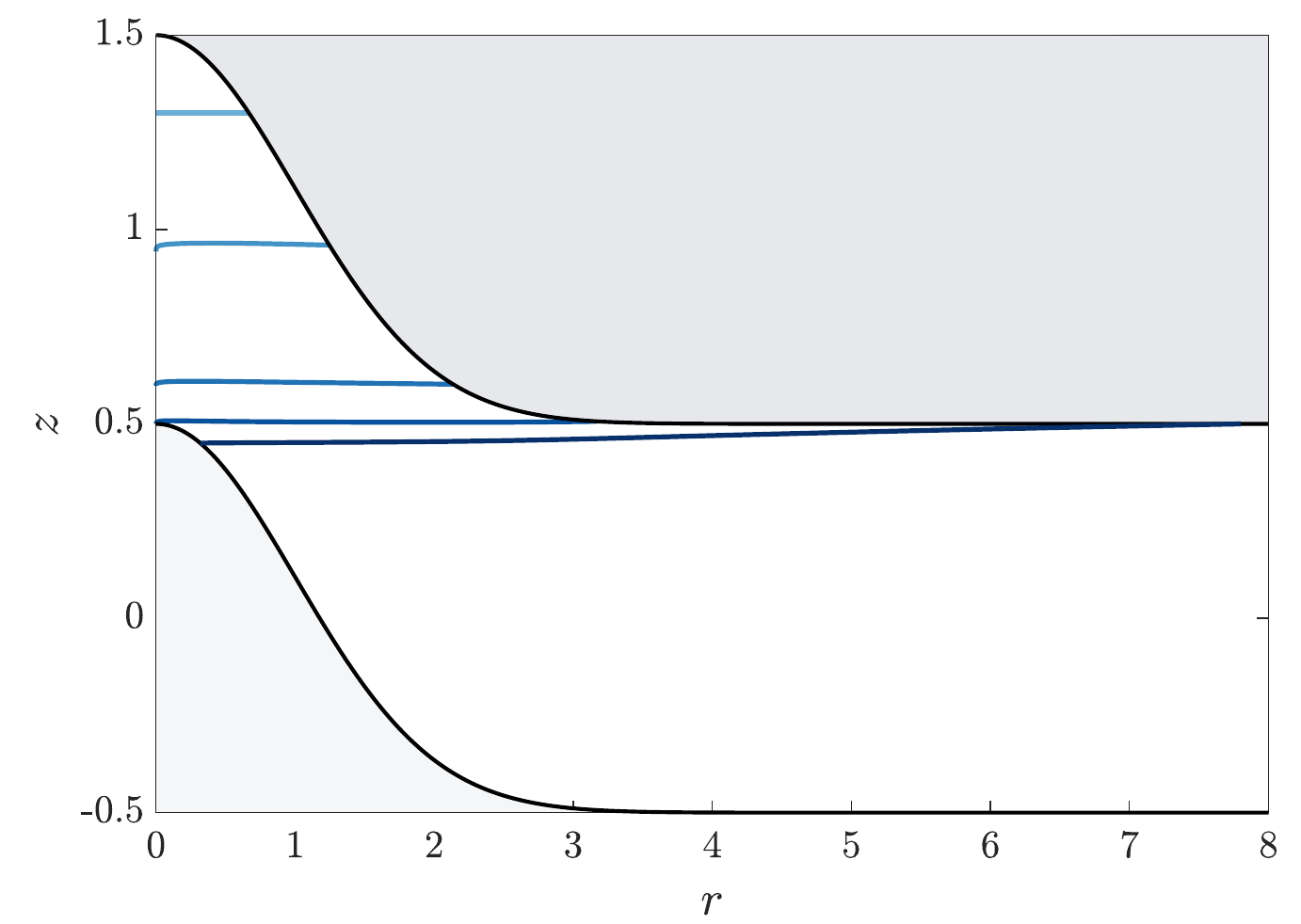}
    \end{subfigure} 
    \hspace{20pt}
    \begin{subfigure}{0.4\textwidth}
        \caption{}
        \includegraphics[width = \textwidth]{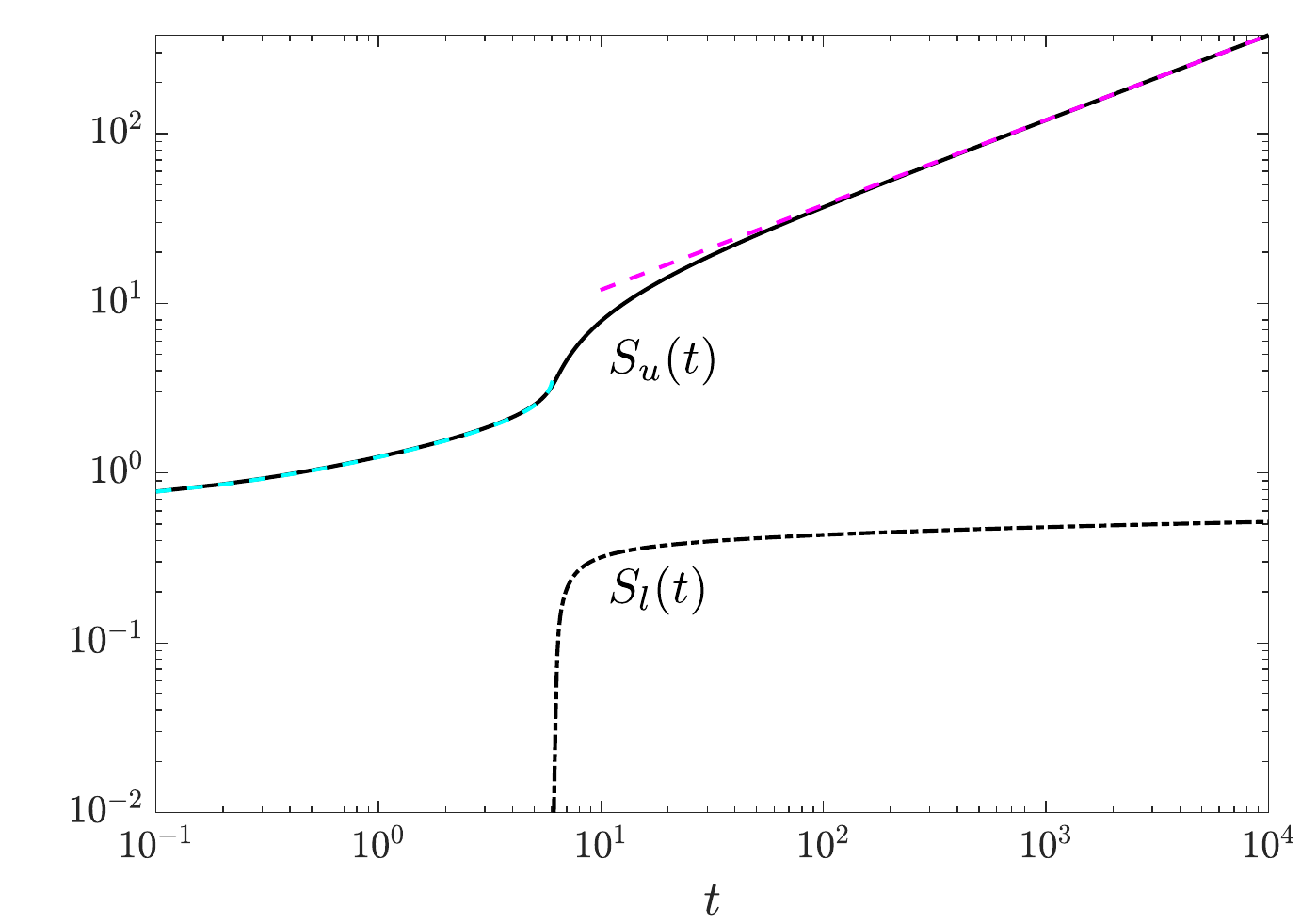}
    \end{subfigure}
    \\
    \begin{subfigure}{0.4\textwidth}
        \caption{}
        \includegraphics[width = \textwidth]{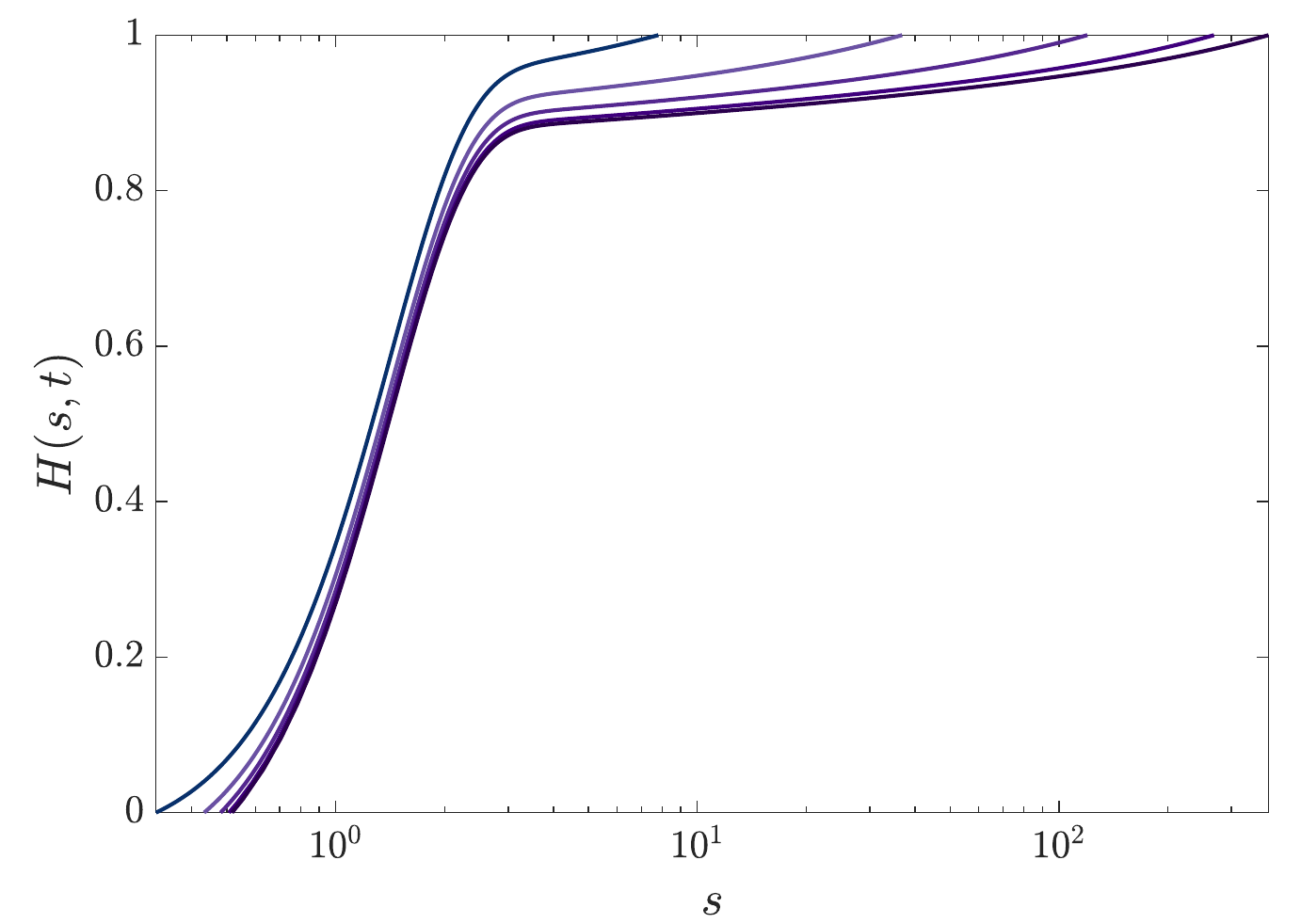}
    \end{subfigure} 
    \hspace{20pt}
    \begin{subfigure}{0.4\textwidth}
        \caption{}
        \includegraphics[width = \textwidth]{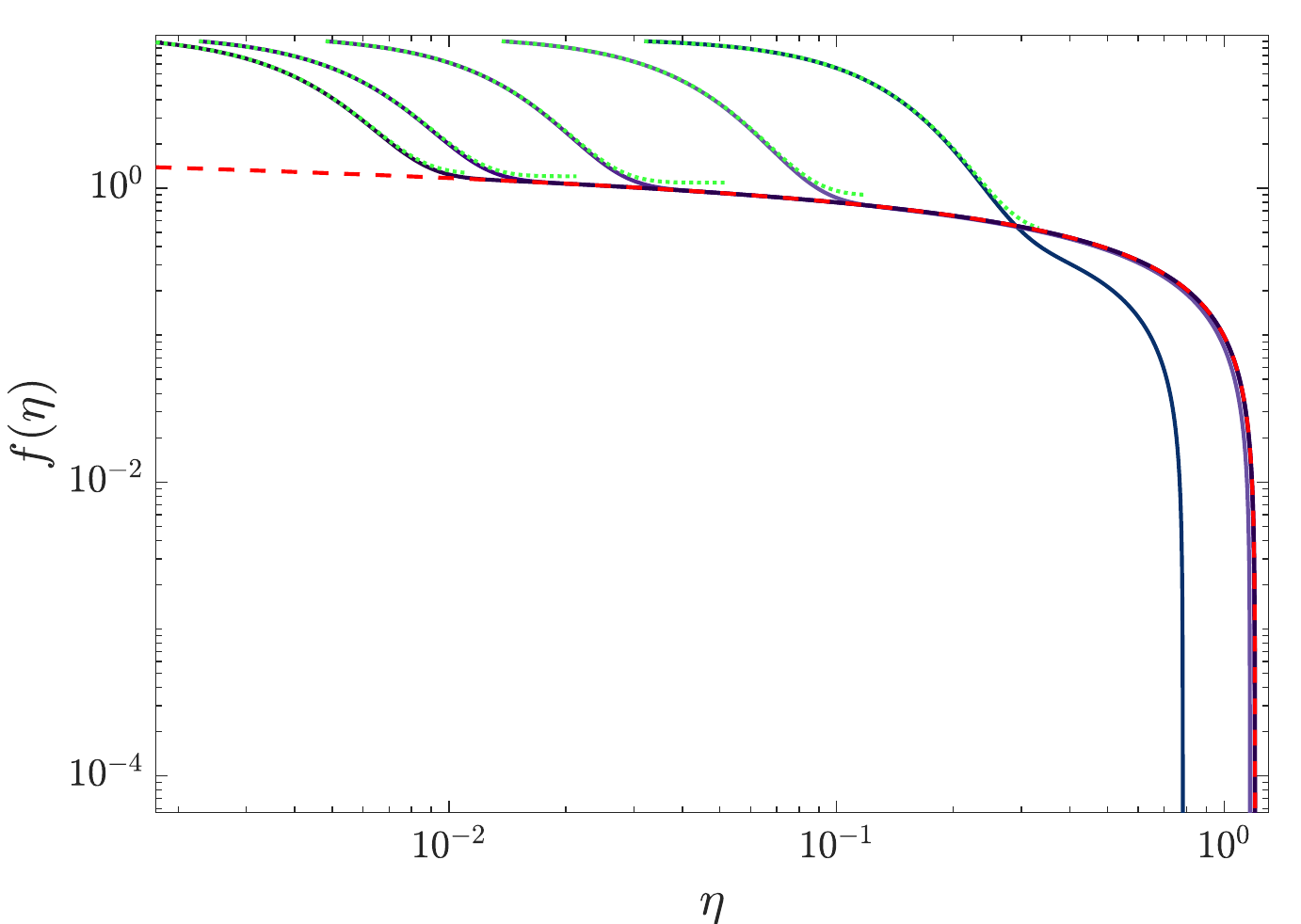}
    \end{subfigure}
    \caption{Numerical solutions of Eqs~(\mainref{eq:Evo2}{33$a$--$e$}) with (\ref{eq:Bellturn}) in a Gaussian channel for $\mathcal{M}=0.1$ and $\lambda=10^{2}$ [$\beta=\mathcal{M}^2\lambda=1$]. (a) The interface is shown in cylindrical coordinates at times $t=\{0,1,4,6,10\}$, with increasing time indicated by a light-to-dark colour gradient. The channel boundaries are shown in black. 
(b) Evolution of the upper (black solid) and lower (black dash-dotted) contact lines. The asymptotic prediction for $S_u$ from Eq.~(\ref{eq:GaussETSu}) (cyan dashed) and the late-time approximation Eq.~(\ref{eq:SuBeta1}) (magenta dashed) are shown for comparison. 
(c) Interface profiles $H(s,t)$ at times $t=\{10,10^{2},10^{3},5\times10^{3},10^{4}\}$. 
(d) The profiles from (c) replotted in the outer self-similar variables of Eqs~(\mainref{eq:GaussOut}{41$a$--$d$}), $\eta=s(\mathcal{M}/t)^{1/2}$ and $H=1-\mathcal{M}f(\eta)$. The dashed red curve shows the outer similarity solution of Eqs~(\mainref{eq:GaussOut}{41$a$--$d$}), obtained via the shooting scheme outlined in Appendix~\ref{App:GaussianAsymp}, while the dotted green curves show the inner solution Eq.~(\ref{eq:GaussLTInnEq}).
}
    \label{fig:Beta=1}
\end{figure}

Increasing $\beta$ to $1$ [Fig.~\ref{fig:Beta=1}] demonstrates the influence of buoyancy on the flow. Prior to the formation of the lower contact line, buoyancy flattens the interface, consistent with the asymptotic solution Eq.~(\ref{eq:GaussETSol}). During this early stage, the upper contact line, $S_u$, advances at a rate set by the injected volume, accurately captured by approximation Eq.~(\ref{eq:GaussETSu}) [Fig.~\mainref{fig:Beta=1}{9(b)}]. Once $S_l$ forms at the origin, the lower contact line rapidly moves to a nearby position and becomes pinned, while a thin gas film spreads along the upper boundary [Figs~\mainref{fig:Beta=1}{9(a,b)}]. Figure~\mainref{fig:Beta=1}{9(c)} shows $H(s,t)$ once the late-time inner–outer structure emerges. In the inner region adjacent to $S_l$, the interface remains locally horizontal, and the contact line remains pinned, as predicted by Eq.~(\ref{eq:GaussLTInnEq}). In the outer region, the balance between injection and buoyancy drives faster spreading of the thin film [compare Fig.~\mainref{fig:Beta=1}{9(b)} to Fig.~\mainref{fig:Beta=1e-2}{8(b)}]. The similarity solution of Eqs~(\mainref{eq:GaussOut}{41$a$--$d$}), obtained by numerically integrating Eqs~(\mainref{eq:ODEsystem}{C10$a$--$c$}) via the shooting method in Appendix~\ref{App:GaussianAsymp}, predicts the upper contact line propagation as
\begin{equation}\label{eq:SuBeta1}
S_u(t) \approx 1.2013 \left(\frac{t}{\mathcal{M}}\right)^{1/2}, \qquad (\beta = 1),
\end{equation}
in close agreement with the numerical solution [Fig.~\mainref{fig:Beta=1}{9(b)}]. Figure~\mainref{fig:Beta=1}{9(d)} shows the interface profiles replotted in similarity coordinates, demonstrating collapse onto the outer similarity solution for all times except the earliest ($t = 10$). The inner solution Eq.~(\ref{eq:GaussLTInnEq}) is also shown, exhibiting good agreement within its range of validity. However, Fig.~\mainref{fig:Beta=1}{9(d)} shows that the logarithmic inner limit of the outer solution in Eq.~(\mainref{eq:GaussOut}{41$c$}) (recalling a boundary condition reported in \citet{lyle_axisymmetric_2005}) is incompatible with the constant outer limit of the inner solution Eq.~(\ref{eq:GaussLTInnEq}), suggesting the existence of a {passive} intermediate spatial region that {we do not investigate further.}

\begin{figure}
    \centering
    \begin{subfigure}{0.4\textwidth}
        \caption{}
        \includegraphics[width = \textwidth]{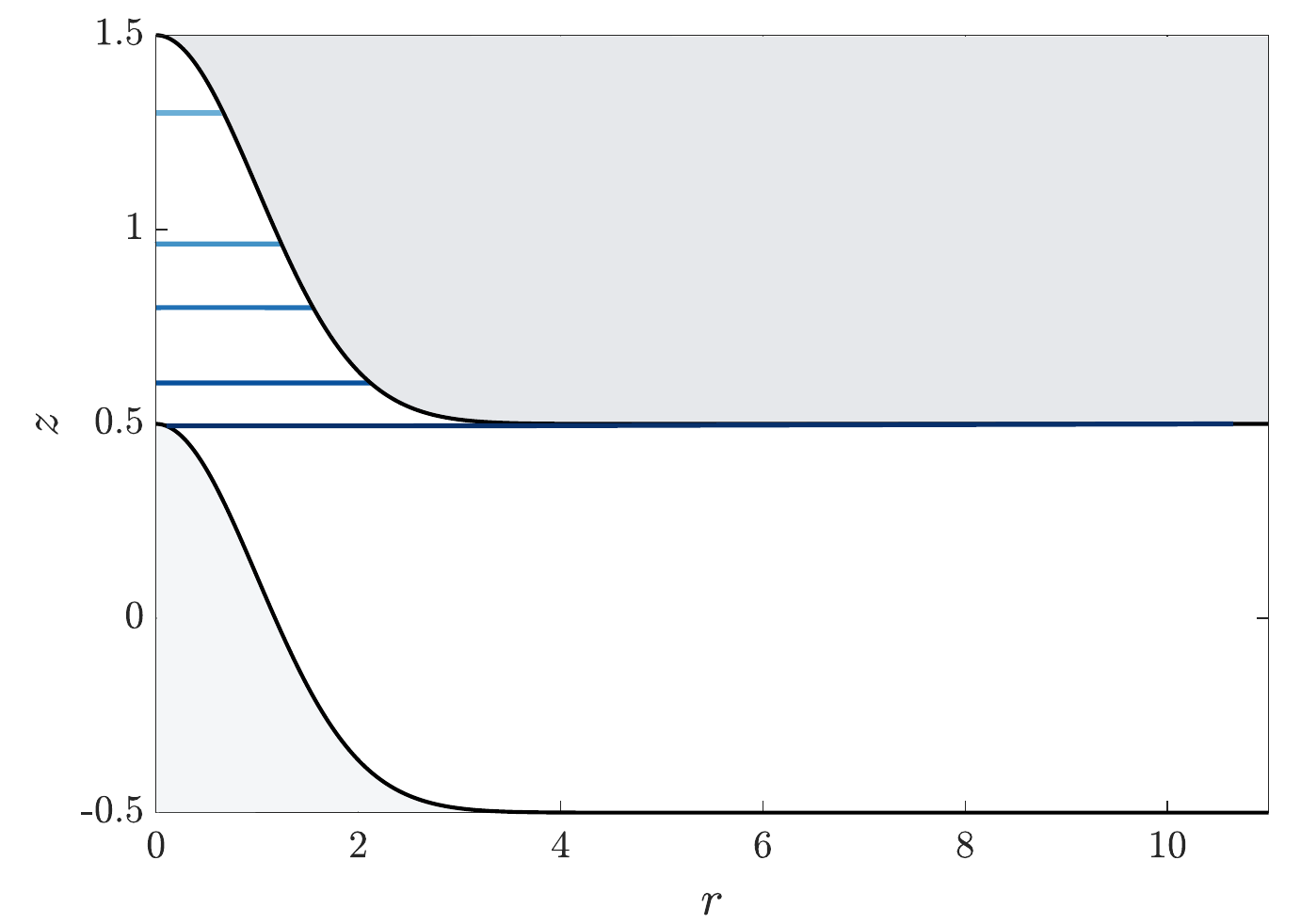}
    \end{subfigure} 
    \hspace{20pt}
    \begin{subfigure}{0.4\textwidth}
        \caption{}
        \includegraphics[width = \textwidth]{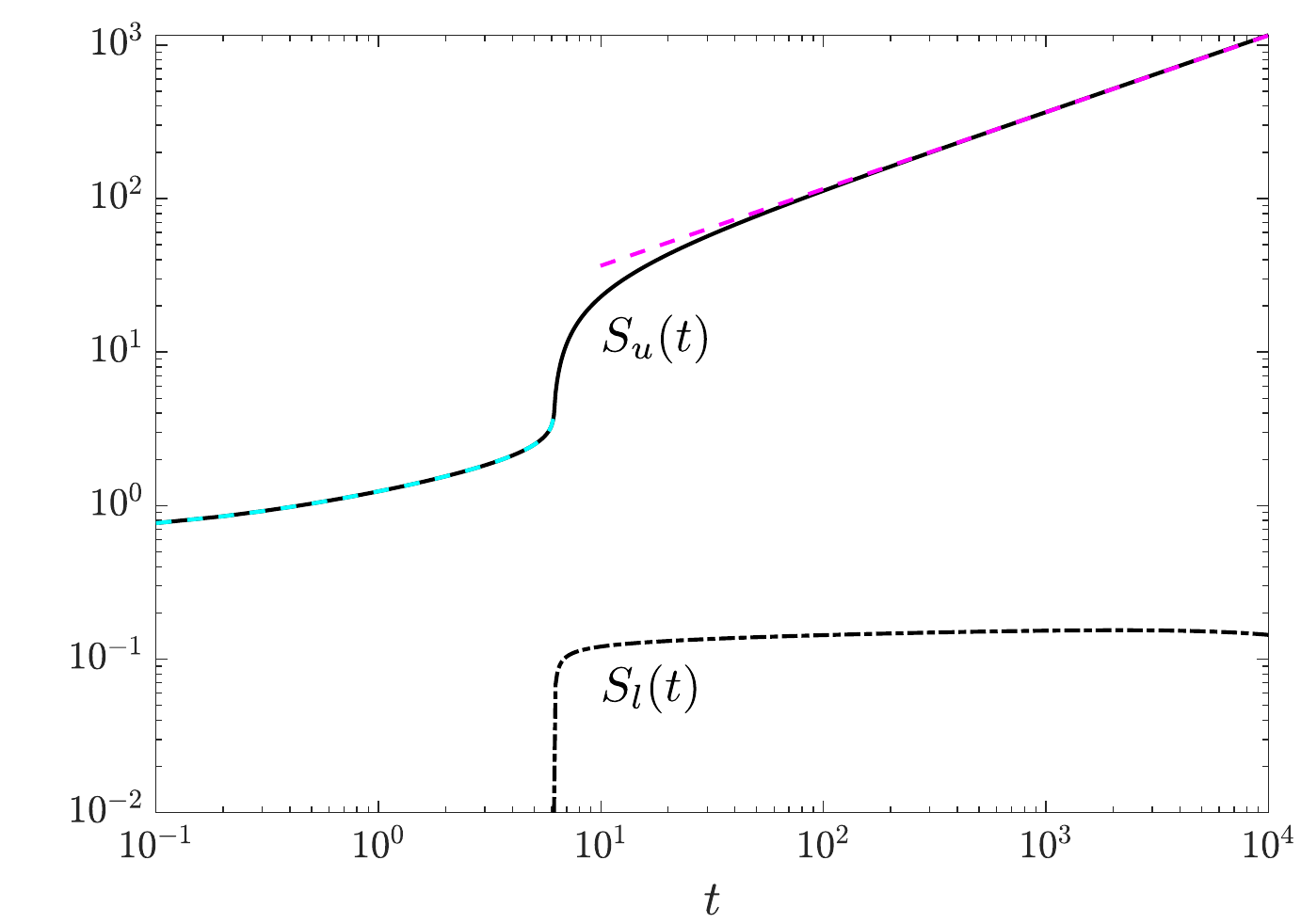}
    \end{subfigure}
    \\
    \begin{subfigure}{0.4\textwidth}
        \caption{}
        \includegraphics[width = \textwidth]{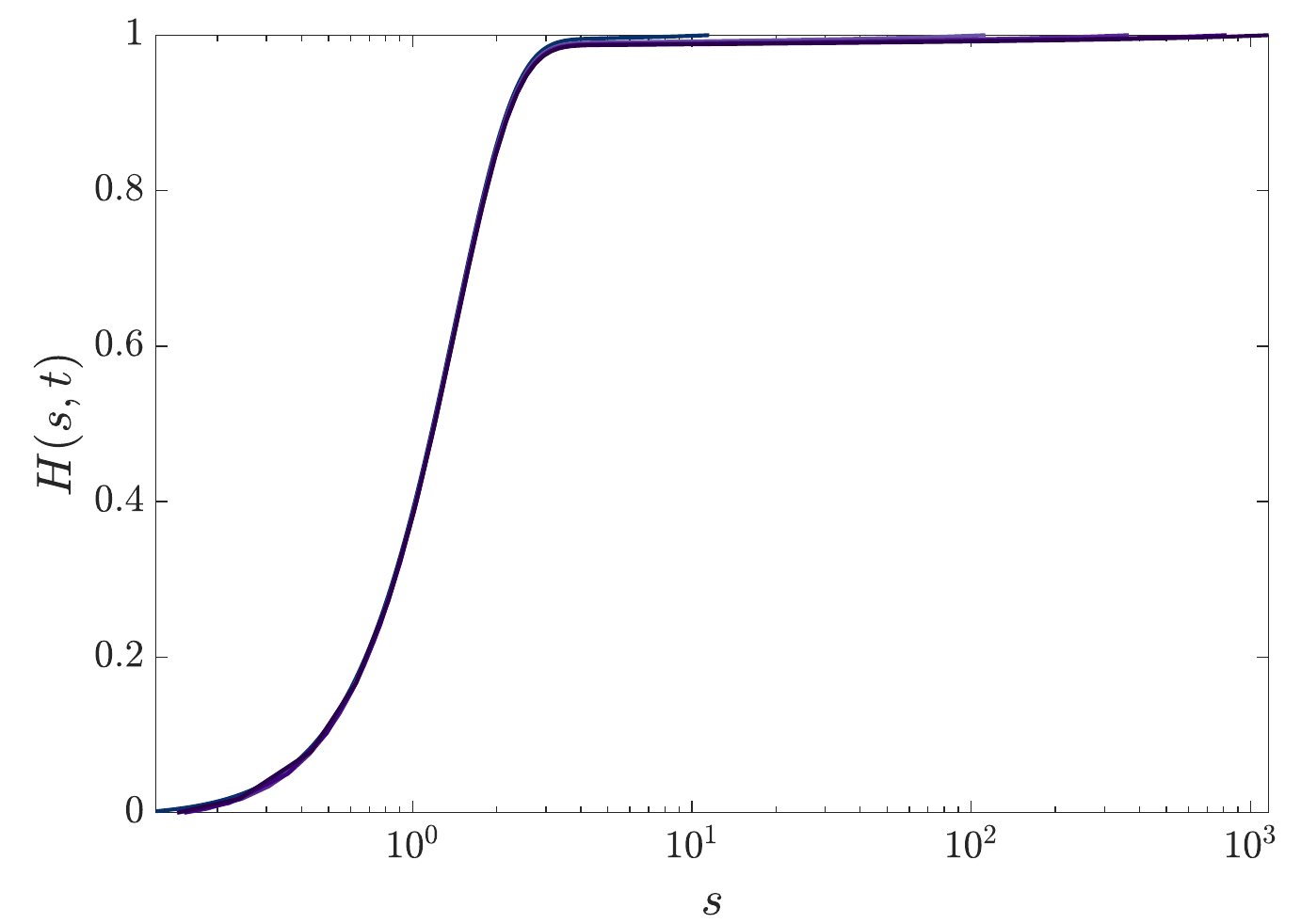}
    \end{subfigure}
    \hspace{20pt}
    \begin{subfigure}{0.4\textwidth}
        \caption{}
        \includegraphics[width = \textwidth]{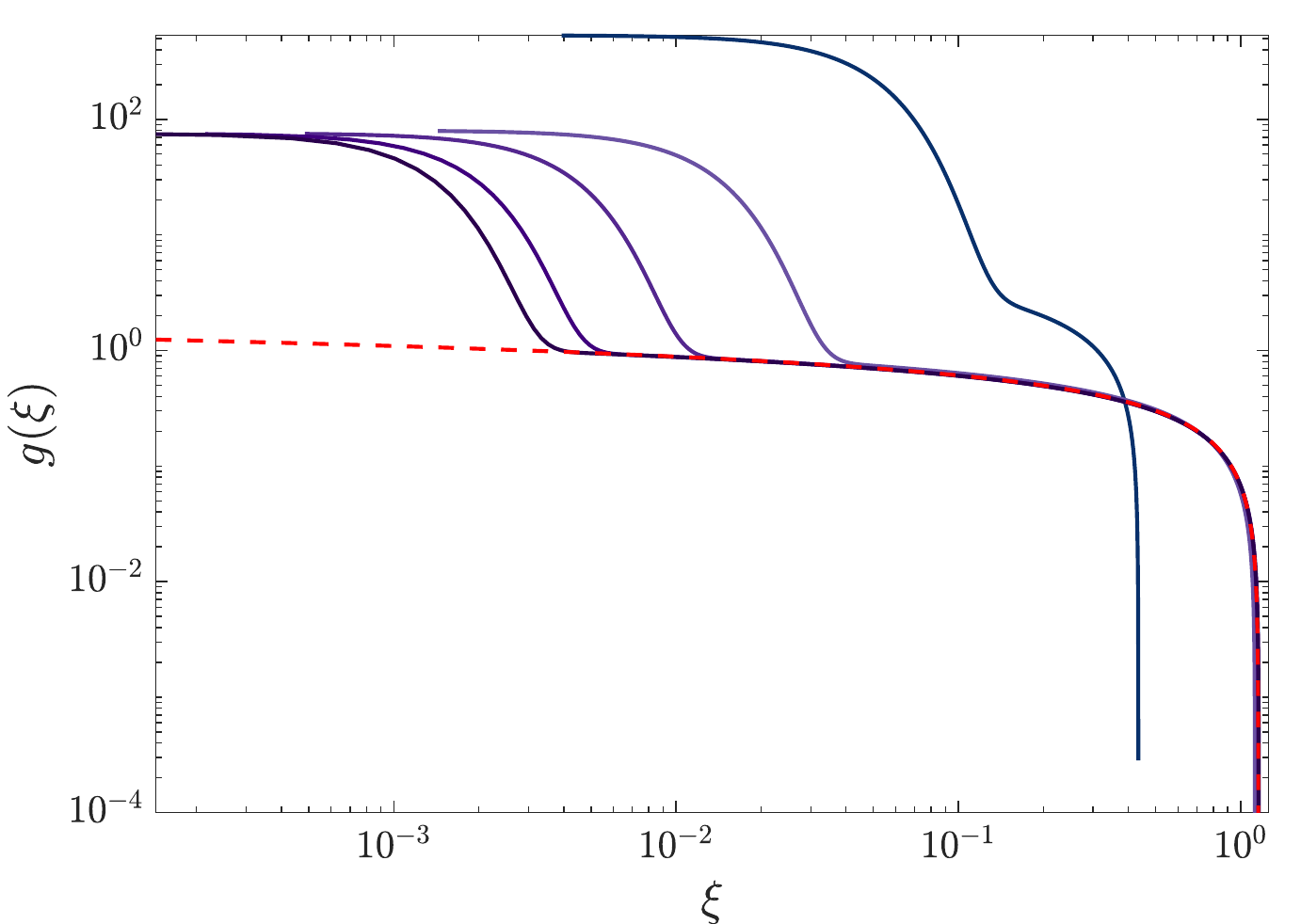}
    \end{subfigure}
    \caption{Numerical solutions of Eqs~(\mainref{eq:Evo2}{33$a$--$e$}) with (\ref{eq:Bellturn}) in a Gaussian channel for $\mathcal{M}=0.1$ and $\lambda=10^{4}$ [$\beta=\mathcal{M}^2\lambda=10^2$]. (a) The interface is plotted in cylindrical coordinates at times  $t = \{0,1,2,4,7\}$. (b) Evolution of the upper (black solid) and lower (black dash-dotted) contact lines. The early time asymptotic prediction for $S_u$ from Eq.~(\ref{eq:GaussETSu}) (cyan dashed), and late-time approximation Eq.~(\ref{eq:LargeBetaSu}) (magenta dashed) are shown for comparison. (c) Interface profiles $H(s,t)$ at times $t=\{7,10^{2},10^{3},5\times10^{3},10^{4}\}$.  
(d) The profiles from~(c) replotted in the self-similar variables of Eqs~(\mainref{eq:PorMed}{C13$a$--$c$}), $\xi = s/(\lambda^{1/4} t^{1/2})$ and $H = 1- \lambda^{-1/2} g(\xi)$. The dashed red curve shows the similarity solution of Eqs~(\mainref{eq:PorMed}{C13$a$--$c$}), computed using a shooting method.
}
    \label{fig:Buoyancy2}
\end{figure}

Increasing the influence of buoyancy further, with $\beta = 10^2$, alters the outer-region dynamics [Fig.~\ref{fig:Buoyancy2}]. Initially, the interface evolution remains similar to the $\beta = 1$ case, and the early-time motion of the upper contact line, $S_u$, is again captured by Eq.~(\ref{eq:GaussETSu}) [Fig.~\mainref{fig:Buoyancy2}{10(b)}]. At later times the inner region near $S_l$ resembles that of $\beta = 1$, although the lower contact line is pinned closer to the origin [Fig.~\mainref{fig:Buoyancy2}{10(b,c)}]. In this parameter regime, injection is subdominant to buoyancy in the outer region, and the film thickness and axial length scale become independent of the viscosity ratio, with $1-H = O(\lambda^{-1/2})$ and $s = O(\lambda^{1/4})$, as suggested by the large-$\beta$ asymptote of $\eta_u$ in Fig.~\mainref{fig:SimSols}{7(b)}. Under these scalings, Eq.~(\mainref{eq:GaussOut}{41$a$}) reduces to the porous medium equation (\mainref{eq:PorMed}{C13$a$}), which describes buoyancy-driven spreading in an unconfined porous layer \citep{lyle_axisymmetric_2005}. While a version of this reduced model was previously obtained by \citet{guo_axisymmetric_2016}, the details of the inner–outer structure have not, to our knowledge, been examined. Numerical integration of Eq.~(\mainref{eq:PorMed}{C13$a$}) using the scheme described in Appendix~\ref{App:GaussianAsymp} yields
\begin{equation}\label{eq:LargeBetaSu}
S_u(t) \approx 1.1552 \lambda^{1/4} t^{1/2}, \qquad (\beta \gg 1),
\end{equation}
which matches closely with the full numerical solution [Fig.~\mainref{fig:Buoyancy2}{10(b)}]. The corresponding similarity solution also shows excellent agreement for all but the earliest time ($t=7$), prior to the establishment of the late-time regime [Fig.~\mainref{fig:Buoyancy2}{10(c)}].

\section{Discussion}
We have shown how the shape of an aquifer plays a central role in determining the rate of spreading of an injected gas current. The two examples we considered (a parabolic and a Gaussian channel) illustrate diverse outcomes.
\subsection{Parabolic channel}
The parabolic channel was taken to be weakly curved, with a dimensionless curvature $\kappa = O(\epsilon)$ along the entire channel. In this geometry, the local slope increases slowly but monotonically with arc length. While the small-slope equations~(\mainref{eq:Evo2}{33$a$--$e$}) do not capture the large slopes that develop at late times, comparison with the full composite model [Figs~\ref{fig:FullEvo} and~\ref{fig:Lambda1Sol}] shows that the reduced-order model retains all essential physics and accurately reproduces the dynamics of the full system before large slopes arise. Even at later times, when slopes become significant, the interface continues to evolve according to the anticipated physical mechanisms, even though discrepancies in interface position and shape may occur due to the approximations of geometric terms such as $r_c \approx s$ and $\varphi \approx -s$ [see Fig.~\mainref{fig:Lambda1Sol}{4(a)} for $\epsilon = 0.1$]. Having established the validity of the small-slope model for spreading in a weakly curved parabolic channel, we adopted this formulation as the principal framework for studying the dynamics.

In a parabolic channel, we identified five temporal regimes in the limit $\mathcal{M} \ll 1$ and $\lambda \ll 1$, corresponding to a highly mobile gas phase and buoyancy forces that are weak relative to injection-induced viscous forces. The asymptotic approximations for the contact lines and associated timescales across each regime are summarised in Table~\ref{Tab:CLs} in dimensional variables, and can be interpreted physically as follows. The timescale for regime \RNum{1} is set by the time required for the injected volume flux to spread over a characteristic arc length $R$, during which the interface and contact lines evolve predominantly under the influence of injection. In this regime the contact lines advance proportionally to $t^{*1/2}$, with the prefactor determined by the injected flux and the fluid viscosities. The upper contact line propagates more rapidly than the lower owing to the formation of a thin gas film near the upper boundary, within which the more mobile gas balances the flux of the more viscous liquid. However, the axisymmetric geometry attenuates the velocity, while the increasing channel slope amplifies the tangential buoyancy component, allowing buoyancy to become relevant.

\begin{table*}
\caption{Summary of the dynamical regimes identified for spreading in a parabolic channel, expressed in terms of dimensional parameters and variables {for $\mu_g\ll \mu_w$ and $ k_0\Delta\rho g h^2 \ll q\mu_g$ }. Column~2 lists the characteristic times that define each regime (with $\sim$ denoting comparable magnitude), while Columns~3 and~4 give the corresponding approximations of the lower and upper contact lines, respectively.  
}
  \begin{ruledtabular}
  \begin{tabular}{cccc} % vertical lines only between body columns
    % Header row (no vertical lines between titles)
    \multicolumn{1}{c}{} &
    \multicolumn{1}{c}{Time} &
    \multicolumn{1}{c}{$S_l^*$} &
    \multicolumn{1}{c}{$S_u^*$} \\
    \midrule
    % Body rows (with vertical lines, no horizontal lines)
    \RNum{1} & $t^* \sim \frac{ h R^2}{q}$ &
    $\left(\frac{\mu_g q }{\pi \mu_w h} \right)^{1/2} t^{*1/2}$ &
    $\left(\frac{\mu_w  q }{\pi \mu_g h} \right)^{1/2} t^{*1/2}$ \\[15pt]

    \RNum{2} & $t^* \sim \frac{\mu_g R^2}{k_0 \Delta \rho g h}$ &
    $\left(\frac{\mu_g q }{\mu_w h} \right)^{1/2} t^{*1/2}$ &
    $\left( \frac{q \mu_w R^2}{2\pi k_0 \Delta \rho g h^2} \right)^{1/2}
    \left[1 - \mathrm{e}^{-\frac{2 k_0 \Delta \rho g h}{\mu_g R^2} t^* } \right]^{1/2}$ \\[15pt]

    \RNum{3} & $t^* \sim \frac{\mu_w R^2 }{k_0 \Delta \rho g h}$ &
    $\left( \frac{q \mu_g R^2}{2 \pi k_0 \Delta \rho g h^2} \right)^{1/2}
    \left[\mathrm{e}^{\frac{2 k_0 \Delta \rho g h}{\mu_w R^2} t^* } - 1\right]^{1/2}$ &
    $\left(\frac{q t^*}{\pi h}\right)^{1/2} \left[1- \mathrm{e}^{-\frac{2k_0 \Delta \rho g h}{\mu_w R^2} t^*}\right]^{-1/2}$ \\[15pt]

    \RNum{4} & $t^* \sim \frac{\mu_w R^2 }{k_0 \Delta \rho g h} \log\left(\frac{\mu_w}{\mu_g}\right)$  &
    $\left( \frac{q \mu_g R^2}{2 \pi k_0 \Delta \rho g h^2} \right)^{1/2}
    \left[\mathrm{e}^{\frac{2 k_0 \Delta \rho g h}{\mu_w R^2} t^* } - 1\right]^{1/2}$   &
    $\left(\frac{q}{h}\right)^{1/2}\left[ t^{*1/2} + 
    \frac{h^2 R}{q} t^{*-1/2} \right]$ \\[15pt]

    \RNum{5} & $t^* \gg \frac{\mu_w R^2 }{k_0 \Delta \rho g h} \log\left(\frac{\mu_w}{\mu_g}\right)$ &
    $\left(\frac{q}{\pi h}\right)^{1/2}\left[ t^{*1/2} - 
    \frac{ \pi h R^2}{q} \zeta_u t^{*-1/2}\right]$ &
    $\left(\frac{q}{h}\right)^{1/2}\left[ t^{*1/2} + 
    \frac{ \pi h R^2}{q} \zeta_l t^{*-1/2} \right]$
  \end{tabular}
  \end{ruledtabular}
\label{Tab:CLs}
\end{table*}

Regime~\RNum{2} corresponds to the onset of buoyancy effects in the gas phase. Buoyancy induces a velocity in the gas of order $U_b \sim \Delta \rho g h k_0/(R \mu_g)$, giving a timescale at which buoyancy begins to influence the flow in the gas, $t^* \sim R/U_b = \mu_g R^2/(k_0 \Delta \rho g h)$. At this stage, buoyancy causes the gas film to thicken, and the upper contact line decelerates exponentially, eventually reaching a plateau at the dimensional position given by the coefficient of the approximation for $S_u^*(t^*)$ from regime \RNum{2} in Table~\ref{Tab:CLs}.

In regime~\RNum{3}, buoyancy effects become significant in the liquid phase. Buoyancy induces a velocity in the liquid of order $U_b \sim \Delta \rho g h k_0/(R \mu_w)$, giving a timescale for buoyancy to influence the liquid, $t^* \sim R/U_b = \mu_w R^2/(k_0 \Delta \rho g h)$. Physically, the liquid responds more slowly than the gas because $\mu_w \gg \mu_g$. Beyond this time, buoyancy drives drainage of the liquid from beneath the gas-filled region, leading to an exponential acceleration of the lower contact line under the combined action of injection and buoyancy. The upper contact line continues to advance slowly under injection, with its motion constrained by the available volume flux, as a substantial fraction of the injected fluid is accommodated by drainage of the underlying liquid. This redistribution of flux is reflected in the prefactor of the exponential appearing in the approximation for $S_u^*(t^*)$ within regime~\RNum{3} (see Table~\ref{Tab:CLs}).

The timescale for regime \RNum{4} follows from that of regime \RNum{3}, scaled by a factor of $\log(\mu_w/\mu_g)$. This factor reflects the additional time required for the liquid to drain almost completely from beneath the gas-filled region, leaving only a thin residual layer of $H^*=O(h \mu_g/\mu_w)$ along the lower boundary. Regime \RNum{5} corresponds to times much greater than those of regime \RNum{4}, by which stage the liquid has completely drained from beneath the gas-filled region, the lower contact line has caught up with the upper, and the interface is approximately horizontal.

\subsection{Gaussian channel}
The second geometry considered was a Gaussian channel, for which both the curvature and slope remain small, with $\kappa = O(\epsilon)$ and $\varphi = O(\epsilon)$. The channel slope decays exponentially with arc length, so that the geometry becomes effectively flat in the far field. Consequently, at late times the boundary geometry influences only the inner region near $S_l$, while the outer dynamics reduce to those of a flat axisymmetric porous channel. In Appendix~\ref{App:GaussianAsymp}, we presented an asymptotic analysis in the small-$\mathcal{M}$ limit, identifying a new distinguished limit in which $\beta \equiv \mathcal{M}^2 \lambda = O(1)$, given by Eqs~(\mainref{eq:GaussInj}{41$a$--$d$}), corresponding to a balance between buoyancy and injection. This limit recovers both the injection-dominated [Eq.~(\ref{eq:GaussInj})] and buoyancy-dominated~[Eqs~(\mainref{eq:PorMed}{C13$a$--$c$})] behaviours reported previously by \citet{guo_axisymmetric_2016} and \citet{nordbotten_similarity_2006}. The boundary between these two regions ($\mathcal{M}^2\lambda = O(1)$) is parallel, in $(\lambda,\mathcal{M})$-parameter space, to the two numerically motivated finite-parameter crossover boundaries presented in figure~7 of \citet{guo_axisymmetric_2016}. These boundaries delimit a finite region in which injection and buoyancy both influence the flow.

The fact that buoyancy can influence the dynamics of the outer region of an axisymmetric geometry in the limit $\mathcal{M} \ll 1$, even when the channel is flat, contrasts with the planar channel, in which \citet{Castellucci_2026} found buoyancy effects to remain subdominant to injection. This difference is caused by geometric attenuation of the injection velocity in an axisymmetric channel: as the radius increases, the injected flux spreads over an increasing cross-sectional area (or, under the small-slope approximation, increasing arc length with $r \approx s$), causing the radial velocity to decay. This mechanism can be understood through a scaling argument. Buoyancy acts to smooth variations in the interface slope by inducing a radial pressure gradient of order $\Delta \rho g H^*_{r^*}$. From the outer scalings for the Gaussian channel [Eqs~(\mainref{eq:GaussOut}{41$a$--$d$})], we found that $1-H = O(\mathcal{M})$ and $r = O(\mathcal{M}^{-1/2})$, therefore the dimensional scale for the gradient of the interface satisfies $H^*_{r^*} \sim \mu_g^{3/2} h/(R \mu_w^{3/2})$, giving a buoyancy-induced pressure gradient of order $\Delta \rho g h \mu_g^{3/2}/(R \mu_w^{3/2})$. The source generates a horizontal velocity of order $q/(r^* h)$. Accounting for the outer region extending over a radial lengthscale of $O(\mathcal{M}^{-1/2})$, this velocity scale becomes $q\mu_w^{1/2}/(\mu_g^{1/2} h R)$. Since the liquid is much more viscous than the gas, it sets the pressure gradient required to sustain this velocity, which is of order $q (\mu_g \mu_w)^{1/2}/(h R k_0)$. Balancing this pressure gradient with the buoyancy-induced pressure gradient yields the condition for buoyancy to influence the outer region, $\beta \equiv k_0 \mu_g \Delta \rho g h^2/(q \mu_w^2) \sim 1$.  

\subsection{Practical implications}
\begin{table*}
    \caption{Table of dimensional and dimensionless model parameters. Fluid properties for hydrogen were obtained from \citet{Castellucci_2026}, and for CO$_2$ from \citet{Bickle2007}. The channel width and radial extent of a representative anticline follow \citet{mortimer_dynamic_2024} and \citet{Harati2023}. Hydrogen injection rates reflect values typically used in numerical simulations due to limited field data \citep{Saeed2024,Harati2023}, while CO$_2$ rates are based on estimates of injection rates at current sites of 1--2 $\si{Mt.yr^{-1}}$ \citep{Bickle2007}. Permeability values correspond to typical ranges of aquifers suitable for storage \citep{Castellucci_2026}.
}
\begin{ruledtabular}
  \begin{tabular}{lccc} % vertical lines only between body columns
    % Header row (no vertical lines between titles)
    \multicolumn{1}{l}{Parameters} &
    \multicolumn{1}{c}{Hydrogen} &
    &
    \multicolumn{1}{c}{CO$_2$} \\
    \hline\\[-7pt]
    Gas density $\rho_g$ ($\si{kg.m^{-3}}$) & $10^{-1}$ -- $10$& & 420 -- 610 \\
     Gas viscosity $\mu_g$ ($\si{kg.m^{-1}.s^{-1}}$) & $10^{-6} $& & 3 -- 5 $\times$ $10^{-5}$\\
     Density difference $\Delta \rho $ ($\si{kg.m^{-3}}$) & $10^3$& & 390 -- 680 \\
     Injection rate $q$ ($\si{m^{3}.s^{-1}}$)& $10^{-1}$ -- 10 & &$10^{-3}$ -- $10^{-1}$\\
    Brine density $\rho_w$ ($\si{kg.m^{-3}}$)& & $10^3$&  \\
         Brine viscosity $\mu_w$ ($\si{kg.m^{-1}.s^{-1}}$) & & $10^{-4}$ &\\
     Channel width $h$ ($\si{m}$)& & 10 -- 50 & \\
     Radial lengthscale $R$ ($\si{m}$)& & $10^2$ -- $10^3$ & \\
     Permeability $k_0$ ($\si{m^2}$)& & $10^{-14}$ -- $10^{-12}$&   \\
    \hline
    Dimensionless parameters: & & &\\
    $\epsilon = h/R$ & & $10^{-2}$ -- $0.5$ &  \\
    $\mathcal{M} = \mu_g/\mu_w$  & $10^{-2}$ & &0.3 -- 0.5 \\
    $\lambda = k_0 \Delta \rho g h^2 / ( q \mu_g)$ & $10^{-3}$ -- 300 & & $8 \times10^{-4}$ -- 600 \\
    $\beta \equiv \mathcal{M}^2 \lambda $   & $10^{-7}$ -- 3 $\times$ $10^{-2}$ & &7$\times 10^{-5}$ -- 200
  \end{tabular}
\label{tab:params}
\end{ruledtabular}
\end{table*}
% \end{comment}

Table~\ref{tab:params} presents order-of-magnitude estimates of the dimensional and dimensionless parameters for the model applied to CO$_2$ and H$_2$ injection into a typical anticline structure. As can be expected, the viscosity ratios for both H$_2$-brine and CO$_2$-brine are in the small-$\mathcal{M}$ regimes. The range of plausible values for $\lambda$ covers many orders of magnitude, reflecting the variations of permeability, channel heights and fluid properties across storage sites as well as the different rates of injection that may be used.

Estimates of $\beta$ for H$_2$ suggest that, in a Gaussian channel, the flow lies in an injection-dominated regime, with late-time spreading described by Eq.~(\ref{eq:GaussInj}). In contrast, the higher viscosity and typically lower injection rates associated with CO$_2$ storage lead to a broader parameter range over which buoyancy effects may influence the dynamics at some storage sites. A buoyancy-dominated regime results in a thinner outer film of gas, and consequently less efficient storage compared with the injection-dominated regime.

Previous studies have shown that, for sufficiently slow injection into anticlines, buoyancy flattens the interface at early times \citep{mortimer_dynamic_2024}. By contrast, at higher injection rates, a long gas finger develops along the upper boundary and propagates rapidly \citep{hagemann_mathematical_2015}, consistent with the behaviour observed in regime \RNum{1} of a parabolic channel and the regime $\beta \ll 1$ in a Gaussian channel. For a parabolic channel, our results indicate that, even at elevated injection rates, buoyancy ultimately restores a flattened interface, albeit at later times, provided that the anticline extends over sufficiently large length scales for this behaviour to manifest. Moreover, the onset of flattening occurs earlier in channels with larger curvature, quantified by $h/R^2$, and with higher permeability $k_0$ (Table~\ref{Tab:CLs}). For underground hydrogen storage, a flat interface is advantageous because it maximises utilisation of the channel volume. Furthermore, curved anticlines possess a spill point, beyond which gas is lost to a neighbouring aquifer \citep{zivar_underground_2021}. Notably, we found that, in the asymptotic regime $\lambda \ll 1$ and $\mathcal{M} \ll 1$, the upper contact line approaches a plateau at $S_u = \left( q \mu_w R^2 / (2\pi k_0 \Delta \rho g h) \right)^{1/2}$, and remains effectively pinned for $ \lambda^{-1} \ll t \ll (\mathcal{M} \lambda)^{-1}$. If the plateau position lies upstream of the spill point, buoyancy may suppress spillage by arresting the advance of $S_u$. Subsequent injection is then accommodated predominantly through drainage of the underlying liquid layer rather than further propagation of the upper contact line, thereby also maximising utilisation of the available channel volume. This mechanism suggests a practical strategy for enhancing storage capacity while maintaining control over current evolution, even at comparatively high injection rates.

Using the parameter ranges listed in Table~\ref{tab:params} we can estimate the order of magnitude of the timescales associated with regimes \RNum{1}–\RNum{5} in a parabolic channel. Consider hydrogen injection at a rate $q = 1~\si{m^3.s^{-1}}$ into an approximately parabolic anticline of permeability $k_0 = 10^{-12}~\si{m.^2}$ and characteristic width $h = 10~\si{m}$, whose vertical coordinate varies by $10~\si{m}$ over a radial length scale $R = 10^{2}~\si{m}$. Using the density difference between hydrogen and brine, together with the hydrogen viscosity listed in Table~\ref{tab:params}, we obtain $\lambda \approx 1$. The resulting timescale for regime \RNum{2} is around one day, while regimes \RNum{3} and \RNum{4} can be expected to arise after approximately $115$ days and $1.5$ years, respectively. At longer times, regime \RNum{5} emerges and the interface will be effectively flattened by buoyancy. Owing to the small viscosity ratio of the hydrogen–brine system ($\mathcal{M}=10^{-2}$), regimes \RNum{4}–\RNum{5} are unlikely to be realised on practical operational timescales when $\lambda$ is sufficiently small. However, this restriction might be relaxed in anticlines with larger curvature, where the dynamics are governed by Eqs~(\mainref{eq:Evo}{30$a$--$e$}) or for CO$_2$ injection, due to its larger viscosity. Furthermore, in practical applications, whether or not the flow reaches these late-time regimes will also depend on the total volume of gas injected into a given storage site. This can be assessed by comparing the time required to inject a given target volume, $t^* = V^*/q$, with the timescales for the different regimes.

Although this section has focused mainly on CO$_2$ and H$_2$ storage, the modelling framework may also provide insight into the storage of other buoyant gases such as compressed air \citep{Sun2023CAEScomparison}. However, gas compressibility and thermodynamic effects are expected to play a more prominent role in the evolution of the gas bubble and the overall energy storage efficiency \citep{Zhang2025CAESreview}. Compressibility effects could be incorporated into the present model using a methodology similar to that developed in \citet{Castellucci_2026}.
\subsection{Final remarks}
In summary, we have extended existing theoretical models for viscous gravity currents in confined porous layers to account for curved channel geometries. By formulating the governing equations in coordinates local to the channel centreline, and exploiting the slenderness of the channel, we constructed a composite approximation that yields the evolution equation~(\mainref{eq:Evo}{30$a$}) for the gas–liquid interface, retaining buoyancy effects exactly. This framework describes the evolution of an injected gas current through thin channels with finite curvature and accommodates large channel slopes. The model is expressed in terms of general geometric functions, so specifying the curvature $\kappa(s)$, inclination angle $\varphi(s)$, and radial coordinate $r_c(s)$ in the full composite equations~(\mainref{eq:Evo}{30$a$--$e$}) allows simulation of the interface evolution from prescribed initial conditions across a broad range of geometries. Using the composite model, we derived a simplified evolution equation~(\mainref{eq:Evo2}{33$a$}) valid for weakly-curved channels with small slopes. Motivated by hydrogen and CO$_2$ storage in dome-shaped anticlines, we used the small-slope model to study gas injection into parabolic- and Gaussian-shaped anticlines.

To isolate the effects of geometry, several simplifying assumptions were made in deriving the models given by Eqs~(\mainref{eq:Evo}{30$a$--$e$}) and Eqs~(\mainref{eq:Evo2}{33$a$--$e$}). In particular, we assumed that the interface remains axisymmetric. However, it is well established that the displacement of a more viscous fluid by a less viscous gas (i.e. $\mathcal{M} < 1$) is susceptible to a viscous fingering instability \citep{Taylor_1958}. Recent work by \citet{hinton_buoyancy_2022} examined the stability of a gas--liquid interface in a flat, axisymmetric channel in the regime $\mathcal{M} < 1$ and $\lambda \ll 1$ using a three-dimensional formulation that allows for azimuthal variation of the interface, $z = H(\theta,r,t)$. They showed that fingering may arise at early times, but once buoyancy segregates the fluids, the fingering instability is suppressed. Once segregation has occurred and the injected gas forms a radially extensive thin film along the upper boundary, as observed in regime \RNum{1} for a parabolic channel and for $\beta \ll 1$ in a Gaussian channel, the total pressure drop is distributed over a larger radial extent. The resulting reduction in the local pressure gradient diminishes its sensitivity to small interfacial perturbations, thereby suppressing the positive-feedback mechanism responsible for viscous fingering. In this self-similar regime, using a linear stability analysis, the authors confirmed that the interface is stable to both axisymmetric and non-axisymmetric perturbations.

Other physical mechanisms may become important under certain conditions. For sufficiently small pore sizes or large interfacial tension, capillary effects may generate a capillary fringe and an associated pressure jump across the interface, thereby invalidating the sharp-interface assumption \citep{zheng_self-similar_2019}. Such effects may additionally promote capillary trapping. If the dissolution timescale is short compared with the characteristic timescale of storage, solubility may significantly influence the evolution by reducing the rate of current migration. This effect can be particularly pronounced in the small-$\mathcal{M}$ regime, where the enhanced interfacial area between the fluids promotes increased mass transfer, as demonstrated by \citet{macminn_co2_2011}. Likewise, if the streamwise pressure drop is sufficiently large, gas compressibility may slow the rate of spreading \citep{Castellucci_2026}. Furthermore, fluid properties depend on pressure and temperature and may therefore vary between aquifers. Spatial variations in temperature may introduce additional dynamical effects, including thermally driven convection or, in the presence of concentration gradients, double-diffusive convection \citep{javaheri2010linear}.

In this study, we focused on gravity currents in weakly curved channels, described by the small-slope approximation Eqs~(\mainref{eq:Evo2}{33$a$--$e$}). However, the full composite model remains valid for channels with stronger curvature, offering a framework to investigate high-curvature, slender geometries. In the more general setting, the term proportional to $\kappa(s)$ in Eqs~(\mainref{eq:Evo}{30$a$,$b$}), which is neglected in the small-slope approximation, may become significant and influence the interface evolution. This term modifies the tangential component of buoyancy and may influence the dynamics. A natural extension would be to consider geometries containing a spill point, beyond which the channel slope becomes positive. In such cases, the injected gas may separate, necessitating the simulation of multiple disconnected bubbles. We leave such extensions to future work.

\section*{Acknowledgements}

% LM acknowledges funding from the EPSRC Industrial Decarbonisation Research and Innovation Centre (IDRIC MIPs 7.4 \& 7.8; EP/V027050/1). 

This work was partially supported by the Engineering \& Physical Sciences Research Council Grant Award Industrial Decarbonisation Research and Innovation Centre (IDRIC MIPs 7.4 \& 7.8; EP/V027050/1) and UKRI3155: APP50312: MaxStoreUK.  For the purpose of open access, the authors have applied a Creative Commons Attribution (CC-BY) licence to any author accepted manuscript version arising.

\section*{Data availability}
The data that support the findings of this article are openly available at \citep{Castellucci2026Figshare}.
\appendix
\section{Numerical methods}
\label{sec:Numerics}
The evolution Eqs~(\mainref{eq:Evo}{30$a$–$e$}) and (\mainref{eq:Evo2}{33$a$–$e$}) are solved using the method of lines after mapping the physical domain $s \in [S_l(t), S_u(t)]$ to a fixed computational domain $\xi \in [0,1]$. The transformed equations and boundary conditions are discretised in $\xi$ using a second-order finite-volume scheme, yielding a system of ODEs for $H$, $S_l$, and $S_u$, which are integrated in time using \MATLAB's stiff solver \texttt{ode15s}. For Eqs~(\mainref{eq:Evo}{30$a$–$e$}), the ODE system is coupled to the geometric relations given by Eqs~(\ref{eq:ParaCurv}–\ref{eq:ParaAngle}). At each time step, $r_c(s)$ is obtained by solving Eq.~(\ref{eq:ParaArc}), and $\varphi(s)$ is computed via numerical integration of Eq.~(\ref{eq:ParaAngle}); these quantities are then used to update the discretised equations. For Eqs~(\mainref{eq:Evo2}{33$a$–$e$}), corresponding to weakly-curved parabolic and Gaussian channels, explicit expressions for the angle to the horizontal, $\varphi^\dag$, are substituted directly into the evolution equations and boundary conditions---Eq.~(\ref{eq:ParaSmallSlope}) for the parabolic case and Eq.~(\ref{eq:Bellturn}) for the Gaussian case.

\section{Parabolic channel: small-$\mathcal{M}$ asymptotics}
\label{App:ParaAsym}
In this appendix, we present an asymptotic analysis of spreading in a parabolic channel, governed by Eqs~(\mainref{eq:Evo2}{33$a$–$e$}), in the limits $\mathcal{M} \ll 1$ and $\lambda \ll 1$. We derive reduced-order models and analytical approximations describing the dynamics across the distinct temporal regimes \RNum{1}–\RNum{5}; within each regime, the solution is further resolved into multiple spatial regions as sketched in Fig.~\ref{fig:HRegions}, each governed by its own reduced-order description.
\subsection{Regime \RNum{1}: formation of a thin gas film}
For $t \ll \lambda^{-1}$, the solution of Eqs~(\mainref{eq:Evo2}{33$a$--$e$}) exhibits two spatial regions, denoted \RNum{1}$_a$ and \RNum{1}$_b$, as shown in Fig.~\ref{fig:HRegions}. Region \RNum{1}$_a$ is characterised by $H = O(1)$ and $s = O\left(\mathcal{M}^{1/2}\right)$. In region \RNum{1}$_b$ a thin film of gas forms along the upper boundary; here $1 - H = O(\mathcal{M})$ and $s = O\left(\mathcal{M}^{-1/2}\right)$.

\textbf{Region \RNum{1}$_a$.} 
Here, we set $s = \mathcal{M}^{1/2} \bar{s}$, $S_l = \mathcal{M}^{1/2} \bar{S}_l(t)$ and $H(s,t) = \bar{H}(\bar{s},t)$. Substituting this into Eq.~(\mainref{eq:Evo2}{33$a$}) and using Eq.~(\ref{eq:ParaSmallSlope}), to leading order, we recover the hyperbolic problem
\begin{equation}\label{eq:R1aEQ}
    \bar{H}_t + \frac{1}{2\pi \bar{s} (1-\bar{H})^2} \bar{H}_{\bar{s}} = 0.
\end{equation}
Here, the buoyancy term appears at O($\lambda$) and is negligible on this early timescale. Physically, the regime $\lambda \ll 1$ corresponds to buoyancy forces being much weaker than injection-induced viscous forces. Thus, for $t \ll \lambda^{-1}$, the interface evolves rapidly under injection, while buoyancy has not yet had sufficient time to influence the flow. The factor $\bar{s}^{-1}$ multiplying the second term in Eq.~(\ref{eq:R1aEQ}) accounts for the attenuation of the fluid velocity in an axisymmetric geometry.
%, arising from the increase in the cross-sectional area with increasing $s$. 
Integrating Eq.~(\ref{eq:R1aEQ}) along characteristics, assuming the initial interface is confined to a region close to the origin $\bar{s} = o(1)$, yields the solution
\begin{equation}\label{eq:R1aSol}
    \bar{H} = 1 - \left(\frac{t}{\pi \bar{s}^2} \right)^{1/2}.
\end{equation}
Setting $\bar{H} = 0$ in Eq.~(\ref{eq:R1aSol}) gives the approximation for the evolution of the lower contact line
\begin{equation}\label{eq:R1Sl}
    \bar{S}_l(t) = \left({t}/{\pi}\right)^{1/2}. 
\end{equation}

\textbf{Region \RNum{1}$_b$.} In the thin-film region we set $s = \hat{s}/\mathcal{M}^{1/2}$, $S_u = \hat{S}_u/\mathcal{M}^{1/2}$ and $H = 1-\mathcal{M} \hat{H}(\hat{s},t)$. Using Eq.~(\ref{eq:ParaSmallSlope}), to leading order, Eq.~(\mainref{eq:Evo2}{33$a$}) then becomes
\begin{equation}\label{eq:R1bEQ}
    \hat{H}_t + \frac{1}{2 \pi \hat{s} (1+\hat{H})^2} \hat{H}_{\hat{s}} = 0.
\end{equation}
Analogously to region \RNum{1}$_a$, the buoyancy term would appear at $O(\lambda)$ in Eq.~(\ref{eq:R1bEQ}). Equation~(\ref{eq:R1bEQ}) has the solution 
\begin{equation}\label{eq:R1bSol}
    \hat{H} = \left(\frac{t}{\pi \hat{s}^2}\right)^{1/2} - 1, \quad \text{for} \quad \hat{s} \leq \hat{S}_u.
\end{equation}
The solution Eq.~(\ref{eq:R1bSol}) matches with Eq.~(\ref{eq:R1aSol}) to leading order in $\mathcal{M}$ as $\hat{s}\rightarrow 0$. Setting $\hat{H} = 0$ in Eq.~(\ref{eq:R1bSol}) gives the solution for the upper contact line
\begin{equation}\label{eq:R1Su}
    \hat{S}_u(t) = \left({t}/{\pi}\right)^{1/2}.
\end{equation}
The solutions Eq.~(\ref{eq:R1bSol}) and Eq.~(\ref{eq:R1aSol}) constitute the axisymmetric analogue of the inner–outer structure describing small-$\mathcal{M}$ spreading in a flat planar channel, as analysed by \citet{Castellucci_2026}.

\subsection{Regime \RNum{2}: initial growth of the thin film of gas}
At later times, with $t = O\left(\lambda^{-1}\right)$, there are again two spatial regions, as shown in Fig.~\ref{fig:HRegions}. The spatial region labelled \RNum{2}$_a$ retains the structure observed in \RNum{1}$_a$, while in region \RNum{2}$_b$ (formerly \RNum{1}$_b$), buoyancy begins to influence the flow, leading to a thickening of the thin gas layer and a corresponding deceleration of the upper contact line.

\textbf{Region \RNum{2}$_a$.} Here, setting $t = \lambda^{-1} \tilde{t}$, $s = (\mathcal{M}/\lambda)^{1/2} s^\prime$ and $S_l = (\mathcal{M}/\lambda)^{1/2} S_l^\prime$ in Eq.~(\mainref{eq:Evo2}{33$a$}) with Eq.~(\ref{eq:ParaSmallSlope}) recovers the same leading-order problem from earlier times given by Eq.~(\ref{eq:R1aEQ}). The lower contact line therefore still evolves according to Eq.~(\ref{eq:R1Sl}), with outer limit $H = 1- (t\mathcal{M}/\pi)^{1/2}/s$ for $(\mathcal{M}/ \lambda)^{1/2} \ll s \ll (\lambda \mathcal{M})^{-1/2}$.

\textbf{Region \RNum{2}$_b$.} In the thin-film region we set $t = \lambda^{-1} \tilde{t}$, $s = (\lambda \mathcal{M})^{-1/2} \tilde{s}$, $S_u = (\lambda \mathcal{M})^{-1/2} \tilde{S}_u$ and $H = 1-\mathcal{M} \tilde{H}(\tilde{s},\tilde{t})$ in Eq.~(\mainref{eq:Evo2}{33$a$}) with Eq.~(\ref{eq:ParaSmallSlope}), yielding
\begin{align}\label{eq:R2bEQ}
    \tilde{H}_{\tilde{t}} + \frac{1-2\pi \tilde{s}^2}{2\pi \tilde{s} (1+\tilde{H})^2}\tilde{H}_{\tilde{s}} =  \frac{2\tilde{H}}{1+\tilde{H}},
\end{align}
to leading order. In this regime, the buoyancy term proportional to $H_s$ in Eq.~(\mainref{eq:Evo2}{33$a$}) is subdominant (at $O(\mathcal{M}^2 \lambda)$), but the slope-induced buoyancy term proportional to $s$ remains at leading order. Consequently, buoyancy and injection act comparably in controlling the interface evolution through Eq.~(\ref{eq:R2bEQ}). Compared with Eq.~(\ref{eq:R1bEQ}), where buoyancy is subdominant, it now modifies the coefficient of $\tilde{H}_s$, adding a term in the numerator that decreases the coefficient with increasing $\tilde{s}$. This reflects a buoyancy-induced reduction in interface advection. Additionally, the right-hand side of Eq.~(\ref{eq:R2bEQ}) contains a term driving growth of $\tilde{H}$ along characteristics, or equivalently decay in $H$. The characteristic equations associated with Eq.~(\ref{eq:R2bEQ}) can be written as
\begin{equation}\label{eq:R2bChar}
    \Dfirst{\tilde{H}}{\tilde{s}} = \frac{4 \pi \tilde{s} \tilde{H}(1+\tilde{H})}{1 - 2\pi \tilde{s}^2} \quad \text{along} \quad \Dfirst{\tilde{s}}{\tilde{t}} = \frac{1- 2 \pi \tilde{s}^2}{2 \pi \tilde{s} ( 1+ \tilde{H})^2}.
\end{equation}
Assuming $\tilde{H} = \tilde{H}_0(\xi)$ at $\tilde{t} = 0$, the first equation for $\tilde{H}$ in Eq.~(\ref{eq:R2bChar}) can be integrated to give
\begin{equation}\label{eq:R2bCharHSol}
    \tilde{H} = \frac{f(\xi)}{1 - 2\pi \tilde{s}^2 - f(\xi)}, \quad \text{where} \quad f(\xi) = \frac{\tilde{H}_0(\xi)(1 - 2\pi \xi^2)}{1 + \tilde{H}_0(\xi)}.
\end{equation}
Equation (\ref{eq:R2bCharHSol}) can then be substituted into the second characteristic equation for $\tilde{s}$ in Eq.~(\ref{eq:R2bChar}), yielding
\begin{equation}\label{eq:R2bChar2}
    \Dfirst{\tilde{s}}{\tilde{t}} = \frac{(1 - 2\pi\tilde{s}^2 - f(\xi))^2}{2 \pi \tilde{s} (1 - 2 \pi \tilde{s}^2)}.
\end{equation}
Integrating Eq.~(\ref{eq:R2bChar2}) gives
\begin{equation}\label{eq:R2bCharsSol}
    \log(1 - 2 \pi \tilde{s}^2 - f(\xi)) - \frac{f(\xi)}{1 - 2 \pi \tilde{s}^2 - f(\xi)} = - 2 \tilde{t}.
\end{equation}
Setting $f(\xi) = 0$ in Eq.~(\ref{eq:R2bCharsSol}) recovers the solution for the upper contact line
\begin{equation}\label{eq:R2Su}
    \tilde{S}_u(t) = \left(\frac{1 - \mathrm{e}^{-2  \tilde{t}}}{ 2 \pi} \right)^{1/2},
\end{equation}
which for $\tilde{t}  \ll 1$ matches with the solution from earlier times given by Eq.~(\ref{eq:R1Sl}) valid for $t \ll \lambda^{-1}$. Equation (\ref{eq:R2Su}) suggests that the upper contact line asymptotes to $\tilde{S}_u \to 1/\sqrt{2\pi}$ for $\tilde{t} \gg 1$. Reverting to regular coordinates, this implies the upper contact line plateaus at
\begin{equation}\label{eq:R2Plateau}
    S_u(t) \approx \left(\frac{1}{2 \pi \mathcal{M} \lambda}\right)^{1/2}, \quad \text{for} \quad t \gg \lambda^{-1}.
\end{equation}

\subsection{Regime \RNum{3}: gas filling and liquid drainage}
For $\lambda^{-1} \ll t \ll (\mathcal{M} \lambda)^{-1}$, the upper contact line remains effectively stationary at Eq.~(\ref{eq:R2Plateau}). During this stage, the liquid continues to drain below the interface and the system transitions from a thin gas layer near the upper boundary to a thin liquid film along the lower boundary. Region \RNum{2}$_b$ splits into two. In the longest region \RNum{3}$_b$, $H = O(1)$ and, with the upper contact line frozen, the interface preserves its spatial profile while its height decreases in time. Region \RNum{3}$_c$ is a short region which connects \RNum{3}$_b$ with the stationary upper contact line; region \RNum{3}$_a$ connects \RNum{3}$_b$ with the lower contact line. These three regions are sketched in Fig.~\ref{fig:HRegions}. For convenience, we scale time with $(\mathcal{M}\lambda)^{-1}$, accommodating the limit $t \ll (\mathcal{M}\lambda)^{-1}$.

\textbf{Region \RNum{3}$_a$.}
In this region, we set $t = (\mathcal{M} \lambda)^{-1} \check{t} $, $s = \lambda^{-1/2} \check{s}$, $S_l = \lambda^{-1/2} \check{S}_l(\check{t})$ and $H(s,t) = \check{H}(\check{s},\check{t})$ in Eq.~(\mainref{eq:Evo2}{33$a$}) and Eq.~(\ref{eq:ParaSmallSlope}). To leading order we then recover the hyperbolic equation
\begin{equation}\label{eq:R3aEQ}
    \check{H}_{\check{t}} + \frac{2 \pi \check{s}^2 (1-\check{H})^2 + 1}{2\pi\check{s}(1-\check{H})^2} \check{H}_{\check{s}} = -2\check{H}.
\end{equation}
At this stage, buoyancy begins to influence the region around the lower contact line through the term proportional to $s$ in Eq.~(\mainref{eq:Evo2}{33$a$}), while the term proportional to $H_s$ remains subdominant at $O(\lambda)$. In Eq.~(\ref{eq:R3aEQ}), the coefficient of the advective term $\check{H}_{\check{s}}$ differs from that in the earlier regions \RNum{1}$_a$ and \RNum{2}$_a$ given by Eq.~(\ref{eq:R1aEQ}) by the addition of a linear factor of $\check{s}$, resulting in an increased advection of $\check{H}$ with arc length. This arises from the hydrostatic pressure gradient that has developed along the centreline of the channel. The additional source term on the right-hand side, $-2 \check{H}$, shows that buoyancy causes a decay of the interface height along characteristics. 

The characteristic equations of Eq.~(\ref{eq:R3aEQ}) can be written as
\addtocounter{equation}{1}
\begin{equation}\label{eq:R3aChar}
    \Dfirst{\check{H}}{\check{t}}  = - 2\check{H}, \quad \text{along} \quad \Dfirst{\check{s}}{\check{t}} = \frac{2 \pi \check{s}^2 (1-\check{H})^2 + 1}{2 \pi \check{s}(1-\check{H})^2}.\tag{B15$a,b$}
\end{equation}
Integrating Eq.~(\mainref{eq:R3aChar}{B15$a$}), with $\check{H} = \check{H}_1(\check{\xi})$ at $\check{t} = 0$, gives
\begin{equation}\label{eq:R3aHSol}
    \check{H} = \check{H}_1(\check{\xi}) \mathrm{e}^{-2\check{t}},
\end{equation}
which shows that the interface height within the inner region begins to decay exponentially on this timescale due to buoyancy. Substituting Eq.~(\ref{eq:R3aHSol}) into Eq.~(\mainref{eq:R3aChar}{B15$b$}) gives
\begin{equation}\label{eq:R3asEQ}
     \Dfirst{}{\check{t}}\left(\check{s}^2\right) - 2\check{s}^2 = \frac{1}{\pi \left(1-\check{H}_1(\check{\xi}) \mathrm{e}^{-2\check{t}}\right)^2}.
\end{equation}
Integrating Eq.~(\ref{eq:R3asEQ}) and applying the initial condition $\check{s} = \check{\xi}$ at $\check{t}=0$, yields
\begin{equation}\label{eq:R3asSol}
    \check{s} = \left[ \check{\xi}^2\mathrm{e}^{2\check{t}} +  \frac{\mathrm{e}^{2\check{t}} - 1}{2 \pi \left(1-\check{H}_1(\check{\xi})\right)\left(1-\mathrm{e}^{-2\check{t}} \check{H}_1(\check{\xi}) \right)} \right]^{1/2}.
\end{equation}
Taking $\check{S}_l(0) = 0$ and setting $\check{H}_1 =0$ in Eq.~(\ref{eq:R3asSol}) then gives an approximation for the lower contact line:
\begin{equation}\label{eq:R3Sl}
    \check{S}_l(\check{t}) = \left( \frac{\mathrm{e}^{2 \check{t}}-1}{2\pi } \right)^{1/2},
\end{equation}
which matches with the solution for $S_l$ from regimes \RNum{1} and \RNum{2}, given by Eq.~(\ref{eq:R1Sl}), for $\check{t} \ll 1$. The solution given by Eq.~(\ref{eq:R3Sl}) captures the acceleration of the lower contact line seen in Fig.~\ref{fig:FullEvo}.

\textbf{Region \RNum{3}$_b$.} Here, we set $s=(\lambda\mathcal{M})^{-1/2} \breve{s}$ and $H = \breve{H}(\breve{s},\check{t})$. To leading order, Eq.~(\mainref{eq:Evo2}{33$a$}) with Eq.~(\ref{eq:ParaSmallSlope}) becomes
\begin{equation}\label{eq:R3bEQ}
    \breve{H}_{\check{t}} + \frac{1}{\breve{s}} \left[ \breve{s}^2 \breve{H} \right]_{\breve{s}} = 0.
\end{equation}
Here the advective injection term and the buoyancy term proportional to $H_s$ that would otherwise appear in Eq.~(\ref{eq:R3bEQ}) are subdominant (at $O(\mathcal{M})$ and $O(\lambda \mathcal{M})$, respectively); only the slope-induced buoyancy contribution remains at leading order. The injected flux is therefore too weak, relative to buoyancy, to drive the interface further, and buoyancy causes an internal rearrangement of the gas. Since the upper contact line is approximately stationary, we seek a spatially uniform solution $\breve{H}=\breve{H}(\check{t})$. Substituting this into Eq.~(\ref{eq:R3bEQ}), integrating and matching to the earlier-time limit from \RNum{2}$_b$, $H\approx1$ to leading order in $\mathcal{M}$, yields
\begin{equation}\label{eq:R3bEQSol}
    \breve{H} = \mathrm{e}^{-2 \check{t}}.
\end{equation}
The solution given by Eq.~(\ref{eq:R3bEQSol}) indicates that $\breve{H} = O(\mathcal{M})$ when $t = O(\log(\mathcal{M}^{-1}) (\mathcal{M} \lambda)^{-1})$, which gives the timescale for the onset of regime \RNum{4}.

\textbf{Region \RNum{3}$_c$.} Close to the upper contact line, the interface adjusts sharply, linking the spatially uniform solution in Eq.~(\ref{eq:R3bEQSol}) to the contact line. In this narrow region, the interface, expressed as $H(s,t)$, is steep; however, in cylindrical coordinates, it corresponds to a nearly flat interface. We denote the location of the advancing front by $S_u(t) \approx \acute{s}_0(\check{t})/(\mathcal{M} \lambda)^{1/2} + (\mathcal{M} \lambda)^{1/2} \acute{s}_1(\check{t})$ and set $s = \acute{s}_0(\check{t})/(\mathcal{M} \lambda)^{1/2} + (\mathcal{M} \lambda)^{1/2} \acute{s}$,  $H(s,t) =  s^2/2 - s_0^2/(2\mathcal{M} \lambda) + \acute{H}(\acute{s},\check{t})$ , where $\acute{s}_0$ is the leading-order front location and $\acute{s}_1$ represents the correction to the upper contact line induced by the dynamics of this short region. In this rescaling, the large geometric contribution to $H$ associated with the curvature of the domain has been subtracted so that $\acute{H}$ captures only the $O(1)$ variation of the interface across the front. Under this rescaling Eq.~(\mainref{eq:Evo2}{33$a$}) with Eq.~(\ref{eq:ParaSmallSlope}) reduces to
\begin{equation}\label{eq:R3cEQ}
    \left[ \left(\acute{H} + \acute{s}_0 \right) \acute{H}_{\acute{s}} \right]_{\acute{s}} = 0,
\end{equation}
to leading order. In the vicinity of the stationary upper contact line, the interface evolves quasi-steadily, with the time derivative being $O(\mathcal{M}\lambda)$. In this region, injection is subdominant (at $O(\mathcal{M})$), while both the slope-induced buoyancy term (proportional to $s$) and the slope-dependent term ($H_s$) contribute at leading order. 

Equation (\ref{eq:R3cEQ}) can be integrated, which after matching with Eq.~(\ref{eq:R3bEQSol}) using $\acute{H} = \mathrm{e}^{-2\check{t}}$ and $\acute{H}_{\acute{s}} = 0$ as $\acute{s} \to 0$, and applying the boundary condition $\acute{H} = 1- \acute{s}_0 \acute{s}_1$ at $\acute{s} = 0$, gives an implicit solution for $\acute{H}$,
\begin{equation}\label{eq:R3cHSol}
    \acute{H} + \mathrm{e}^{-2 \check{t}} \log \left(\frac{\acute{s}_0 \acute{s} + \acute{H}-\mathrm{e}^{-2\check{t}}}{1 - \mathrm{e}^{-2\check{t}}} \right) = 1 -\acute{s}_0(\check{t}) \acute{s}_1(\check{t}) .
\end{equation}
Accounting for the volume contributions from all three regions \RNum{3}$_a$--\RNum{3}$_c$, we can write the volume balance Eq.~(\ref{eq:Mass}) as
\begin{equation}\label{eq:R3Vol}
    \frac{S_l^2}{2} +  \int_{S_l}^{s_0} s ( 1- \mathrm{e}^{-2\mathcal{M}\lambda t})\, \mathrm{d} s +  \int_{s_0}^{S_u} s ( 1-H) \, \mathrm{d} s = \frac{t+ V_0}{2\pi}.
\end{equation}
Here, the first term is the contribution from \RNum{3}$_a$, the second from \RNum{3}$_b$ and the third from \RNum{3}$_c$. Using Eq.~(\ref{eq:R3Sl}) and integrating the second term, Eq.~(\ref{eq:R3Vol}) can be written as
\begin{equation}\label{eq:R3Vol2}
     \frac{1 - \mathrm{e}^{-2\mathcal{M}\lambda t}}{4 \pi \lambda} + \frac{1}{2}\left(s_0^2-\frac{\mathrm{e}^{2\lambda t}- 1}{2\pi \lambda}\right) \left(1 -\mathrm{e}^{-2\mathcal{M} \lambda t} \right) + 2 \int_{s_0}^{S_u} s ( 1-H) \, \mathrm{d} s =  \frac{t+V_0}{2\pi}.
\end{equation}
Substituting the rescaled variables for \RNum{3}$_c$ into this expression, retaining the leading order terms and reverting back to regular coordinates reveals that the leading order approximation of the front position and the upper contact line is then given by
\begin{equation}\label{eq:R3Su}
    S_u(t) \approx \left(\frac{t}{\pi(1 - \mathrm{e}^{-2\mathcal{M} \lambda t})} \right)^{1/2}.
\end{equation}
For $t \ll (\mathcal{M} \lambda)^{-1}$, Eq.~(\ref{eq:R3Su}) shows that the upper contact line is approximately stationary at $S_u \approx (2 \mathcal{M} \lambda)^{-1/2}$, consistent with Eq.~(\ref{eq:R2Su}). At later times, $t \gg (\mathcal{M} \lambda)^{-1}$, Eq.~(\ref{eq:R3Su}) predicts $S_u \approx (t/\pi)^{1/2}$. This confirms that, over the intermediate window $\lambda^{-1} \ll t \ll (\mathcal{M} \lambda)^{-1}$, the upper contact line remains effectively frozen at $S_u \approx (2 \mathcal{M} \lambda)^{-1/2}$.

\subsection{Regime \RNum{4}: evolution of a thin liquid film}
The solution in Eq.~(\ref{eq:R3bEQSol}) shows that $\breve{H} = O(\mathcal{M})$ when $t = O\big(\log(\mathcal{M}^{-1}) (\mathcal{M} \lambda)^{-1}\big)$, signalling the onset of the new regime in which the upper contact line resumes motion. As before, the domain splits into three spatial regions, illustrated in Fig.~\ref{fig:HRegions}. For convenience, we retain the timescale $\check{t} = \mathcal{M}\lambda t$ in this regime.

\textbf{Region \RNum{4}$_a$.} Here, we obtain a simplification of Eq.~(\ref{eq:R3aEQ}) by introducing $H = \mathcal{M} \check{H}(\check{s},\check{t})$. Substituting this into Eq.~(\ref{eq:R3aEQ}) gives
\begin{equation}\label{eq:R4aEQ}
    \check{H}_{\check{t}} + \frac{2 \pi \check{s}^2 + 1}{2 \pi \check{s}} \check{H}_{\check{s}} = - 2 \check{H},
\end{equation}
to leading order. Integrating along the characteristics of Eq.~(\ref{eq:R4aEQ}), yields the solution
\addtocounter{equation}{1}
\begin{equation}\label{eq:R4aEQSol}
    \check{H} = \check{H}_1(\check{\xi}) \mathrm{e}^{- 2 \check{t}}, \qquad
    \check{s} = \left[\frac{ \left(1 + 2\pi \check{\xi}^2\right)\mathrm{e}^{2 \check{t}} - 1 }{2 \pi} \right]^{1/2}.\tag{B28$a,b$}
\end{equation}
Setting $\check{\xi} = 0$ in Eq.~(\mainref{eq:R4aEQSol}{B28$b$}) then specifies $S_l$, yielding the same functional form as in Eq.~(\ref{eq:R3Sl}).

\textbf{Region \RNum{4}$_b$.} Setting $H = \mathcal{M} \breve{H}(\breve{s},\check{t})$ in Eq.~(\ref{eq:R3bEQ}) leaves the equation unchanged
\begin{equation}\label{eq:R4bEQ}
    \breve{H}_{\check{t}}+ \breve{s}\breve{H}_{\breve{s}} = - 2\breve{H}.
\end{equation}
However, as the upper contact line has resumed motion, the interface must now depend on both $s$ and $t$. Integrating along the characteristic equations of Eq.~(\ref{eq:R4bEQ}) and matching with the earlier-time solution Eq.~(\ref{eq:R3bEQSol}) yields
\addtocounter{equation}{1}
\begin{equation}\label{eq:R4bEQSol}
    \check{H} = \mathrm{e}^{-2\check{t}} \quad \text{along} \quad \breve{s} = \xi \mathrm{e}^{\check{t}}.\tag{B30$a,b$}
\end{equation}
Comparison of Eq.~(\mainref{eq:R4bEQSol}{B30$b$}) with Eq.~(\mainref{eq:R4aEQSol}{B28$b$}) shows that the characteristics in region \RNum{4}$_a$ expand more rapidly owing to the additional influence of injection. Consequently, we expect regions \RNum{4}$_a$ and \RNum{4}$_b$ to eventually merge.

\textbf{Region \RNum{4}$_c$.} Downstream of the thin liquid film, the interface continues to advance as a moving front located at $s = (t/\pi)^{1/2}$, as suggested by Eq.~(\ref{eq:R3Su}). Here, we set $t = (\mathcal{M} \lambda)^{-1} \check{t}$, $s = s_0(\check{t})/(\mathcal{M}\lambda)^{1/2} + (\mathcal{M}\lambda)^{1/2}\mathring{s}$, $S_u = \mathring{S}_u(\check{t})$ and $H = s^2/2 - s_0^2/(2\mathcal{M} \lambda) + \mathcal{M} \mathring{H}(\mathring{s},\check{t})$. Substituting into Eq.~(\mainref{eq:Evo2}{33$a$}) with Eq.~(\ref{eq:ParaSmallSlope}), to leading order, we obtain
\begin{subequations}
\label{eq:R4cEQ}
    \begin{align}
        0 &= \left( s_0 \mathring{s} \mathring{H}_{\mathring{s}} -\frac{\mathring{s}}{2\pi(1 - s_0 \mathring{s})}\right)_{\mathring{s}}, \\
        \mathring{S}_u &\approx \frac{s_0}{(\mathcal{M}\lambda)^{1/2}} + \frac{(\mathcal{M}\lambda)^{1/2}}{s_0},
    \end{align}
\end{subequations}
where $s_0(\check{t})$ is to be determined by global conservation of volume. In this short region Eq.~(\mainref{eq:R4cEQ}{B31$a$}) shows that the interface adjusts quasi-steadily through the buoyancy term in response to the propagation of the front. 

Since the regions away from the front are characterised by a thin liquid film of $H = O(\mathcal{M})$, we can write the volume constraint Eq.~(\ref{eq:Mass}) as
\begin{equation}\label{eq:R4Vol}
    \int_0^{s_0} s \,\mathrm{d}s + \int_{s_0}^{S_u} s(1-H) \,\mathrm{d}s = \frac{t + V_0}{2\pi}.
\end{equation}
Substituting the rescaled variables for this region into Eq.~(\ref{eq:R4Vol}), integrating and retaining the leading order terms reveals
\begin{equation}\label{eq:R4s0}
    s_0 = \left({\check{t}}/{\pi}\right)^{1/2}.
\end{equation}
Combining Eq.~(\ref{eq:R4s0}) with Eq.~(\mainref{eq:R4cEQ}{B31$b$}) yields the approximation for the upper contact line:
\begin{equation}\label{eq:R4Su}
    \mathring{S}_u(\check{t}) \approx \left(\frac{\check{t}}{\pi \mathcal{M}\lambda}\right)^{1/2} + \left(\frac{\pi \mathcal{M}\lambda}{\check{t}}\right)^{1/2},
\end{equation}
which is consistent with the solution given by Eq.~(\ref{eq:R3Su}) for $t \gg (\mathcal{M} \lambda)^{-1}$, to leading order in $t$. Substituting Eq.~(\ref{eq:R4s0}) into Eq.~(\mainref{eq:R4cEQ}{B31$a$}), we obtain
\begin{equation}\label{eq:R4cEQ2}
    0 = \left( \check{t}  \mathring{s} \mathring{H}_{\mathring{s}}  - \frac{\check{t}^{1/2}  \mathring{s}}{2 \pi(1 - \check{t}^{1/2} \mathring{s})} \right)_{\mathring{s}}.
\end{equation}
Integrating Eq.~(\ref{eq:R4cEQ2}) twice and matching with Eq.~(\mainref{eq:R4bEQSol}{B30$a$}), yields the solution 
\begin{equation}\label{eq:R4cHSol}
    \mathring{H} = \frac{\mathrm{e}^{-2 \check{t}}}{\mathcal{M}} - \frac{1}{2 \check{t}} \log(\pi^{1/2}-\check{t}^{1/2} \mathring{s}), \qquad (0 \leq \mathring{s} < \check{t}^{-1/2}).
\end{equation}
Equation (\ref{eq:R4cHSol}) can be combined with solutions given in Eqs~(\mainref{eq:R4aEQSol}{B28$a$}) and (\mainref{eq:R4bEQSol}{B30$a$}) to recover a uniformly valid approximation for $H$.

\subsection{Regime \RNum{5}: evolution of the moving front}
As the liquid film continues to drain, the interface height decays exponentially according to Eq.~(\mainref{eq:R4aEQSol}{B28$a$}). The lower contact line solution in Eq.~(\ref{eq:R3Sl}) increases exponentially in time, whereas the upper contact line given by Eq.~(\ref{eq:R4Su}) increases algebraically. Consequently, the lower contact line will eventually catch up with the advancing front. Volume conservation ensures that the leading-order location of the front is still $(t/\pi)^{1/2}$. This coalescence occurs when the lower contact line Eq.~(\ref{eq:R3Sl}) matches the position of the moving front. Equating these expressions determines the time, $t_c$, at which the thin liquid film has completely drained and the interface evolves as a single continuous moving front. The corresponding time is then given by the root of the transcendental equation
\begin{equation}\label{eq:R5t}
\lambda t =  \left( \frac{\mathrm{e}^{2 \mathcal{M} \lambda t} - 1}{2} \right).
\end{equation}
Beyond this time ($t > t_c$), we set $t = (\mathcal{M} \lambda)^{-1} \check{t}$ and rescale into a region local to the front with $s = \check{t}^{1/2}/(\mathcal{M}\lambda)^{1/2} + \left[(\mathcal{M}\lambda)^{1/2}/\check{t}^{1/2} \right]\zeta$ and $H = f(\zeta)$. Substituting these into Eq.~(\mainref{eq:Evo2}{33$a$}), to leading order we recover
\begin{equation}
    \left[f (f_{\zeta} -1) \right]_{\zeta} = 0.
\end{equation}
Integrating this equation yields
\addtocounter{equation}{1}
\begin{equation}\label{eq:R5Sol1}
    f = \zeta - \zeta_l, \qquad \zeta_u - \zeta_l = 1,\tag{B39$a,b$}
\end{equation}
where $\zeta_l$ and $\zeta_u$ are to be determined by conservation of volume. To leading order, volume conservation, Eq.~(\ref{eq:Mass}), becomes 
\begin{equation}\label{eq:R5Vol}
    \zeta_l + \int_{\zeta_l}^{\zeta_u} ( 1- \zeta + \zeta_l) \, \mathrm{d}\zeta = \frac{V_0}{2\pi},
\end{equation}
Integrating this expression and using Eq.~(\mainref{eq:R5Sol1}{B39$b$}) gives
\begin{equation}
    \zeta_l = \frac{V_0}{2\pi} - \frac{1}{2}, \qquad \zeta_u = \frac{V_0}{2\pi} + \frac{1}{2}.
\end{equation}
The interface height can then be written as
\begin{equation}\label{eq:R5H}
    H = \left(\frac{t}{\pi}\right)^{1/2} s - \frac{t}{\pi} + \frac{1}{2} - \frac{V_0}{2\pi},
\end{equation}
and the contact lines propagate according to 
\addtocounter{equation}{1}
\begin{equation}\label{eq:R5CLs}
S_l(t) = \left(\frac{t}{\pi}\right)^{1/2} - \left(\frac{\pi}{4 t}\right)^{1/2}\left(1-\frac{V_0}{2\pi}\right), \qquad S_u(t) =  \left(\frac{t}{\pi}\right)^{1/2} + \left(\frac{\pi}{4 t}\right)^{1/2}\left(1+\frac{V_0}{2\pi}\right).\tag{B43$a,b$}
\end{equation} 

\section{Gaussian channel: asymptotics}
\label{App:GaussianAsymp}
Here, we present an asymptotic analysis of a Gaussian-shaped channel in the limit $\mathcal{M} \ll 1$. We first consider the distinguished limit $\beta \equiv \mathcal{M}^2 \lambda \sim O(1)$ in which buoyancy and injection balance. This distinguished limit serves as the organising regime, from which additional asymptotic sublimits may subsequently be recovered.
\subsection{Early times}
At early times when $H(0,t) > 0$, in the limit $\mathcal{M} \ll1$ with $\beta = O(1)$, or equivalently $\lambda \sim \mathcal{M}^{-2}$, % the channel curvature allows 
the interface can relax to its energetically preferred configuration: a horizontal interface in physical space. This interface subsequently translates downward at a rate set by the injection strength. Taking this limit in Eq.~(\mainref{eq:Evo2}{33$a$}) with Eq.~(\ref{eq:Bellturn}), yields
\begin{equation}\label{eq:GaussET}
\left[ s H (1-H)\bigl(H_s -s \mathrm{e}^{-s^2/2}\bigr) \right]_s = 0,
\end{equation}
to leading order. Here the time derivative is subdominant at $O((\lambda\mathcal{M})^{-1}$), while the injection term enters at $O(\lambda^{-1})$. Physically, the interface adjusts quasi-steadily in response to the imposed injection. While $H(0,t) > 0$ and $S_l(t) = 0$, Eq.~(\ref{eq:GaussET}) has the solution
\begin{equation}\label{eq:GaussETSol}
H = 1 + \mathrm{e}^{-S_u^2/2} - \mathrm{e}^{-s^2/2},
\end{equation}
which corresponds to a horizontal interface in physical space. The evolution of the upper contact line follows from the global volume constraint Eq.~(\ref{eq:Mass}). Substituting Eq.~(\ref{eq:GaussETSol}) into Eq.~(\ref{eq:Mass}) and integrating yields the transcendental relation
\begin{equation}\label{eq:GaussETSu}
1 - \mathrm{e}^{-S_u^2/2}\left(1 + \frac{S_u^2}{2}\right) = V_0 + t ,
\end{equation}
determining the early-time evolution of $S_u$ prior to the formation of the lower contact line.

\subsection{Late times}
At later times, once the lower contact line has formed, {inner and outer} spatial regions emerge when
$\beta \equiv \mathcal{M}^2 \lambda \sim O(1)$. The inner region near $S_l$ consists of a horizontal interface, and in the outer region a thin film of gas spreads over long length scales.
\subsubsection{Inner region}
Once the lower contact line has formed, a short inner region where $s = O(1)$ continues to be described by Eq.~(\ref{eq:GaussET}). Integrating this equation, but now applying the boundary condition $H = 0$ at $s = S_l(t)$, yields the solution
\begin{equation}\label{eq:GaussLTInnEq}
    H = \mathrm{e}^{-S_l^2/2} - \mathrm{e}^{-s^2/2},
\end{equation}
which is a horizontal interface in physical space, with the lower contact line $S_l(t)$ so far undetermined. Equation (\ref{eq:GaussLTInnEq}) has the limit $H \approx \mathrm{e}^{-S_l^2/2}$ for $s \gg 1$.

\subsubsection{Outer region}
\label{App:GaussOut}
Analogously to region \RNum{1}$_b$ in a parabolic channel, the outer region consists of a thin gas film that spreads along the upper boundary. We therefore adopt the same scalings as in region \RNum{1}$_b$, setting $s = \hat{s}/\mathcal{M}^{1/2}$, $S_u = \hat{S}_u/\mathcal{M}^{1/2}$ and $H = 1 - \mathcal{M}\hat{H}(\hat{s},t)$ in Eqs~(\mainref{eq:Evo2}{33$a$--$e$}) together with Eq.~(\ref{eq:Bellturn}).
To leading order, the evolution of the film is governed by a mixed nonlinear hyperbolic--parabolic equation,
\begin{equation}\label{eq:GaussLTOut}
    \hat{H}_{\hat{t}} + \frac{1}{2 \pi \hat{s}(1+\hat{H})^2}\,\hat{H}_{\hat{s}} = \frac{\beta}{\hat{s}} \left(\frac{\hat{s} \hat{H} \hat{H}_{\hat{s}}}{1+\hat{H}} \right)_{\hat{s}},
\end{equation}
which is analogous to Eq.~(\ref{eq:R1bEQ}), but with buoyancy effects retained. Equation (\ref{eq:GaussLTOut}) admits a similarity solution. Introducing the similarity variables $\eta = \hat{s}/t^{1/2}$ and $\hat{H} = f(\eta)$ in Eq.~(\ref{eq:GaussLTOut}), produces an ODE for $f(\eta)$:
\begin{subequations}\label{eq:GaussLTOut2}
    \begin{align}
        \left(\frac{1}{2\pi \eta(1+f)^2} - \frac{\eta}{2} \right) f_\eta &= \frac{\beta}{\eta} \left(\frac{\eta f f_{\eta}} {1+f}\right)_\eta, \quad &&(0 \leq \eta \leq \eta_u), \\
        f &= 0, \quad &&(\eta = \eta_u),
    \end{align}
\end{subequations}
where $\eta_u$ determines the upper contact line location through $\hat{S}_u = \eta_u t^{1/2}$. The constant $\eta_u$ is determined by global volume conservation Eq.~(\ref{eq:Mass}); in the similarity variables, this constraint can be expressed as
\begin{equation}\label{eq:SimMass}
    \int_0^{\eta_u} \eta f \, \mathrm{d} \eta = \frac{1}{2\pi}.
\end{equation}

Equation (\mainref{eq:GaussLTOut2}{C6$a$}) has the following asymptotic limits:
\begin{subequations}\label{eq:GaussLTOutLims}
    \begin{align}
        f &= \left(-\frac{\log(\eta)}{{\pi\beta }}\right)^{1/2}, \quad &&(\eta \to 0), \\
        f &= \frac{1}{2\beta} \left(\eta_u-\eta \right) \left( \frac{1}{\pi \eta_u} - \eta_u \right), \quad &&(\eta \to \eta_u).
    \end{align}
\end{subequations}
The logarithmic limit Eq.~(\mainref{eq:GaussLTOutLims}{C8$a$}) of the outer solution is incompatible with the inner solution Eq.~(\ref{eq:GaussLTInnEq}), indicating the presence of a passive transition layer that allows matching between them, which we do not investigate further. To reformulate Eq.~(\ref{eq:GaussLTOut2}) as a first-order system, we introduce the dependent variables
\begin{equation}
    y_1(\eta) = f, \quad y_2(\eta) = f_{\eta}, \quad y_3(\eta) = \int_{\eta}^{\eta_u} \eta^{\prime} y_1\,  \mathrm{d} \eta^{\prime},
\end{equation}
where $y_3$ represents the cumulative contribution to the mass integral from $\eta$ to the upper contact line. Substitution of these variables into Eq.~(\ref{eq:GaussLTOut2}), together with the mass constraint Eq.~(\ref{eq:SimMass}), yields the first-order system
\begin{subequations}\label{eq:ODEsystem}
    \begin{align}
        y_{1,\eta} &= y_2, \\
        y_{2,\eta} &=  \frac{y_2}{\beta y_1} \left( \frac{1}{2\pi\eta(1 + y_1)} -\frac{\eta(1 + y_1)}{2}\right)- \frac{y_2}{\eta}- \frac{y_2^2}{y_1}+ \frac{y_2^2}{1 + y_1}, \\
        y_{3,\eta} &= -\eta y_1.
    \end{align}
\end{subequations}
The limits in Eqs~(\mainref{eq:GaussLTOutLims}{C8$a,b$}), together with the mass constraint in Eq.~(\ref{eq:SimMass}) then provide the boundary conditions for the system Eqs~(\mainref{eq:ODEsystem}{C10$a$--$c$}):
\begin{subequations}\label{eq:ODEsystemLimits}
    \begin{align}
        y_1 = \frac{(\eta_u - \eta)(1-\pi \eta_u^2) }{2 \pi\beta \eta_u} , \quad y_2 &= -\frac{(1-\pi \eta_u^2)}{2 \pi\beta \eta_u}, \quad y_3 = - \frac{\left(\eta_u-\eta\right)^2 (1-\pi \eta_u^2)}{4 \pi \beta }, \quad (\eta \to \eta_u), \\
        y_3 &= \frac{1}{2\pi}, \quad (\eta \to 0).
    \end{align}
\end{subequations}
The system Eqs~(\mainref{eq:ODEsystem}{C10$a$--$c$}) and (\mainref{eq:ODEsystemLimits}{C11$a,b$}) constitute an eigenvalue problem for the unknown eigenvalue $\eta_u$ and can be solved using a shooting method. The equations are initialised near $\eta=\eta_u$ using the limits in Eq.~(\mainref{eq:ODEsystemLimits}{C11$a$}) and integrated towards $\eta=0$. We integrate on the domain $ \delta \leq \eta \leq \eta_u - \delta$ for $\delta \ll 1$, since Eq.~(\mainref{eq:ODEsystem}{C10$b$}) is singular at $\eta = 0$ and $\eta = \eta_u$. The value of $\eta_u$ is adjusted iteratively until the mass constraint Eq.~(\mainref{eq:ODEsystemLimits}{C11$b$}) is satisfied. 

Setting $\beta \ll 1$ in Eq.~(\ref{eq:GaussLTOut2}), we recover the solution obtained by \citet{guo_axisymmetric_2016} and others
\begin{equation}\label{eq:GaussInj}
    f(\eta) = \left(\frac{1}{\pi \eta^2}\right)^{1/2} - 1, \quad \eta_u = \frac{1}{\sqrt{\pi}}.
\end{equation}
Here, buoyancy no longer influences the leading-order dynamics of the outer region; the flow is therefore predominantly driven by injection.

For $\beta \gg 1$, setting $f= \beta^{-1/2} \xi_u^2 \,g(x \equiv \xi/\xi_u)$ and $\eta=\beta^{1/4}\xi$, where $\xi_u = \beta^{-1/4} \eta_u$ in Eq.~(\ref{eq:GaussLTOut2}) and Eq.~(\ref{eq:SimMass}), yields
\begin{subequations}\label{eq:PorMed}
    \begin{align}
        0 &= x^2 g_{x} + 2\left( x g g_{x} \right)_{x}, \quad &&(0 < x \leq 1), \\
        g &= 0, \quad &&(x = 1),\\
        \xi_u &= \left( \int_0^1 x g \, \mathrm{d} x \right)^{-1/4},
    \end{align}
\end{subequations}
to leading order. Equation (\mainref{eq:PorMed}{C13$a$}) is the similarity form of the well-studied porous–medium equation for flow in a deep, effectively unconfined channel subject to a constant injection \citep{lyle_axisymmetric_2005}. In this limit the flow is primarily buoyancy-driven, with the contribution from injection entering at $O(\beta^{-1/2})$. Eqs~(\mainref{eq:PorMed}{C13$a$--$c$}) were addressed by \citet{guo_axisymmetric_2016} for buoyancy-driven flow in an axisymmetric, flat channel. As shown in \citet{lyle_axisymmetric_2005}, Eq.~(\mainref{eq:PorMed}{C13$a$}) has the limits
\begin{subequations}\label{eq:PorMedLimits}
    \begin{align}
        g &= C \left( - \log(x) \right)^{1/2}, \quad &&(x \to 0), \\
        g &= \frac{1}{2}(1-x), \quad &&(x \to 1),
    \end{align}
\end{subequations}
for a constant $C$. Equation (\mainref{eq:PorMed}{C13$a$}) can be solved numerically by integrating inwards from $x=1$ and using Eq.~(\mainref{eq:PorMedLimits}{C14$b$}) to initialise.

\end{document}